%% file: 00_pump.tex
\documentclass[twocolumn]{article}
\usepackage[utf8]{inputenc}
\usepackage[T1]{fontenc}
\usepackage[UKenglish]{babel}
\usepackage[english]{varioref}
\usepackage[nottoc]{tocbibind}
\usepackage[toc,page]{appendix}
\usepackage{caption}
\usepackage{amsmath}
\usepackage{amsthm}
\usepackage{amssymb}

\usepackage{chngcntr}  

\usepackage{cuted}

\usepackage{mathrsfs}
\usepackage{geometry}
\usepackage{float}
\usepackage[caption = false]{subfig}

\usepackage{graphicx}
\usepackage{array}

\usepackage{xcolor}

\usepackage{hyperref}
\hypersetup{
    colorlinks = True,
    linkcolor = blue,
    citecolor = magenta
}
\usepackage[font = small,labelfont = bf]{caption}
\usepackage{booktabs}

\usepackage{csquotes}

\usepackage{authblk}
\usepackage{subfiles}
\usepackage{graphbox}

\usepackage[table, svgnames, dvipsnames]{xcolor}
\definecolor{deepskyblue}{RGB}{0,191,255}

\usepackage{tikz}

\usetikzlibrary{arrows.meta, positioning, shapes.misc, calc}

\definecolor{MyGreenish}{RGB}{150,210,104}
\definecolor{MyLightGreenYellow}{RGB}{215, 241, 145}
\definecolor{MyYellowOrange}{RGB}{253, 202, 121}
\definecolor{MyReddishOrange}{RGB}{246, 122, 73}
\definecolor{Silver}{RGB}{220,220,220}

\usepackage{pdflscape} 
\usepackage{booktabs}   
\usepackage{ragged2e}   
\usepackage{calc}       

\usepackage{multirow}

\usepackage{titlesec}

\titleformat*{\section}{\bfseries}
\titleformat*{\subsection}{\itshape}
\titleformat*{\subsubsection}{\itshape}

\titleformat{\paragraph}[runin]
  {\normalfont\itshape}   
  {\theparagraph.}       
  {1em}
  {}

\newcommand{\absavg}[1]{\langle\,|#1|\,\rangle}

\newcommand{\GammaTor}{\Gamma_{\phi}}

\newcommand{\vVec}{\mathbf{V}}
\newcommand{\vTorVec}{\mathbf{V}_{\phi}}
\newcommand{\vTor}{V_{\phi}}
\newcommand{\vTorBC}{V_{\phi}^{\text{BC}}}
\newcommand{\vTorSep}{V_{\phi}^{\text{SEP}}}
\newcommand{\vTorInfty}{V_{\phi}^{\infty}}
\newcommand{\vTorWall}{V_{\phi}^{\text{dome}}}
\newcommand{\vPolVec}{{\mathbf{V}}_{\theta}}

\newcommand{\vR}{{V}_{r}}
\newcommand{\vRVec}{{\mathbf{V}}_{r}}
\newcommand{\vZ}{{V}_{z}}
\newcommand{\vZVec}{{\mathbf{V}}_{z}}

\newcommand{\pTor}{p_{\phi}}
\newcommand{\pPol}{p_{\theta}}
\newcommand{\pGain}{\mathcal{G}_{p}}

\newcommand{\BEIZ}{\mathbb{P}_{\circlearrowleft}}

\newcommand{\TplasmaBC}{T^{\text{BC}}}

\newcommand{\uHatTorVec}{\hat{\mathbf{u}}_{\phi}}

\newcommand{\lambdaPfr}{\lambda_{\text{PFR}}}
\newcommand{\lambdaDuct}{\lambda_{\phi}}
\newcommand{\lambdaWall}{\lambda_{\text{dome}}}

\newcommand{\DoT}{\text{DoT}}
\newcommand{\Dzero}{\text{D}}
\newcommand{\DzeroMol}{\Dzero_2}
\newcommand{\DzeroEq}{\Dzero_{\text{eq}}}

\newcommand{\GammaCirc}{\Gamma_{\circlearrowleft}}

\newcommand{\accT}{\alpha_t}
\newcommand{\accN}{\alpha_n}

\newcommand{\etaCapture}{\eta_{\text{r}}}

\title{\textbf{Harnessing Toroidal Neutral Flows to Enhance Divertor Particle Exhaust}}

\author[1]{M. Moscheni*}
\author[1]{A. Herrmann}
\author[2]{J.D. Lore}
\author[3,4]{M. Kryjak}
\author[1]{C. Soika}
\author[5]{R. Kembleton}
\author[1]{S. Lazerson}
\author[1]{K. Revel}
\author[1]{the Gauss Fusion Team}

\affil[1]{Gauss Fusion GmbH, Parkring 29, 85748 Garching bei M\"unchen, Germany}
\affil[2]{Oak Ridge National Laboratory, Oak Ridge, TN 37831, United States of America}
\affil[3]{United Kingdom Atomic Energy Authority, Culham Campus, OX14 3DB Abingdon, United Kingdom of Great Britain and Northern Ireland}
\affil[4]{University of York, York Plasma Institute, Department of Phyics, Engineering and Technology, YO10 5DD York, United Kingdom of Great Britain and Northern Ireland}
\affil[5]{RI Research Instruments GmbH, Friedrich-Ebert-Stra{\ss}e 75, 51429 Bergisch Gladbach, Germany}

\date{}
\begin{document}

\twocolumn[
\begin{@twocolumnfalse}

\maketitle

\begin{abstract}

In 1991 Reiter \textit{et al.} (1991 \textit{Plasma Phys. Control. Fusion} \textbf{33} 1579) considered the onerous exhaust requirements of ITER, and wrote: ``The vacuum pumping problem of a fusion reactor will probably require some novel solution''. Here we show that a toroidally-oriented pump inlet can passively exploit intrinsic neutral flows to reduce back-flow, raise duct pressure, and ultimately improve particle-exhaust performance.
Drawing on previous experimental observations and SOLPS-ITER edge-plasma simulations, we consolidate the evidence for a plasma-imprinted, multi-species toroidal neutral ``wind'' in detached tokamak divertors. We isolate the underlying mechanism in a prototypical divertor private-flux region using a database of two-dimensional direct simulation Monte Carlo (DSMC) calculations. The ordered neutral motion is recovered---with a strong toroidal alignment, kilometre-per-second velocities, and persistence up to several centimetres across slip-to-transitional rarefied regimes ($\mathrm{Kn} \simeq0.02$--$2$).
We then assess the consequences of capturing this ordered motion by using a second database of idealised proof-of-principle DSMC simulations. Compared to the traditional poloidal arrangement, a toroidally-oriented pump inlet reduces back-flow by up to $20\%$ for deuterium and up to $33\%$ for helium at $10\%$ concentration. Partial pressures in the toroidal exhaust path are enhanced across the database---nominally by a factor of $1.78\pm0.04$ for deuterium and $2.00\pm0.05$ for helium. For fixed throughput, this implies a reduction in the required effective pumping speed and corresponding hardware.
More broadly, these results motivate explicit retention of toroidal neutral momentum in divertor and sub-divertor modelling, and dedicated studies of neutral aerodynamics---including in stellarators, where an analogous directional imprinting is expected to occur.

\end{abstract}

\small{*email: \textcolor{blue}{matteo.moscheni@gauss-fusion.com}}

\vspace{1em}

\end{@twocolumnfalse}
]

\section{Introduction}\label{sec:intro}

Among the challenges that must be addressed to realise nuclear fusion energy is the robust, reactor-compatible control of the plasma--wall interaction (PWI) and of its many coupled consequences \cite{donne2019, PITTS2019100696, stangeby2000plasma, Militello_2022}. At present, divertor detachment has emerged as the most promising operating regime for reducing plasma heat and particle loads to material surfaces \cite{Krasheninnikov_Kukushkin_2017, Soukhanovskii_2017}. In detached conditions, coordinated injection of fuel species---deuterium and tritium---and extrinsic impurities---typically noble gases---promotes particle, momentum, and energy losses in the edge plasma \cite{Kallenbach_2013}. The plasma is extinguished, and neutral particles populate the plasma-facing region. Although such neutrals represent unconfined fuel or ash, and therefore do not directly contribute to fusion power production, their presence is essential, in particular in the divertor \cite{YOU2022113010}---where the inlets to the pumping exhaust system are located.

While detachment alleviates divertor material constraints, the associated requirements on particle throughput, exhaust, and the fuel cycle remain severe in both tokamaks and stellarators \cite{Barbarino_2020, IAEA_TECDOC_2049_2024}---and have recently been shown to be even more demanding than previously anticipated \cite{Meschini_Moscheni_2026}. Reactor-relevant studies indicate the need for high fuel---including tritium---puffing rates \cite{Moscheni_2026}, imposing an onerous burden on the pumping system. Lower-throughput scenarios may be conceivable, but they risk degrading helium ash exhaust \cite{PITTS2019100696, IAEA_TECDOC_2049_2024, Park_2024}, which must be continuously maintained to balance helium production in the burning plasma \cite{Reiter_1991}. Taken together, these constraints imply large effective pumping capacities implemented through bulky, non-shielding hardware, embedded within a highly integrated and radioactive environment where space, access, shielding, and material compatibility are all severely limited \cite{Pitcher_1997, YOU2022113010}.

Particle exhaust therefore constitutes a central challenge arising from PWI in magnetic-confinement fusion devices \cite{Eich_2026}. If a detailed understanding of \textit{plasma} flows is ``crucial for designing an effective divertor geometry and divertor pumping system'' \cite{Asakura_2000}, a role at least as critical must be assigned to \textit{neutral} flows. Ultimately, it is the rarefied neutral population---not the plasma itself---that is captured, transmitted, and removed by the vacuum pumps. Neutral transport therefore forms the kinetic bottleneck of the exhaust pipeline, linking plasma-edge physics, divertor geometry, rarefied-gas dynamics, and engineering pumping performance.

Over several decades of divertor optimisation, a wide range of geometric features has been developed to passively facilitate both detachment and exhaust \cite{Loarte_2001, Loarte_2007}. These include strong two-dimensional \textit{poloidal} shaping of the divertor volume, \textit{poloidal} optimisation of pump-inlet placement \cite{Sang_2021, YU2024101826, BI2023113916, HOLM2024101782}, \textit{poloidal} closure through baffles and domes \cite{FEVRIER2021100977, Sang_2017, Reimerdes_2022, KUKUSHKIN2007308, Cowley_2026}, and horizontal or vertical target plates designed to tailor \textit{poloidal} neutral pathways and exhaust \cite{SHAFER2019487, GROTH2015471, Moulton_2018, Loarte_2001}. Many plasma-facing features are necessarily constrained to be as toroidally-symmetric as possible, in order to avoid localised heat loads \cite{PITTS201760, Looby02012022}. By contrast, the same restriction is not fundamental for non-plasma-contacting divertor structures, including baffles, domes, and divertor-facing pump-inlet ducts. Nevertheless---and even in stellarators \cite{Grigull_2001, Renner_2002, Wolf_2019, Feng_2021, Ohyabu_1994, Bader_2017, Bader_2018, Garcia_2023, Garcia_2025}, their design space has remained largely restricted to two-dimensional, axisymmetric concepts. The deliberate use of three-dimensional, macroscopic, \textit{toroidal} shaping to influence neutral transport has therefore remained essentially unexplored.

This omission is important because---in addition to unavoidable toroidally-localised features \cite{Lore_2015, LORE2015515, Xie_2025}---neutral motion in detached divertor conditions is not necessarily isotropic, nor merely pressure-driven \cite{Stangeby_1993, Stacey_1995}. A concentrated body of work from the mid-1990s to early 2000s reported ordered \textit{toroidal} neutral motion in the divertor: a macroscopic, directed rarefied-gas flow with characteristic velocities of order kilometres per second \cite{Pitcher_1997, Krasheninnikov_1996, Knoll_1996, Vahala_1996, Vahala_1997, Knoll_1998, PITCHER19991009, Goetz_1999, Welch_2001}. Since then, this observation has largely receded from the divertor-exhaust discourse, reappearing only through sporadic numerical \cite{ROZHANSKY2022101316, GIACOMIN2022111294} or diagnostic \cite{Gradic_2018} results, and without being systematically exploited as a particle-exhaust principle. Consequently, despite decades of exhaust-system studies, reactor pumping concepts have continued to be scoped primarily within two broad philosophies \cite{Loarte_2001, Havlickova_2014, Overskei_1981, TEXTORTEAM1989115, Maingi_1999, Bader}---neither explicitly using the directed toroidal component of neutral motion as an input to pump-inlet geometry.

In the present work, we examine this possibility. We propose and numerically test a toroidally-oriented pumping paradigm that treats ordered toroidal neutral motion as a neutral-transport resource, rather than as a feature to be averaged away in axisymmetric exhaust design. The objective is not to deliver an integrated reactor pump design, but to establish a meaningful proof-of-principle.

The paper is structured as follows. Section~\ref{sec:evidence} collates and critically analyses the available theoretical, numerical, and experimental evidence for the existence of toroidal neutral flows, including its limitations. Section~\ref{sec:philosophies} reviews the established pumping philosophies, highlighting their distinguishing features, advantages, and limitations. In section~\ref{sec:tfp}, we introduce the toroidal-flow pumping concept and its relation to ordered neutral motion. Section~\ref{sec:methods} describes the numerical methods used to isolate the relevant physics in proof-of-principle rarefied-gas simulations. The results are presented in section~\ref{sec:results} and discussed in section~\ref{sec:discussion}, with particular attention to the neutral-plasma interactions, and integrated exhaust performance in different types of reactor. Finally, section~\ref{sec:conclusion} summarises the conclusions of this work and outlines the key developments required to advance this approach further---a first step along a new research line centred on rarefied divertor aerodynamics.

\section{Motivating evidence for toroidal neutral flows in fusion devices}\label{sec:evidence}

This section reviews the theoretical, experimental, and numerical evidence for toroidal neutral flows in fusion devices. The aim is to establish the physical basis for the development of the new concept, which hinges on a consequential but still under-recognised aspect of plasma--neutral dynamics.

\subsection{Theoretical and phenomenological considerations}\label{sec:evidence__theory}

A number of studies have discussed the existence of strong, ordered, toroidal neutral motion in the scrape-off layer (SOL) and divertor of tokamak plasmas \cite{Pitcher_1997,LABOMBARD1997149,Lipschultz01042007,BRUNNER2013S1196}. This is sometimes referred to as a toroidal neutral ``wind'' \cite{BRUNNER2013S1196}. The key physical elements underlying this phenomenon are summarised below, expanding on the reasoning of \cite{Pitcher_1997,Vahala_1996, Vahala_1997,Krasheninnikov_1996}.

In toroidal magnetic confinement devices, the magnetic field is predominantly toroidal, with typical values $B_{\phi}/B \gtrsim 0.9$ in the divertor region (see, e.g., figure~2 of \cite{Frerichs_2024}). In a fluid description, plasma transported across the separatrix by turbulence subsequently flows primarily along magnetic field lines \cite{Walkden2022Turbulence,QUADRI2024101756,Oliveira_2022}. Along each flux tube and sufficiently close to the detachment front, or to the divertor target if in attached conditions, the parallel plasma flow must be directed towards the divertor over the full radial extent of the SOL \cite{Loarte_2001}. Flow reversal can occur upstream \cite{Neuhauser_1984,KRASHENINNIKOV1992899,Zito_2025}, and may be particularly pronounced for higher-$Z$ impurity species. However, the Bohm--Chodura sheath criterion enforces a net flow towards the target for each plasma species sufficiently close to the sheath \cite{Bohm_1949, Chodura_1982, Riemann_1991, Stangeby_1995}.

Although drifts introduce asymmetries and poloidal structure \cite{Stangeby_1996,Chankin_2015,Hammond_2019,Kaveeva_2020,Killer_2025,Kriete_2023}, and temperature and finite-ion-orbit effects may further broaden the local velocity distribution \cite{Chang_2017}, the dominant component of the mean plasma momentum remains toroidal---reflecting the helicity of the magnetic field. As a result, the net plasma momentum incident on the divertor region is strongly toroidally-directed, with opposite toroidal signs expected at opposite strike points \cite{Pitcher_1997}.

Under detached conditions, plasma momentum is primarily transferred to neutrals through charge exchange (CX) and recombination (RC) \cite{SCHNEIDER1999175,Isler_1999,Asakura_2000,Loarte_2001,KRASHENINNIKOV1992899}, anisotropising the neutral velocity distribution \cite{Stacey_1995} and its pressure tensor \cite{Stangeby_1993}. In both CX and RC---``non-collided'' neutrals in the terminology of \cite{Stacey_1995}, the electron contribution to momentum is negligible. The newly formed neutral retains nearly the full incoming ion velocity. By conservation of momentum, neutrals formed in detached plasmas necessarily acquire a directed \textit{toroidal} velocity, superimposed on the neutral random thermal motion. This non-collided, primary, component is therefore expected to be ubiquitous wherever detachment occurs, largely independent of the toroidal extent or geometric complexity of the tokamak---or stellarator~\cite{Grigull_2001, Renner_2002, Wolf_2019, Feng_2021, Ohyabu_1994, Bader_2017, Bader_2018, Garcia_2023, Garcia_2025}---divertor under consideration.

As noted by Knoll \textit{et al.} \cite{Knoll_1996}, since neutrals move across the magnetic field much more freely than ions, this in turn produces significant cross-field secondary transport of parallel momentum to ``collided'' neutrals \cite{Stacey_1995}, owing to elastic (EL) collisions. A collective and spatially-extended toroidal neutral motion can therefore arise.

\begin{figure*}
    \centering
    \includegraphics[width = 0.75\textwidth]{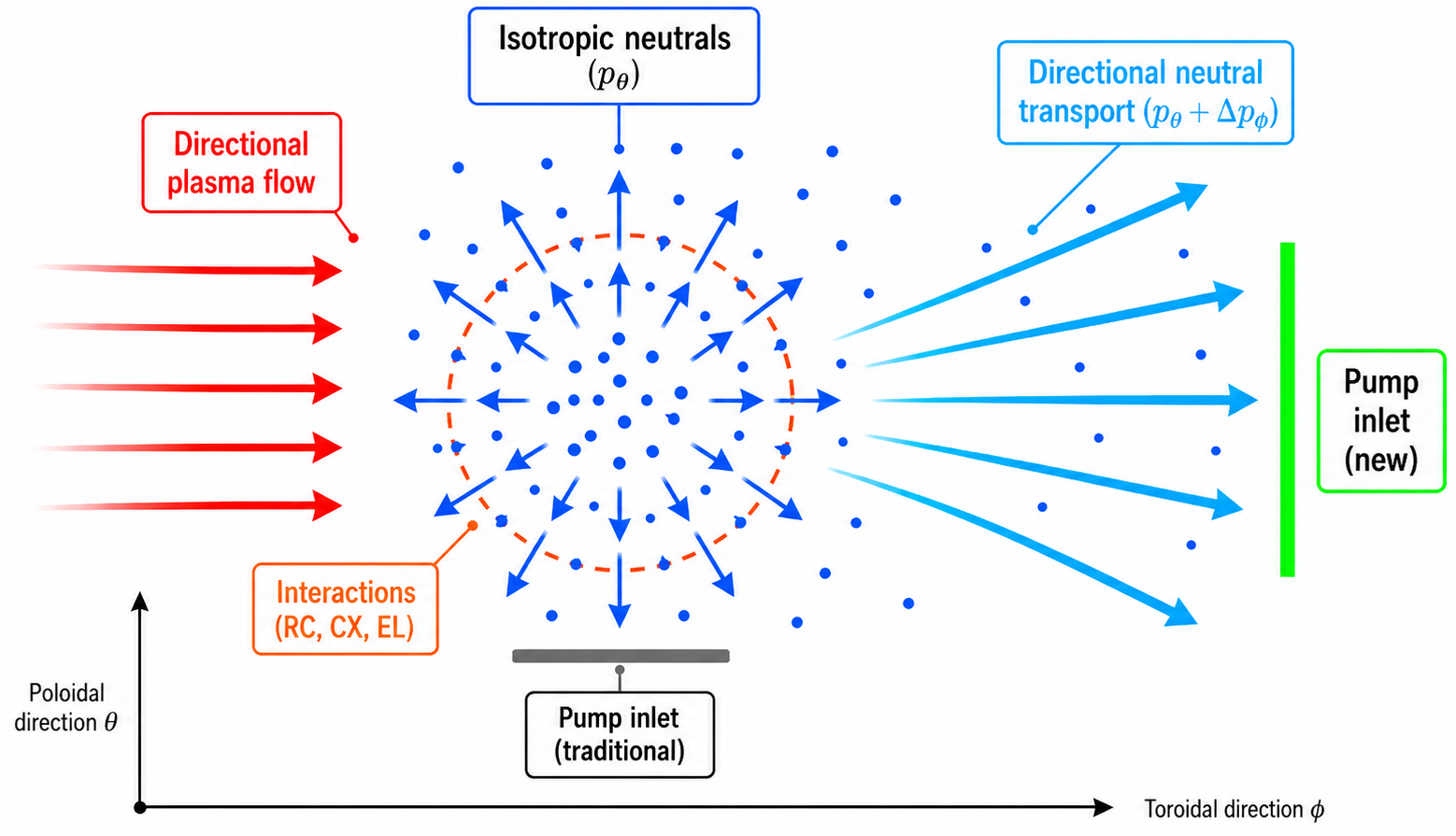}
    \caption{Simplified schematic of toroidal neutral flow generation in the divertor (not to scale). Isotropic, pseudo-random neutral thermal motion carries a pressure $p_{\theta}$, tapped by the traditional, poloidal arrangement of pump inlets. However, ordered plasma momentum---primarily aligned with the magnetic field---is transferred to neutrals through charge exchange (CX), recombination (RC) and elastic collisions (EL). A toroidal neutral flow is generated, which carries a surplus $\Delta p_{\phi}$ which a toroidally-oriented pump inlet can access (lime).}
    \label{fig:overview_wind}
\end{figure*}

An illustrative sketch of the overall dynamics is pictured in figure \ref{fig:overview_wind}.

Impurity species follow the same qualitative dynamics with quantitative differences. RC remains effective in detached plasmas, whereas impurity CX and EL with deuterium ions are generally less efficient because of mass-difference and cross-section effects. Neutral--neutral EL collisions, however, remain significant and provide a mechanism for flow entrainment \cite{Zito_2025}---expected to increase in larger machines. This favours the propagation of toroidally-directed motion from the majority hydrogenic species to minority impurity neutrals.

Experimentally separating toroidally-directed collided/primary and non-collided/secondary populations is inherently challenging. However, the possibility of diagnosing their combined net effect was already postulated by Pitcher and Stangeby \cite{Pitcher_1997}---and, as discussed in the following section, such measurements were then actually accomplished.

\subsection{Experimental evidence}\label{sec:evidence__experiments}

Experimental evidence for ordered toroidal neutral flows has been obtained in different tokamaks. In Alcator C-Mod \cite{Greenwald_2014}, directed toroidal motion of neutral atomic deuterium was diagnosed during plasma detachment using Doppler spectroscopy by Pitcher \textit{et al.} \cite{PITCHER19991009}, Goetz \textit{et al.} \cite{Goetz_1999} and by Welch \textit{et al.} \cite{Welch_2001}. More recently, toroidal neutral flows have also been measured in ASDEX Upgrade \cite{Gradic_2018}, albeit under attached plasma conditions.

Overall, in these experiments the neutral flow velocity was found to exceed $1~\text{km}\,\text{s}^{-1}$ and reach up to $\sim 5~\text{km}\,\text{s}^{-1}$ along the quasi-toroidal magnetic field direction in the vicinity of the divertor. Although this velocity represents only a fraction of the plasma sound speed ($\gtrsim 10 \, \rm km \, s^{-1}$), it is unambiguously distinguishable from random thermal motion and constitutes clear evidence of ordered neutral dynamics.

With the critical absence of fully-resolved two-dimensional experimental measurements---and with further caveats listed in appendix \ref{apx:evidence_limitations__experiments}, modelling provides a natural and necessary complement to interpret existing observations.

\subsection{Numerical evidence}\label{sec:evidence__simulations}

\subsubsection{Neutral hydrogenic species}\label{sec:evidence__simulations__fuel}

Early numerical studies already indicated that neutral deuterium in fusion edge plasmas can develop strong, ordered flows. Using coupled fluid-plasma/fluid-neutral modelling, Knoll \textit{et al.} \cite{Knoll_1996,Knoll_1998} demonstrated that non-trivial toroidal neutral flow patterns can arise and convect significant energy in a laminar plasma description \cite{Vahala_1996, Vahala_1997}. These works, while not addressing toroidal exhaust directly, established that neutral dynamics in the divertor are not purely diffusive and can support strongly directed motion\footnote{Interestingly, a parallel neutral velocity fixed to a high fraction of the ions' can also be implemented as a target-plate boundary condition \cite{HOLM2024101782}, e.g. in UEDGE \cite{ROGNLIEN1992347}.}.

More recent turbulence-resolving simulations recover the same underlying dynamics. In GBS simulations of low-power TCV plasmas, Giacomin \textit{et al.} \cite{GIACOMIN2022111294} report parallel neutral-deuterium velocities approaching $10~\mathrm{km}\,\mathrm{s}^{-1}$ (figure\footnote{Computed from $0.3\,c_{\mathrm{s}0}$, for the quoted plasma sound speed $c_{\mathrm{s}0}\simeq41~\mathrm{km}\,\mathrm{s}^{-1}$.}~17c in \cite{GIACOMIN2022111294}). Similarly, Coroado \textit{et al.} \cite{Coroado_2022} show fast ordered flows for molecular ions $\mathrm{D_2^+}$ (figure~2 in \cite{Coroado_2022}). These confirm that even sophisticated turbulent edge modelling produces strongly directed neutral and molecular-ion motion, reinforcing the conclusion that such behaviour is a robust consequence of plasma--neutral momentum coupling.

The workhorse of edge plasma modelling and interpretation remains the state-of-the-art suite SOLPS-ITER \cite{XavierBONNIN2016, SOLPSITERGitHub, SOLPSZenodoCommunity}, which combines multi-species plasma fluid dynamics with fully-kinetic neutral transport through the EIRENE Monte Carlo code \cite{kotov_reiter_kukushkin_bochum,Reiter2019EIRENE,Borodin_2022}. The code has undergone extensive verification \cite{Ghoos_2019,Boeyaert_2023}, validation \cite{Zito_2025, HORSTEN2025101842, Wigram_2026, Tonello_2026}, benchmarking \cite{Wu_2020, Wiesen_2025}, and cross-comparison with independent edge codes \cite{Moscheni_2022,Moscheni_2025, RUBINO2026115759, Rivals_Soledge3x}. Within this well-established framework, SOLPS-ITER simulations provide further compelling numerical evidence for the emergence of ordered toroidal neutral flows under reactor-relevant conditions.

In ITER, Rozhansky \textit{et al.} \cite{ROZHANSKY2022101316} report strong ordered parallel neutral velocities around the divertor strike points, with magnitudes in the range $3$--$8~\text{km}\,\text{s}^{-1}$ for neutral densities in $1$--$15\times10^{20}~\text{m}^{-3}$ (figure~3 in \cite{ROZHANSKY2022101316}).

Further specialised insight is provided by SOLPS-ITER simulations of Lore \textit{et al.} \cite{Lore_2022} for ITER. To quantify the directionality of neutral flows, the degree of toroidality (DoT) of a neutral species $\alpha$ is defined as:
\begin{equation}\label{eq:DoT}
    \DoT(\alpha) = \frac{\vTor(\alpha)}{\sqrt{\|\vTorVec(\alpha)\|^2 + \|\vPolVec(\alpha)\|^2}},
\end{equation}
where $\vTor(\alpha) = \vVec(\alpha) \cdot \uHatTorVec$ is the signed toroidal component of the mean particle speed and $\vPolVec(\alpha) = \vRVec(\alpha) + \vZVec(\alpha)$ the poloidal velocity vector\footnote{Stored in the \texttt{fort.46} output variables \texttt{v*den*} for the different directions and species, to be divided by the species mass and particle density (\texttt{p*den*}) \cite{SOLPSITERGitHub}.}. The quantity $\DoT \in [-1,1]$ therefore represents the fractional projection of the total mean speed along the toroidal direction.

\begin{figure*}
    \centering
    
    \subfloat[]{\includegraphics[width = 0.325\textwidth]{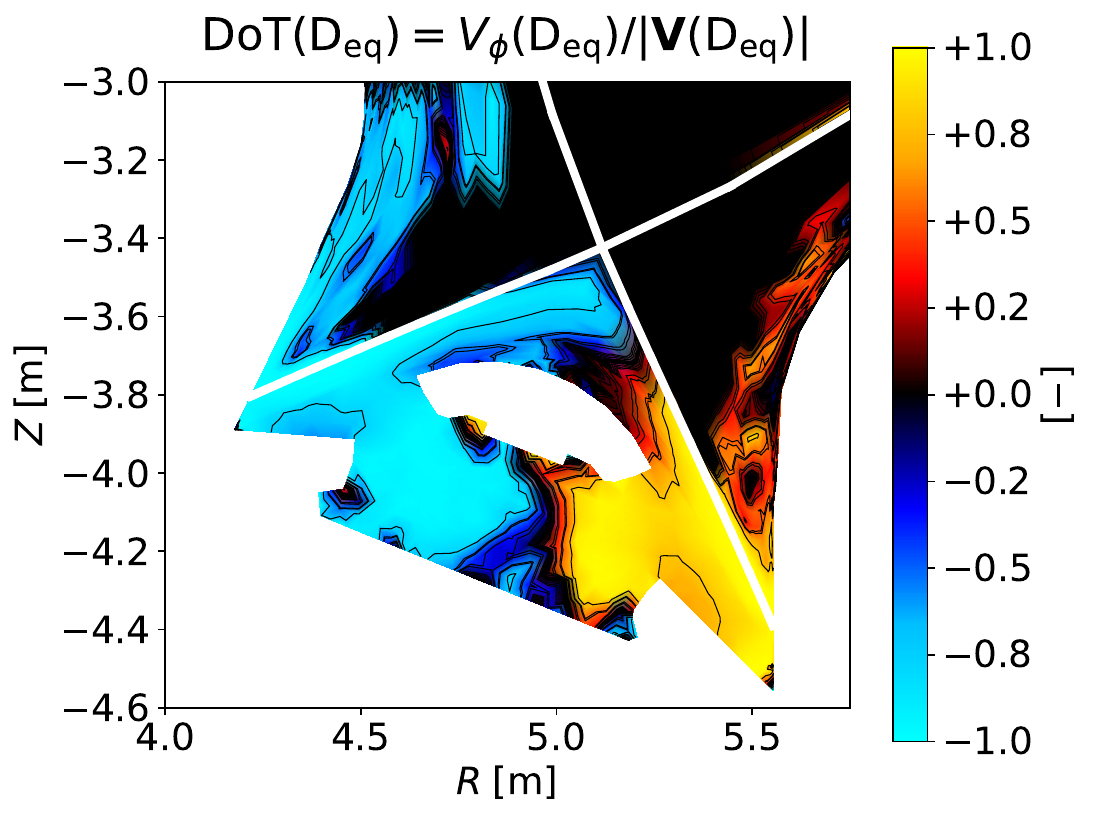}\label{fig:ITER_123013_Dplus2_D2_toroidality}}
    \hspace{0.075cm}
    \subfloat[]{\includegraphics[width = 0.325\textwidth]{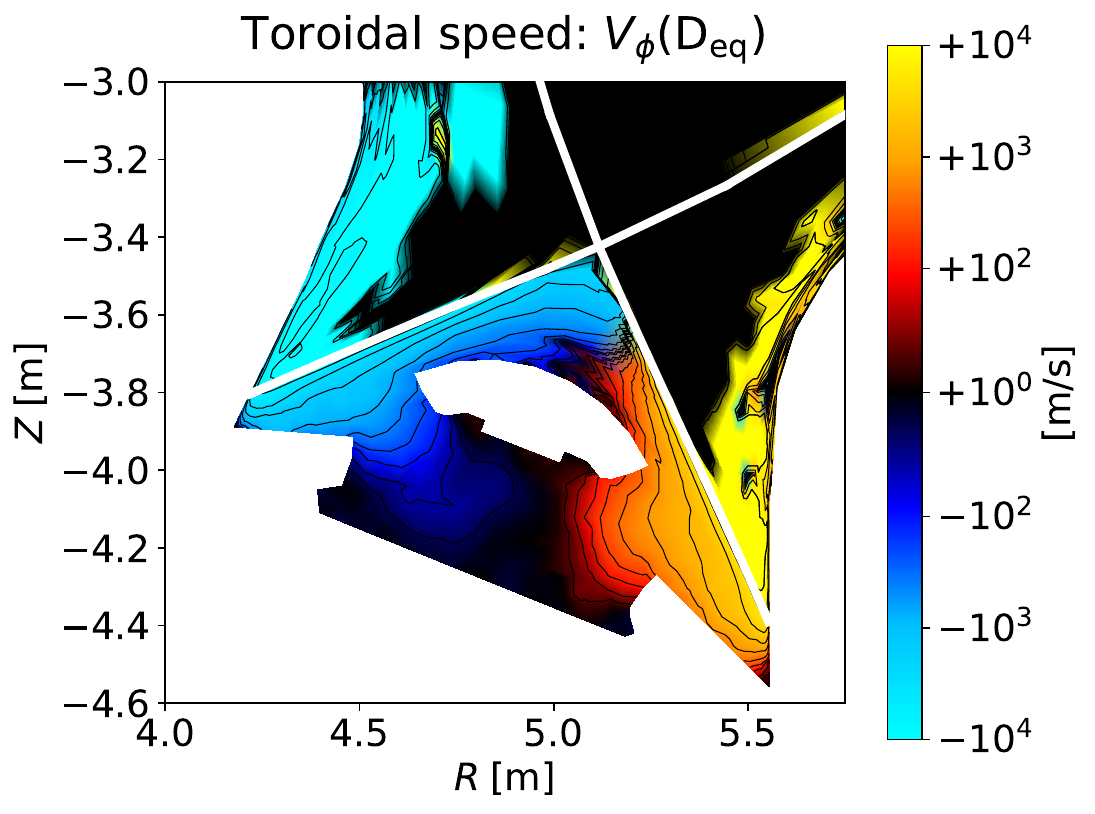}\label{fig:ITER_123013_Dplus2_D2_velocity_z}}
    \hspace{0.075cm}
    \subfloat[]{\includegraphics[width = 0.325\textwidth]{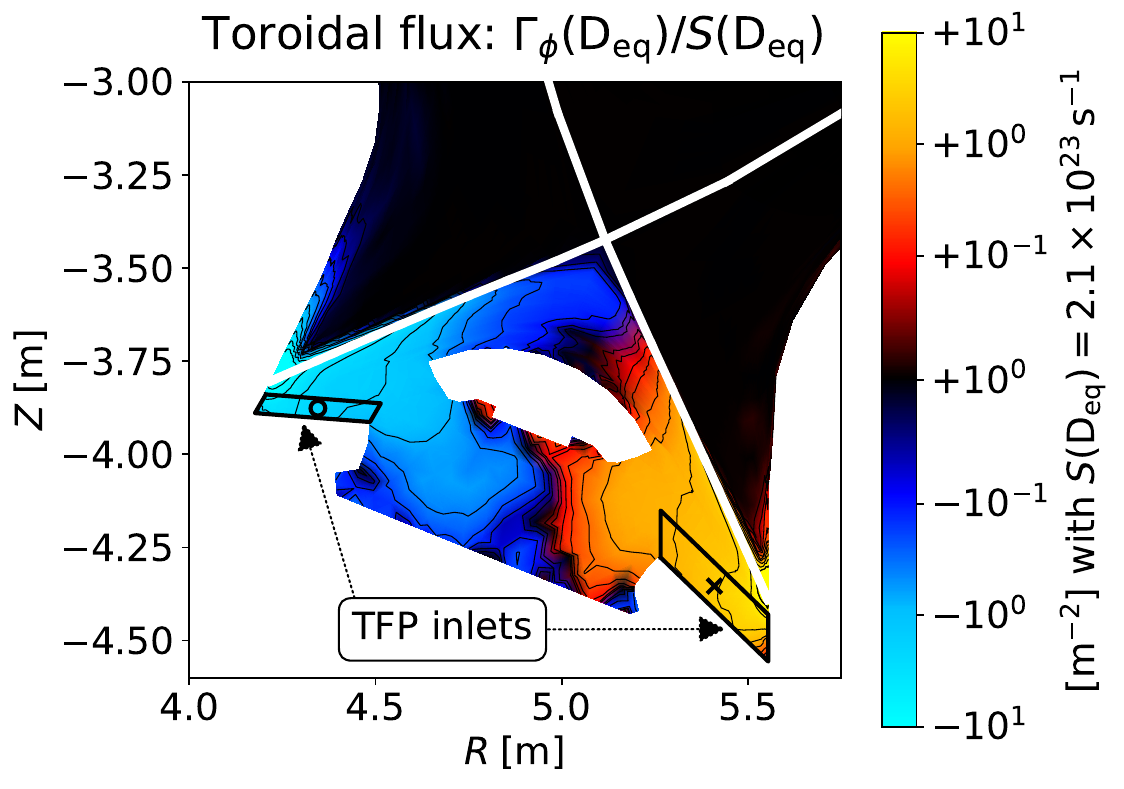}\label{fig:ITER_123013_Dplus2_D2_pflux_z}}\\
    
    \subfloat[]{\includegraphics[width = 0.325\textwidth]{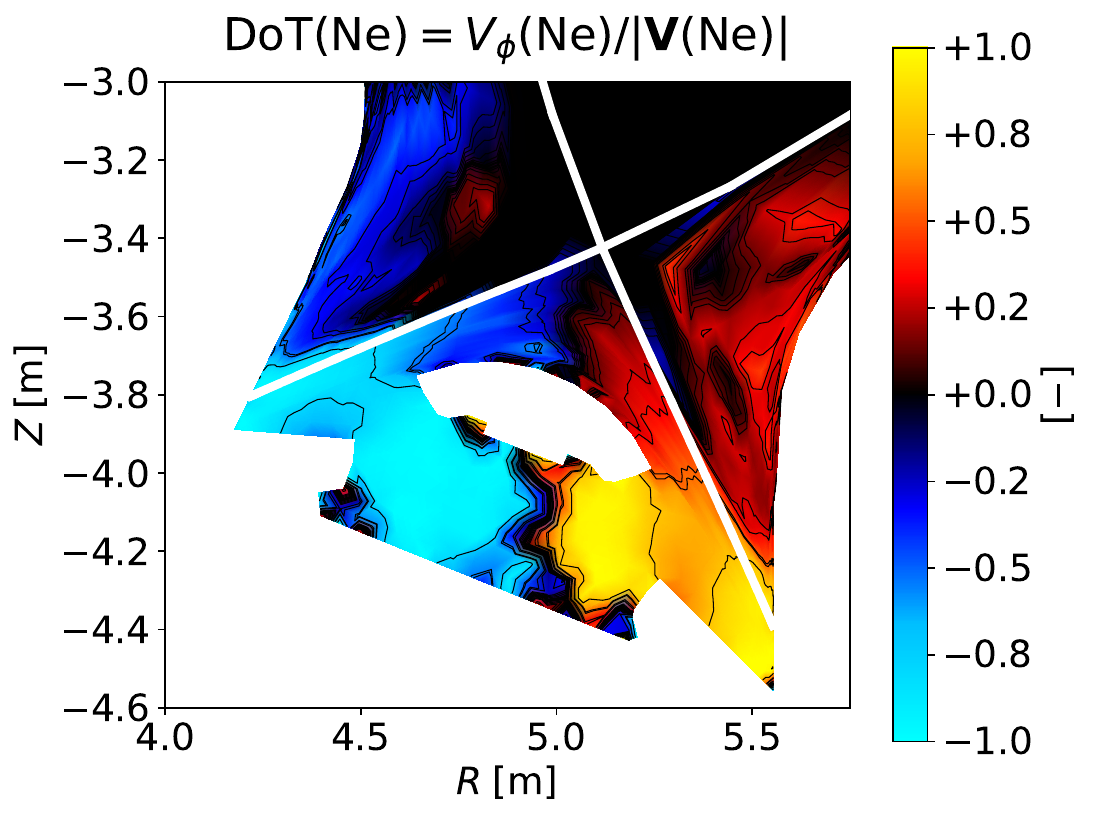}\label{fig:ITER_123013_Ne_toroidality}}
    \hspace{0.075cm}
    \subfloat[]{\includegraphics[width = 0.325\textwidth]{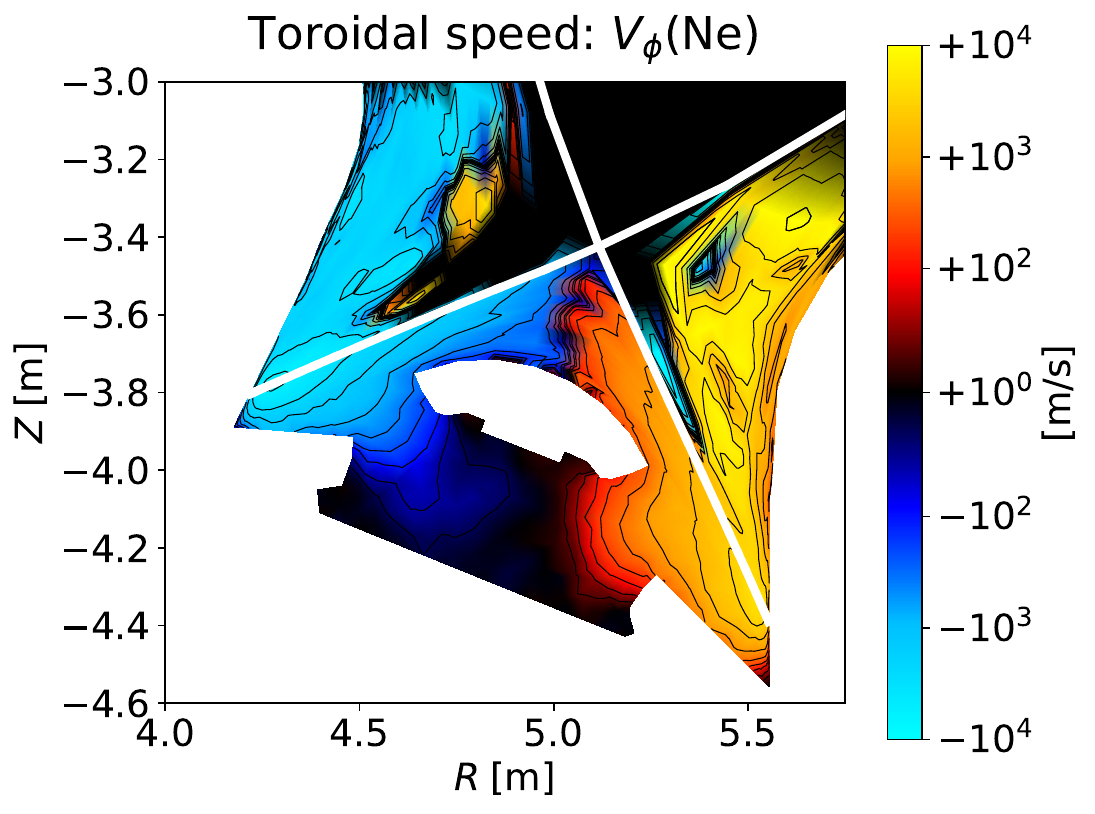}\label{fig:ITER_123013_Ne_velocity_z}}
    \hspace{0.075cm}
    \subfloat[]{\includegraphics[width = 0.325\textwidth]{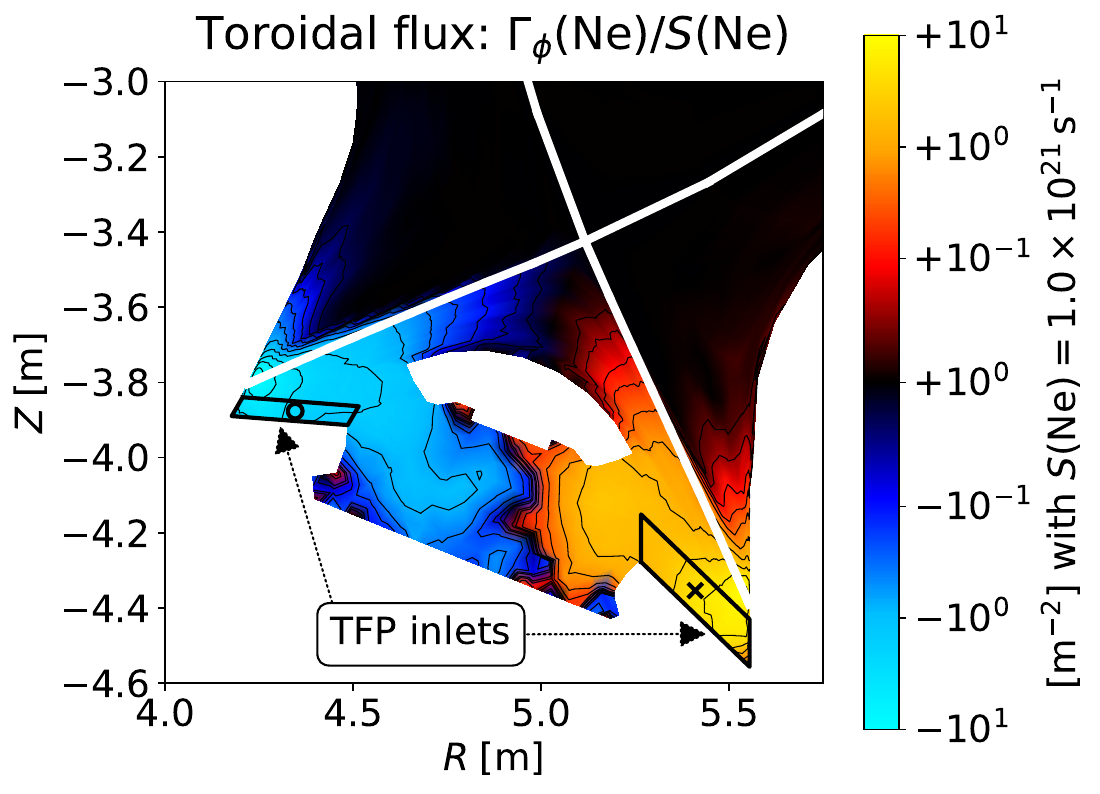}\label{fig:ITER_123013_Ne_pflux_z}}\\
    
    \subfloat[]{\includegraphics[width = 0.325\textwidth]{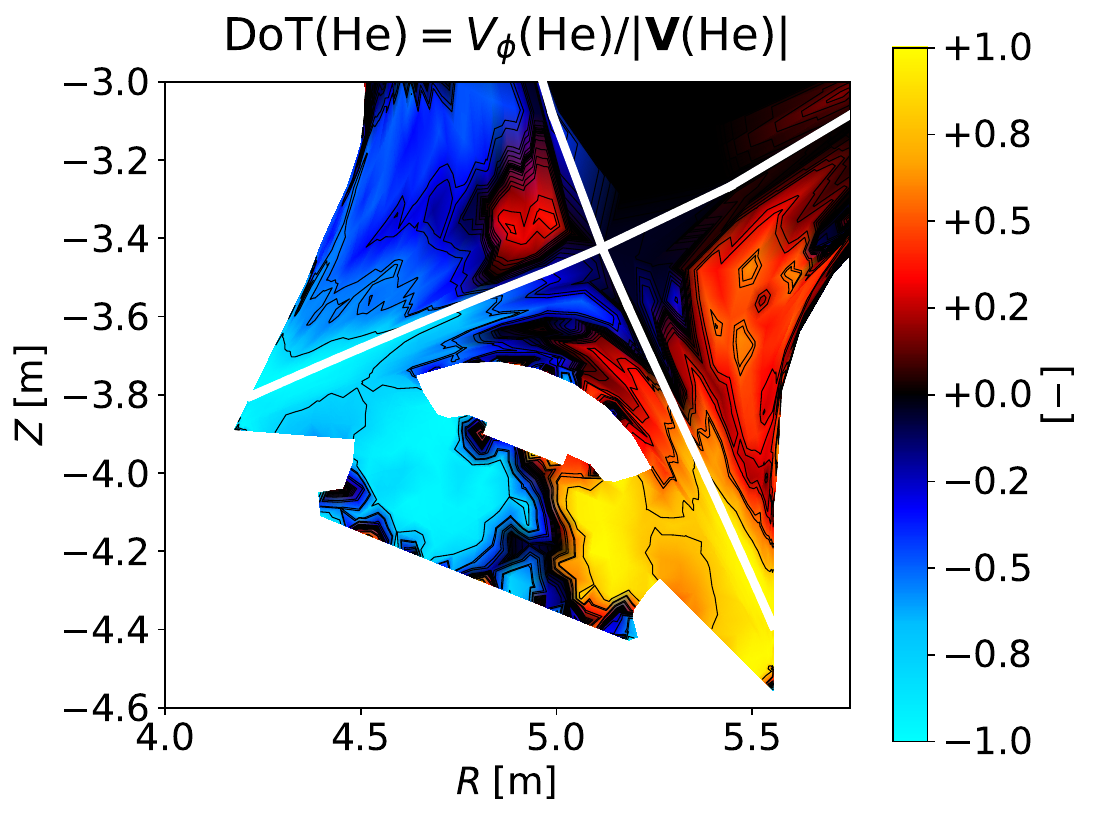}\label{fig:ITER_123013_He_toroidality}}
    \hspace{0.075cm}
    \subfloat[]{\includegraphics[width = 0.325\textwidth]{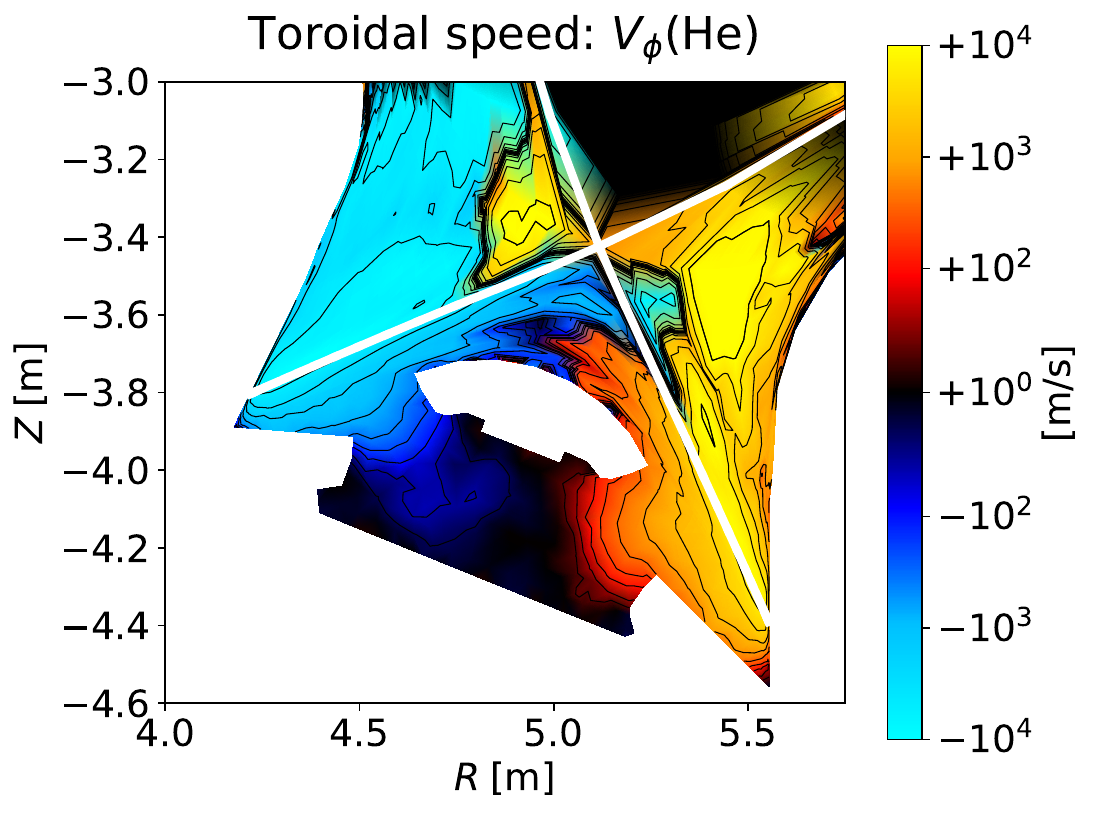}\label{fig:ITER_123013_He_velocity_z}}
    \hspace{0.075cm}
    \subfloat[]{\includegraphics[width = 0.325\textwidth]{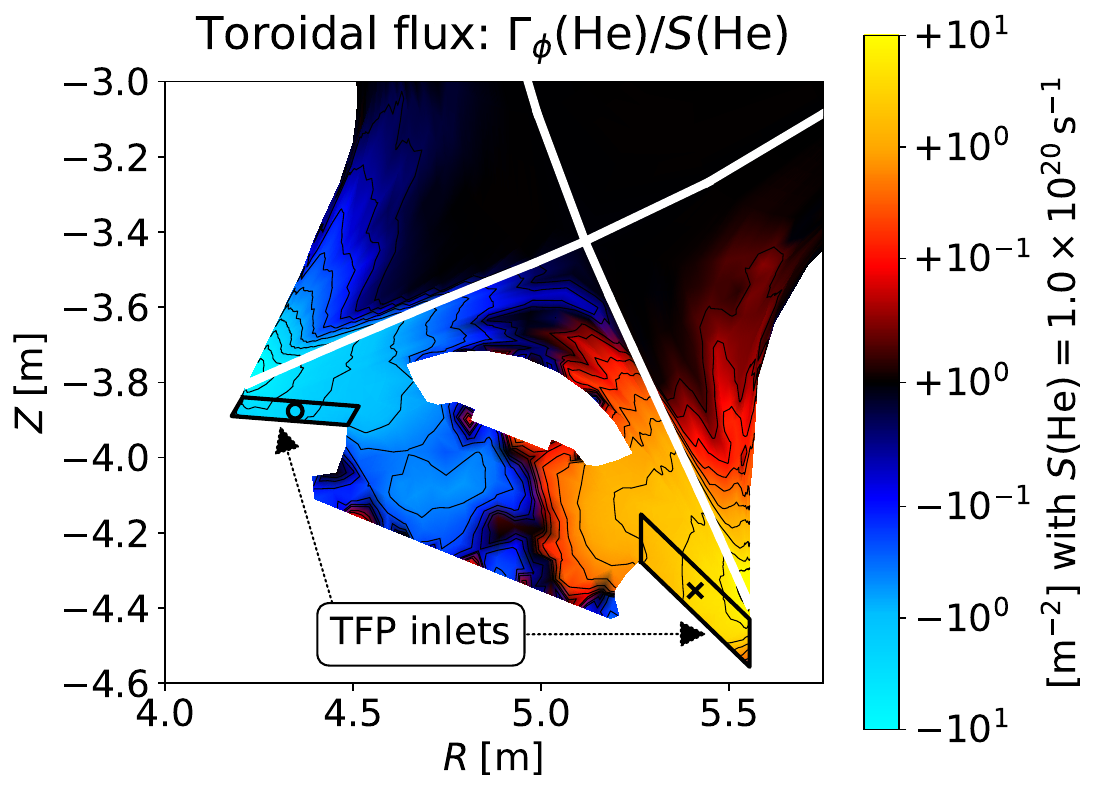}\label{fig:ITER_123013_He_pflux_z}}
    
    \caption{SOLPS-ITER simulation for ITER of Lore \textit{et al.} \cite{Lore_2022}---neon-seeded, detached at 3\% separatrix impurity concentration. From top to bottom: atomic-equivalent deuterium, neon and helium. Source rates: $\Gamma_\mathrm{D} = 1.95 \times 10^{23} \, \mathrm{D \, s^{-1}}$ puff and $10^{22} \, \rm D \,s^{-1}$ unburnt core fuelling; $\Gamma_\mathrm{Z} = 10^{21} \, \mathrm{Ne \, s^{-1}}$ neon seeding; $10^{20} \, \mathrm{He \, s^{-1}}$ helium production rate. From left to right: degree of toroidality (DoT); toroidal speed; normalised toroidal particle flux. An illustrative placement of the presented concept with counter-oriented toroidally-facing inlets is sketched in black (right).}
    \label{fig:ITER_123013}
\end{figure*}

Let us firstly consider a representative neon-seeded ITER simulation in \cite{Lore_2022}, partially detached and with 3\% separatrix impurity concentration. Figure \ref{fig:ITER_123013} presents a new post-processing of the original simulation output \cite{Lore_2022}. The first row illustrates the main output quantities for atomic-equivalent deuterium $\DzeroEq$---whereby $\Dzero$ and $\DzeroMol$ are combined\footnote{E.g. $n({\DzeroEq}) = n({\Dzero}) + 2 \, n({\DzeroMol})$ while $p({\DzeroEq}) = p({\Dzero}) + p({\DzeroMol})$ is the sum of the partial pressures, etc.}. Analogous conclusions separately hold for $\Dzero$ and $\DzeroMol$.

The poloidal distribution of the degree of toroidality $\DoT \, [-]$ (left), toroidal velocity $\vTor \, [\rm m/s]$ (middle), and toroidal particle flux $\GammaTor = n \vTor \, [\rm m^{-2} \, s^{-1}]$ normalised to the corresponding particle source $S \, [\rm s^{-1}]$ (right, with details in caption) paint a compelling picture in the PFR.

For hydrogenic species and including neutral-neutral collisions, several of the 14 default reactions implemented in EIRENE \cite{kotov_reiter_kukushkin_bochum,reiterAMJUEL,reiterHYDHEL} are capable of generating toroidally-directed neutral deuterium, resulting in $|\DoT(\DzeroEq)|\gtrsim0.75$ throughout most of the private flux region.

The corresponding toroidal velocities---widespread across the PFR---reach several kilometres per second, consistent with experimentally-measured values (section \ref{sec:evidence__experiments}). The sign of $\vTor$ reverses across the vertical mid-plane, vanishing around the centre-line and giving rise to two counter-streaming toroidal neutral fluxes. This feature agrees with theoretical \cite{Pitcher_1997} and experimental \cite{Gradic_2018} observations.

Crucially, the existence of a well-defined sign of both $\DoT$ and $\vTor$ is incompatible with purely disordered or diffusive neutral motion, and instead constitutes a clear signature of momentum imprinting from the anisotropic plasma flow.

The magnitude of the normalised toroidal particle flux is substantial. The inverse quantity, $S/|\GammaTor| \, [\rm m^2]$, corresponds to the minimum, cumulative, toroidally-facing surface that would be required to intercept a particle flux equal to the total neutral source in the chamber. In the present simulations this required capture area is of order $\lesssim 1\,\mathrm{m}^2$, in a macroscopic domain within the PFR. This underscores the potential relevance of toroidal neutral transport for particle exhaust. Further comments in this respect are given in section \ref{sec:discussion__integration}.

\subsubsection{Neutral impurity species}\label{sec:evidence__simulations__impurities}

The second and third rows in figure \ref{fig:ITER_123013} show the corresponding results for neon, representing an extrinsically-seeded impurity, and helium---intrinsic fusion ash. Despite their minority concentration, both impurity species exhibit the same qualitative dynamics observed for hydrogenic neutrals---indicating that impurities are incorporated into the ordered toroidal flow.

In these simulations \cite{Lore_2022}, neutral--neutral collisions are included through a BGK approximation \cite{Chernyak2010couette, Torrilhon_2015, Pfeiffer2018}, allowing both intra- and inter-species collisional coupling. Impurity neutrals can therefore acquire toroidal momentum, primarily, via electronic recombination\footnote{Implemented through the ADAS dataset \texttt{scd96} \cite{adas}.}  and, secondarily, through entrainment by the majority hydrogenic neutral wind. The relative strength of the two channels is investigated in section \ref{sec:discussion__interactions}.

\subsubsection{Toroidal neutral wind trend with increasing detachment}\label{sec:evidence__simulations__trends}

Next we tackle the question of how toroidal neutral winds evolve as the degree of detachment is increased. Physical expectations suggest a non-linear response, arising from competing effects.

On the one hand, deeper detachment is accompanied by a reduction of the ion flow speed, as illustrated for example in figures~13 and 21 of \cite{Effenberg_2019}. In addition, deeper detachment tends to promote higher neutral densities, which are unfavourable from the perspective of neutral velocity sustainment: for a given ion momentum budget, a denser, ``heavier'' neutral population implies a lower average neutral velocity. Both effects therefore suggest a decrease of the toroidal neutral speed with increasing detachment.

On the other hand, deeper detachment increases plasma--neutral coupling and therefore the efficiency with which ion momentum is transferred to neutrals in the first place \cite{Lipschultz_2001, Krasheninnikov_1996}. Moreover, increasing detachment shifts the detachment front progressively farther from the targets \cite{KRASHENINNIKOV1999251}, as shown in figure~7 of \cite{Lore_2022}. This increases the volume over which toroidal neutral motion can be generated and sustained, while reducing the relative importance of wall losses and frictional damping \cite{PITCHER19991009}. Qualitatively, this resembles flow in a toroidal duct, whose conductance increases with the available cross-sectional area \cite{Litovoli_2026}.

Once more, the net result can be explored quantitatively using the SOLPS-ITER simulations of Lore \textit{et al.} \cite{Lore_2022}. We focus on the scan where the deuterium fuelling rate is kept fixed while neon seeding is gradually increased \cite{Lore_2022}, since it exhibits the richest phenomenology and the widest variation in degree of detachment. With the space-averaging procedure described in appendix~\ref{apx:solps_average}, figure \ref{fig:ITER_123013_detachment_trend} shows the trends of the main quantities of interest around both inner and outer strike points---crucially, only within the PFR.

\begin{figure*}
    \centering
    \subfloat[]{\includegraphics[width = 0.975\textwidth]{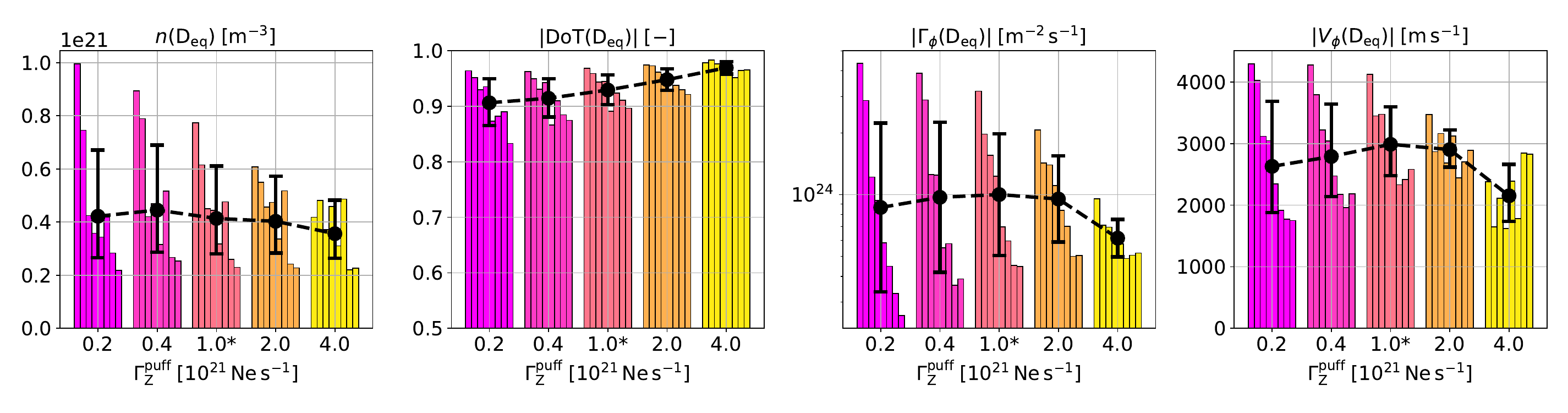}\label{fig:detachment_trend_Deq}}\\
    \subfloat[]{\includegraphics[width = 0.975\textwidth]{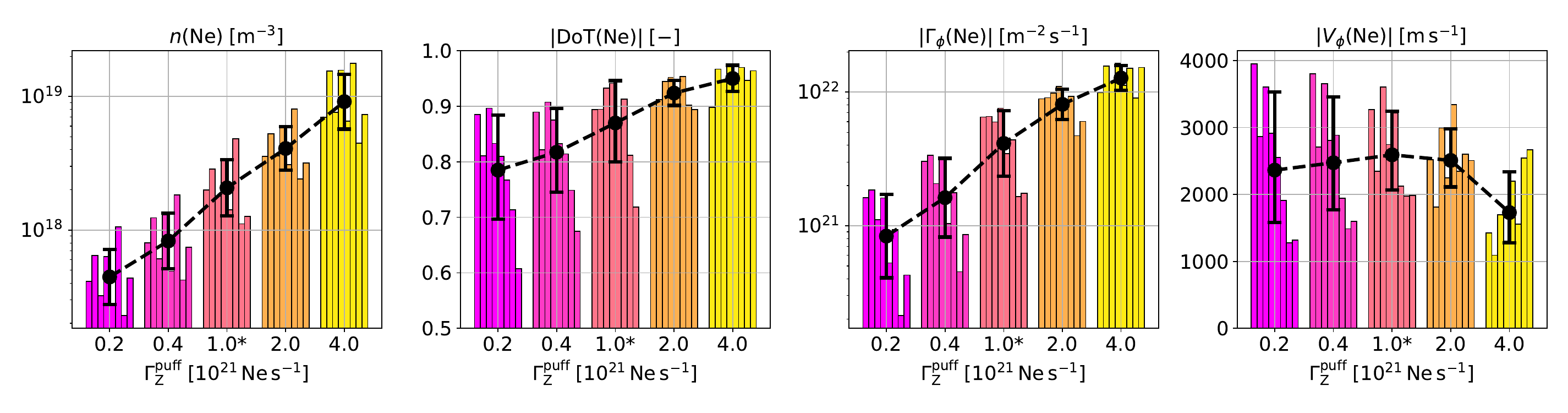}\label{fig:detachment_trend_X}}\\
    \subfloat[]{\includegraphics[width = 0.975\textwidth]{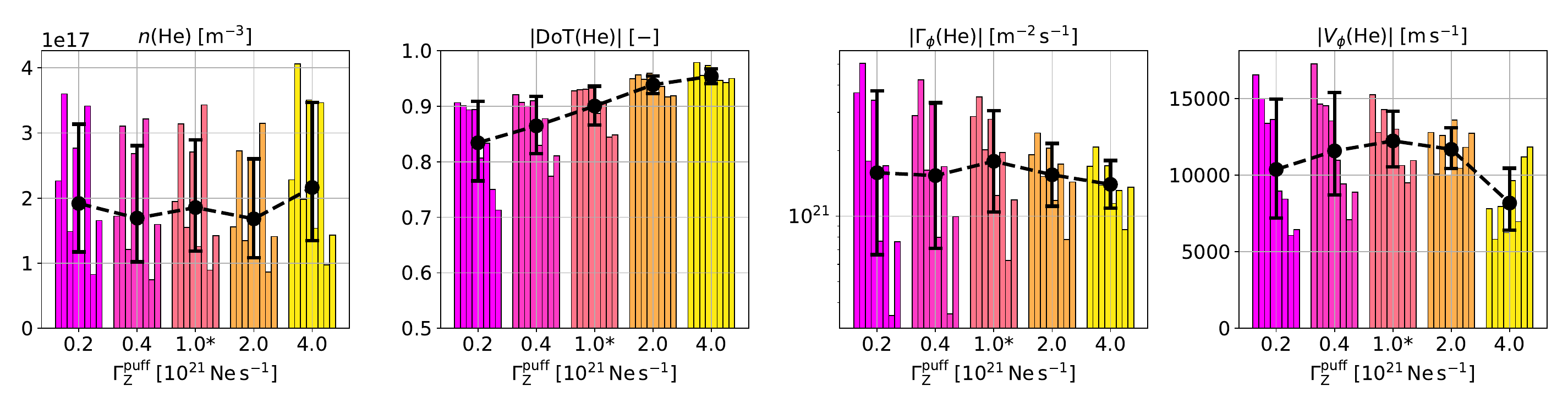}\label{fig:detachment_trend_He}}
    \caption{Trends of toroidal neutral-wind quantities with increasing neon seeding in the ITER of Lore \textit{et al.} \cite{Lore_2022} (``triangle series''), space-averaged around strike points within the private-flux region (appendix \ref{apx:solps_average}). The simulation depicted in figure \ref{fig:ITER_123013} is asterisked. The toroidal particle flux is here reported non-normalised.}
    \label{fig:ITER_123013_detachment_trend}
\end{figure*}

The degree of toroidality increases monotonically with detachment, approaching unity over much of the analysed region. For all cases with neon seeding $\leq 2\times10^{21}~\mathrm{Ne}\,\mathrm{s}^{-1}$, the toroidal velocities remain approximately stable on average\footnote{The helium speed differential is noted, and commented upon in appendix \ref{apx:evidence_limitations__simulations}.}, while the spatial cohesion of the wind increases. In practice, the toroidal neutral wind strengthens farther from the strike points, up to distances of order 30~cm (appendix \ref{apx:solps_average}), as suggested by the narrowing of the variation range indicated by the black vertical bars. As a consequence, the toroidal particle fluxes, already identified as substantial in magnitude in section \ref{sec:evidence__simulations__fuel}, trend equivalently.

Interestingly, these benefits vanish at the highest neon seeding rate. In this case, the peak perpendicular heat flux falls to only $0.4~\mathrm{MW}\,\mathrm{m}^{-2}$ and the separatrix impurity concentration grows to $6\%$ \cite{Lore_2022}. Under such extreme conditions, the afore-mentioned non-linear detrimental effects of strong momentum dissipation appear to prevail.

Appendix~\ref{apx:ITER_123013_detachment_trend__d_ne_series} shows the corresponding overview for scan in which fuel puffing and impurity seeding are increased together at fixed ratio \cite{Lore_2022}. The same beneficial conclusions are recovered for the toroidal neutral wind---with no drop at the highest puffing rates.

Overall---and before hitting unworkable impurity concentrations---the simulations indicate a favourable trend of toroidal neutral-wind formation with increasing detachment.

\subsection{Synthesis and implications}\label{sec:evidence__synthesis}

Taken together, the theoretical arguments, experimental observations, and numerical results reviewed above converge on a coherent physical picture. The available evidence remains heterogeneous and subject to well-known limitations---most notably sparse diagnostics in experiments and incomplete validation in modelling (appendix \ref{apx:evidence_limitations}). Yet, these caveats do not invalidate the central conclusion: \textit{across machines, species, and modelling frameworks, ordered toroidal neutral flows exist}.

Their implications for particle exhaust, however, have not yet been systematically explored---and are hence discussed next.

\section{Established pumping philosophies}\label{sec:philosophies}

The importance of particle exhaust can hardly be overstated. An overview of the associated challenges is provided in \cite{Meschini_Moscheni_2026}, where the onerous burden imposed by matter injection on particle exhaust and fuel-cycle requirements in reactor-relevant regimes is quantified. These results substantiate the concerns recently raised by the IAEA on this topic \cite{IAEA_TECDOC_2049_2024}.

A clear discussion of established pumping philosophies is therefore essential. Existing approaches are seldom compared on equal footing \cite{Takenaga_2001}, yet such a comparison is needed to identify their respective strengths and limitations. It also provides the basis for introducing a third pumping concept, which can be regarded as a hybrid of the two established philosophies.

\subsection{Fundamentals}\label{sec:philosophies__generalities__fundamentals}

A necessary condition for successful steady-state particle exhaust is that a net particle flux matching the total effective source rate is channelled toward the pumping system, independently for each neutral species. Once particles enter the vacuum ducting network, they establish a finite neutral density and therefore a pressure gradient. This is the fundamental driver of rarefied exhaust flow toward the pump hardware, ideally maintained at the lowest pressure throughout the network \cite{Van_Oost_2023}.

The transport mechanism by which neutral particles are transferred from the divertor volume to the \textit{inlet} of the pump ducting network---thereby contributing to the maximisation of the inlet pressure---defines what is referred to here as the pumping ``philosophy.''

To compare different philosophies, we use the removal efficiency $\etaCapture$ \cite{TEXTORTEAM1989115, Wenzel_2022, Wenzel_2024, Bader, Overskei_1981, Ohyabu_1992}---defined as the pumped flux normalised to the plasma flux reaching the divertor, $\etaCapture = \Gamma^{\mathrm{pump}}/I^{\mathrm{sat}}$. This metric is useful because it is reported across different studies and enables meaningful comparisons whenever sufficiently large differences are observed \cite{Reiter_1991}. Its caveats and limitations are discussed in appendix~\ref{sec:philosophies__generalities_capture_efficiency}.

Ultimately, the \textit{key implication} is that, in the words of Pitcher and Stangeby \cite{Pitcher_1997}: ``a high neutral pressure in the vicinity of the divertor reduces the required pump duct size, which is \textit{a significant engineering advantage when it comes to neutron shielding in a reactor}'' (italics added). The relevance of this constraint is illustrated, for example, by figure~3 of \cite{CROFTS2022113121}, which demonstrates the impact of a vacuum duct in an irradiated reactor environment---a $\sim$200-fold increase in neutronic dose rate compared to corresponding shielded volumes after one month of operation.

\subsection{Indirect pressure-driven pumping (IPDP)}\label{sec:philosophies__ipdp}

Developed in close synergy with the understanding of divertor detachment, what we may here term ``indirect pressure-driven pumping'' (IPDP) has long been regarded as the reference philosophy for particle exhaust in magnetic confinement fusion devices \cite{Loarte_2001}.

In IPDP, the transport mechanism from the recycling source at the plasma strike points toward, though not exclusively to, the pump duct inlets is primarily established by a pressure gradient in the poloidal plane. This gradient, typically of order a few to several pascals, is commonly generated by ``poloidal divertor shaping'', i.e. by passive geometric measures such as baffles \cite{FEVRIER2021100977, Sang_2017, Reimerdes_2022, Cowley_2026}, dome structures \cite{KUKUSHKIN2007308}, vertical target configurations \cite{SHAFER2019487, GROTH2015471, Moulton_2018, Loarte_2001} which enable plasma screening/plugging/self-baffling \cite{Reiter_1991, PITTS2019100696, Cowley_2026}. This is superimposed on the disordered, random-walk-like thermal motion of neutrals (see, for instance, figure~25b of \cite{Loarte_2001} and figure~1 of \cite{Havlickova_2014}). For this reason, the pumping mechanism is indirect: neutrals are not actively guided toward the pump inlets, but rely on pressure-driven, pseudo-random diffusion to reach them---whether located in the private flux region (PFR) \cite{Federici_2019_EU_DEMO, VAROUTIS2019120, Tantos_2024, BI2023113916} or in the scrape-off layer (SOL) \cite{YU2024101826, Henderson_2025}.

This leads to the central limitation of IPDP. Weakly directional transport can result in neutral leakage away from the pump inlets and toward the low-pressure upstream regions \cite{Gao_2023}. At the same time, significant back-flow phenomena can occur \cite{Reiter_1991}. Such behaviour has been numerically observed, for example, in JT-60U \cite{Takenaga_2001, SAKASAI2001957}, in ASDEX Upgrade \cite{Zito_2025, Senichenkov_2025}, and in ITER \cite{Day_2016}: neutrals poloidally channelled from one target toward the sub-divertor region are seen to backflow at the other target. Similar effects have been reported in simulations of the W7-X stellarator \cite{Varoutis_2024}, where up to 70\% of the particles crossing a pump inlet are found to return toward their region of origin. This mechanism should be distinguished from parasitic back-flow through gaps \cite{HAUER2009903}, which can in principle be mitigated by improved hardware design and gap closure. By contrast, \textit{back-flow at the pump inlet is a physics limitation of the transport mechanism itself} \cite{Binder_2016}.

For further indicative insight, the removal efficiency $\etaCapture$ (appendix~\ref{sec:philosophies__generalities_capture_efficiency}) is applied to lower single-null detached plasmas in vertical target configuration. SOLPS-ITER simulations yield an average value of $(5 \pm 1)\%$ across species \cite{Subba_2021, Moscheni_2022, Moscheni_2025, bonnin_2026_20431676, kaveeva_2026_20325809, Lore_2022, Zito_2025}. In the W7-X stellarator \cite{Haak_2023}, values from $<1\%$ up to at most 3\% have been reported \cite{Wenzel_2022}. These results substantiate the low removal efficiency that can characterise the IPDP approach.

Overall, IPDP represents a mature and well-established pumping philosophy and is therefore envisaged as the baseline solution for most reactor designs \cite{GLEASONGONZALEZ20141042, GLEASONGONZALEZ2016693, VAROUTIS2019120, BI2023113916, Park_2024}. In the following, it will be referred to as the ``traditional'' pumping approach. At the same time, it fundamentally relies on pseudo-random transport and the corresponding inefficiencies, which this study aims to address directly.

\subsection{Direct flow-driven pumping (DFDP)}\label{sec:philosophies__dfdp}

In contrast to IPDP, what we may here call ``direct flow-driven pumping'' (DFDP) exploits the preferential direction of plasma flow in the poloidal plane. A fraction of the neutrals reflected at the divertor retains memory of the incident plasma-flow direction---generating ``fast'', specularly-reflected particles \cite{Varoutis2023DIVGASDEMO, Eckstein2009Reflection, Moulton_2018, LIVADIOTTI2020100675, ECKSTEIN1984550, Reiter2019EIRENE}. DFDP therefore places pump inlets close to the plasma strike points and aligns them with the dominant reflected-neutral trajectories \cite{Overskei_1981, TEXTORTEAM1989115, Corbett_1991, Loarer_1995, Loarte1997Controlled, Loarte_2001, Maingi_1999, Komori_2005, Bader, Lore_2019}. The objective is to maximise ballistic, or first-flight, capture before multiple scattering events isotropise the neutral motion into the random-walk regime characteristic of IPDP \cite{Loarte_2001}. In addition to the angular distribution of the recycled neutral population, the geometrical acceptance of the pump inlet plays a central role. In the collisionless limit, this is set by the solid angle subtended by the inlet as seen from the strike-point region \cite{Klepper_1993, Maingi_1999, Lore_2019}.

Pump limiter concepts \cite{Overskei_1981, TEXTORTEAM1989115} represent an extreme realisation of this philosophy: part of the SOL plasma is effectively routed within the pump inlet itself. This enables measured removal efficiencies approaching 25\% \cite{Overskei_1981, TEXTORTEAM1989115}, up to a factor of four larger than values typically achieved with IPDP (section \ref{sec:philosophies__ipdp}). It was suggested that even higher values could be attainable, potentially approaching 50\% \cite{LIPSCHULTZ1984441}. From a modelling perspective, the simulations adopted in \cite{Komori_2005} and \cite{Bader} for pump-limiter-like configurations yields estimated removal efficiencies of 50--60\% and 13\%, respectively, further illustrating---albeit in a simplified framework---the potential of DFDP. 

Other examples of DFDP have been reported, notably in DIII-D \cite{Klepper_1993, Maingi_1999}. In particular, Unterberg \textit{et al.}~\cite{Unterberg_2010} compare two configurations with distances between the outer strike point and the pump inlet of 6~cm and 4~cm. The removal efficiency increases by about a factor of two as this distance decreases, from $(4 \pm 0.5)\%$ to $(8 \pm 0.5)\%$.

The most compelling evidence for DFDP-like benefits is arguably provided by the JET corner configuration discussed by Loarte \cite{Loarte_2001}. In figure~25a of \cite{Loarte_2001}, an increase in particle exhaust capability is measured in the corner configuration. Here, directed neutral trajectories favour access to the sub-divertor, compared with the vertical outer target configuration---representative of IPDP. The different neutral trajectories are also illustrated in figure~2 of \cite{Moulton_2018} and figure~1 of \cite{Cowley_2026}. Loarte \cite{Loarte_2001} reports approximately a factor of two increase in sub-divertor pressure in the corner configuration\footnote{Interestingly, even a non-corner horizontal arrangement retains a measurable advantage over the vertical configuration.}, together with a comparable increase in helium enrichment (figure~31 of \cite{Loarte_2001}) also quoted by Groth \textit{et al.}~\cite{Groth_2001, Groth_2002}. The same corner configuration is also shown to improve core plasma performance \cite{Joffrin_2017, Frassinetti_2021}, demonstrating that enhanced exhaust can have beneficial consequences at the plasma-system level.

Despite these encouraging results, DFDP features several critical shortcomings.

The favourable values obtained in pump limiter configurations should be interpreted with caution. Pump limiters have not been shown to improve particle exhaust as a whole \cite{LIPSCHULTZ1984441}. This provides a useful reminder that the removal efficiency $\etaCapture$ alone is not a sufficient figure of merit for particle exhaust performance (appendix \ref{sec:philosophies__generalities_capture_efficiency}).

At a more fundamental level, DFDP places a disproportionate emphasis on particle exhaust at the expense of other critical divertor functions. The results of Unterberg \textit{et al.} \cite{Unterberg_2010} illustrate the sensitivity to the distance of the strike point from the inlet \cite{Takenaga_2001, WIESEN20173}. Maintaining the proper distance would demand precise geometric alignment---costly and difficult to guarantee in a reactor environment, where diagnostic access and real-time control capabilities are expected to be limited \cite{BIEL2022113122, Mazon_2025}.

The alignment requirement also effectively constrains the divertor geometry, favouring horizontal target arrangements \cite{Loarte_2001, Groth_2001}. In this configuration, reflected neutrals avoid crossing the plasma divertor leg, where plasma--neutral interactions would rapidly isotropise their motion \cite{KAVEEVA2023101424, Lore_2019}. Unless combined with tight baffling, such as in STEP's Super-X outer divertor \cite{Henderson_2025}, horizontal targets are generally associated with reduced power exhaust capability and deteriorated detachment access \cite{Loarte_2001}.

Finally, even in circumstances where detachment can be accessed, the principle underpinning DFDP is compromised, as discussed by Bader \textit{et al.} \cite{Bader}. Under detached conditions, plasma--surface interaction at the strike point is weakened, while the increased neutral density strengthens neutral--neutral collisionality. Both effects blur the poloidal directionality of the reflected neutral population, thereby impairing first-flight capture and undermining the core mechanism on which DFDP relies.

\subsection{Synthesis}\label{sec:philosophies__synthesis}

A preliminary---albeit unsurprising---conclusion is that leveraging directed transport rather than pseudo-random motion can be advantageous for particle exhaust.

More generally, however, the two established philosophies highlight a fundamental tension between robustness and efficiency, already reported by Reiter \textit{et al.} \cite{Reiter_1991}. IPDP tends to prioritise compatibility with the broader set of divertor functions, while typically resulting in modest removal efficiency. DFDP, conversely, can achieve higher removal efficiency---but is accompanied by more stringent constraints.

Table \ref{tab:capture_efficiencies} summarises representative---albeit indicative---values and provides a compact comparison between the two philosophies. This tension motivates the exploration of alternative pumping approaches.

\begin{table*}[t]
    \centering
    \caption{Representative removal efficiencies, $\etaCapture$, for a high-level comparison between established pumping philosophies. As explained in appendix \ref{sec:philosophies__generalities_capture_efficiency}, values are indicative and and do \textit{not} directly reflect the overall exhaust efficiency. Asterisked entries are experimental measurements.}
    \label{tab:capture_efficiencies}
    \begin{tabular}{l l l l l}
        \toprule
        Philosophy & Refs. & Machine(s) & $\etaCapture \, [-]$ \\
        \midrule
        \multirow{2}{*}{IPDP}
            & \cite{Subba_2021, Moscheni_2022, Moscheni_2025, bonnin_2026_20431676, kaveeva_2026_20325809, Lore_2022, Zito_2025} & AUG, DTT, EU-DEMO, ITER, JET & $(5 \pm 1)\%$ \\
            & \cite{Takenaga_2001} & JT-60U* & $\lesssim 3\%$ \\
             & \cite{Wenzel_2022} & Wendelstein 7-X* & $0.5$--$3\%$ \\
        \midrule
        \multirow{4}{*}{DFDP}
             & \cite{Komori_2005} & LHD & $50$--$60\%$ \\
             & \cite{LIPSCHULTZ1984441} & Alcator DCT & $10$--$50\%$ \\
             & \cite{Overskei_1981, TEXTORTEAM1989115} & TEXTOR* & $25\%$ \\
             & \cite{Bader}          & Infinity Two & $13\%$ \\
             & \cite{Unterberg_2010} & DIII-D*  & $8\%$ \\
        \bottomrule
    \end{tabular}
\end{table*}

\section{Toroidal Flow Pump: the TFP}\label{sec:tfp}

\subsection{Motivation and conceptual gap}\label{sec:tfp__motivation}

None of the established pumping philosophies explicitly harnesses the ordered toroidal neutral flows described in section \ref{sec:evidence}. Several factors may explain this absence.

From a scientific perspective, previous investigations of neutral flows and their toroidal motion have mainly been pursued in the context of possible improvements to \textit{power} exhaust, rather than \textit{particle} exhaust \cite{Vahala_1996, Knoll_1996}.

From a modelling and expertise perspective, the separation between edge plasma and neutral modelling on the one hand, and vacuum duct and pumping-system modelling on the other, may have further limited the exploration of such concepts. The former is typically treated as an edge-plasma physics problem, while the latter is more often addressed as a rarefied-gas-dynamics and vacuum-engineering topic. Edge plasma--neutral simulations are commonly truncated at, or near, the pump inlet---where the exhaust system is represented as a boundary condition \cite{Moscheni_2022, Moscheni_2025}. Conversely, the downstream vacuum ducts, being non-plasma-facing volumes, are often modelled separately and with different numerical tools \cite{TANTOS2025115021, Tantos_2022, Tantos_2024}. Although exceptions exist \cite{KAVEEVA2021101030, Bonelli_2017}, the common artificial partitioning of the problem space may have obscured pumping concepts that rely on a continuous treatment of neutral transport across the plasma-facing and pumping domains, such as toroidal flow pumping.

\subsection{The TFP: general features}\label{sec:tfp__features}

The toroidal flow pump (TFP) is the proposed pumping concept in which a pump inlet is oriented to intercept ordered toroidal neutral motion in the divertor or private-flux-region volume \cite{Moscheni_2025_patent}. It can be viewed as a hybrid approach, ``indirect flow-driven pumping'' (IFDP): the concept retains the compatibility with divertor functions and vertical targets of IPDP, while seeking to exploit neutral-flow directionality in a manner conceptually related---but directionally orthogonal---to DFDP.

In its simplest form, schematically represented in figure~\ref{fig:overview_wind}, a toroidally-facing pumping inlet is placed in a region where toroidal neutral flow persists, for instance in the private flux region (figure~\ref{fig:ITER_123013}). The inlet intercepts part of this directed neutral population and redirects it toward the exhaust duct, while remaining displaced from the near-plasma region. The transmission of drifting velocity distributions through apertures is a non-trivial kinetic problem and remains an active topic of fundamental \cite{Binder_2016, SATO201960, YUAN2024113366, Sepahi_2018, Manela_Gibelli_2020} and applied \cite{ROMANO2021225, ZHENG2023223, ZHANG202451, YAKUNCHIKOV2025102, Almeida2021A, Ruetten_2013, Miansari_2020} research. Nevertheless, the physical intuition relevant here is straightforward: if the mean neutral velocity has a component directed into the duct, the probability of back-flow from the inlet is reduced \cite{Binder_2016}.

At system level, this mechanism suggests a possible exhaust advantage. Let $Q=pS^{\rm eff}$ be the throughput to be exhausted and $S^{\rm eff}$ the effective pumping speed. A toroidal approach ($\phi$) does not rely solely on the static, approximately isotropic pressure $\pPol$ exploited by a traditional poloidal pressure-driven arrangement ($\theta$). Instead, the directed neutral motion contributes an additional momentum-flux-related contribution $\Delta \pTor > 0$ to the inlet. This may be summarised as
\begin{equation}\label{eq:throughput_gain}
    \begin{gathered}
        Q = p_{\theta} S^{\rm eff}_{\theta}
          = (p_{\theta} + \Delta p_{\phi}) S^{\rm eff}_{\phi} \,,
        \\
        \Rightarrow
        S^{\rm eff}_{\phi}
        =
        S^{\rm eff}_{\theta}
        \frac{p_{\theta}}{p_{\theta}+\Delta p_{\phi}}
        \leq S^{\rm eff}_{\theta} \, .
    \end{gathered}
\end{equation}

An approximate numerical quantification of $\Delta\pTor$ and of the effective gain is targeted by the present study. Equation~(\ref{eq:throughput_gain}) suggests that a reduction of the effective pumping speed required downstream may be achieved, although not necessarily one-to-one. In light of section~\ref{sec:philosophies__generalities__fundamentals}, such an increase would translate into a practical advantage in conductance-limited systems, where higher inlet pressure eases duct and pumping constraints \cite{Pitcher_1997}.

The TFP is not mutually exclusive with established pumping concepts. It could in principle be implemented as an auxiliary exhaust path in configurations that otherwise rely on IPDP. Physically, the mechanism is passive: the toroidal neutral flow is generated by plasma--neutral processes. The proposed inlet orientation attempts to capture part of a directed particle population that would otherwise merely circulate toroidally or be redistributed by subsequent collisions.

The following sections describe the modelling strategy used to isolate this mechanism and quantify its effect under idealised conditions. The results should therefore be interpreted as a physics-based proof-of-principle for toroidal-flow capture, rather than as an integrated design assessment of a reactor pumping system.

\section{Methods}\label{sec:methods}

\subsection{Direct simulation Monte Carlo}\label{sec:methods__dsmc}

Self-consistent treatments of plasma--neutral interactions, discussed in section~\ref{sec:evidence__simulations}, are available in coupled edge-plasma codes such as SOLPS-ITER \cite{kotov_reiter_kukushkin_bochum, XavierBONNIN2016, Moscheni_2022}. These tools, however, are currently formulated primarily in poloidal geometry and rely on reduced descriptions of neutral-neutral collisions via linearised BGK approximations \cite{Chernyak2010couette, Torrilhon_2015, Pfeiffer2018}. This motivates the use of all-Knudsen-number direct simulation Monte Carlo (DSMC) transport models in the present context, particularly when flexibly scoping geometries, rarefaction, and collisional momentum redistribution are central to the physics. Detailed discussions on the importance of DSMC for divertor and sub-divertor neutral transport can be found in \cite{Tantos_2022, Varoutis_2024, Tantos_2024}.

In the present work, the SPARTA DSMC code \cite{Plimpton_2019, SPARTA, Gallis_2017, Gu_Barber_John_Emerson_2019} is employed with particular focus on the plasma--neutral interface. This region lies at the boundary of applicability of stand-alone neutral kinetic modelling. However, the private flux region contains, by definition, little plasma compared with the surroundings \cite{Brida_2025, Subba_2021}, and can therefore be treated to leading order as a neutral-dominated domain in which plasma effects enter through physically-motivated boundary conditions.

Accordingly, stand-alone DSMC is used here as a \textit{demonstrative and scaling-oriented} approach complementary to---and not substitutive of---EIRENE's capabilities.

\subsection{Simulation set and rationale}\label{sec:methods__simulation_set}

This section clarifies the purpose and scope of each two-dimensional simulation class. Taken together, they form a coherent, simplified proof-of-principle---progressing from the underlying physics mechanism to its possible engineering exploitation.

\begin{itemize}
    \item \textit{Physics proof-of-principle} (``PoP-Phys''). 
    The first set of simulations is designed to demonstrate, in a minimalistic and controlled manner, that plasma-like boundary conditions alone constitute a sufficient condition for the emergence of ordered toroidal neutral flow. A nominal baseline case is first defined, and is then extended into a simulation database for systematic exploration. Regression analyses allow identifying the parameters controlling the magnitude and spatial persistence of toroidal neutral flow, and its robustness. The typical computational cost is 40 CPUh per case, up to a maximum of 1200 CPUh.

    \item \textit{Exhaust proof-of-principle} (``PoP-Exh''). 
    The second set of simulations addresses---in an idealised manner---the particle exhaust implication of PoP-Phys, namely whether interception of an ordered toroidal neutral wind can translate into a tangible advantage for particle exhaust. As for PoP-Phys, a nominal reference case is first considered and then extended into a simulation database to test the robustness of toroidal-flow-assisted capture. All cases feature an indicative 10\% helium concentration, to approximately assess whether majority- and minority-species dynamics remain comparable. The typical computational cost is 300 CPUh per case.
    
\end{itemize}

Further details are provided in the following sections, and in appendix \ref{apx:setup} and \ref{apx:numerics}. Unless otherwise specified, the same modelling and numerical practices of PoP-Phys are then employed in the subsequent PoP-Exh simulations.

\subsection{Physics proof-of-principle setup}\label{sec:methods__pop_phys}

\subsubsection{Geometry}\label{sec:methods__pop_phys__geometry}

Similarly to the DIVGAS simulations of Varoutis \textit{et al.} \cite{VAROUTIS2019120} and of Tantos \textit{et al.} \cite{Tantos_2022}, our domain represents the PFR of a nominal lower single-null divertor configuration---which is sketched in figure \ref{fig:pop_phys_domain_sketch}. However, no sub-divertor volume is here included. 

The plasma-facing boundaries are formed by the separatrix (SEP) divertor legs, extending from the strike points at the inner (left) and outer (right) targets up to their convergence at the magnetic X-point. The remaining boundaries correspond to wall surfaces enclosing the PFR.

The radial width and vertical height are $L_r$ and $L_z$, respectively. The domain is then assumed to be a straightened infinite torus along the third dimension. Its equivalent hydraulic diameter is defined as
\begin{equation}\label{eq:equivalent_diameter}
    d_h = \frac{4A}{P} ,
\end{equation}
where $A$ is the cross-sectional area and $P$ is the corresponding wetted perimeter. Regions of interest are defined in appendix \ref{apx:methods__pop_phys__geometry__rois}, and simplifications in appendix \ref{apx:methods__pop_phys__geometry__no_pump} and \ref{apx:methods__pop_phys__geometry__simplification}.

\input{00_PoP_Phys_domain}

\subsubsection{Interaction models}\label{sec:methods__pop_phys__interactions}

\paragraph{Neutral species and collisions}\label{sec:methods__pop_phys__interactions__neutrals}

In traditional vacuum pump duct models \cite{VAROUTIS2019120, Tantos_2022, Varoutis_2024, Tantos_2024, TANTOS2025115021}, only deuterium molecules are considered, consistently with the sub-divertor and vacuum-duct regimes analysed in such studies.

For the present application, deuterium atoms are instead used as the baseline neutral species. This choice is motivated by the focus on the near-plasma volume, where plasma-driven molecular dissociation enhances the atomic population \cite{Lore_2022, Moscheni_2022, Moscheni_2025, Subba_2021}. Molecular deuterium is omitted as a modelling simplification. This choice is not expected to qualitatively affect the conclusions: $\DzeroMol$ is found to attain toroidal speeds comparable to those of atomic deuterium (section~\ref{sec:evidence__simulations__fuel}). Omitting $\DzeroMol$ hence remains conservative with respect to the total directed momentum, because molecules would contribute twice the mass at comparable velocity. Surface recombination of atoms into molecules, D--$\DzeroMol$ momentum exchange, and molecular mass effects should nevertheless be assessed in more detail in future work.

Neutral--neutral collisions are one of the main differences between the present SPARTA DSMC simulations and EIRENE simplified treatment. While EIRENE approximates neutral--neutral collisions through a linearised BGK operator \cite{Chernyak2010couette, Torrilhon_2015, Pfeiffer2018}, SPARTA resolves them kinetically using the variable soft sphere (VSS) formulation described in appendix~\ref{apx:numerics__collisions}.

For atomic deuterium, a reference diameter $\sigma_{\Dzero} = 2.81\times10^{-10}\,\mathrm{m}$ is adopted from \cite{Tantos_2024}. The corresponding mean free path\footnote{For a 90:10 D:He mixture, mean free paths are computed as follows: $\lambda_{\rm D}^{\mathrm{mfp}} = \left[ n \left(0.9 \sigma_{\rm DD}\sqrt{2} + 0.1 \sigma_{\rm DHe}\sqrt{1+m_{\rm D}/m_{\rm He}} \right)\right]^{-1} $ and $ \lambda_{\rm He}^{\mathrm{mfp}} = \left[ n \left( 0.1\sigma_{\rm HeHe}\sqrt{2} + 0.9 \sigma_{\rm DHe}\sqrt{1+m_{\rm He}/m_{\rm D}} \right) \right]^{-1}$ where $n = n_{\rm D} + n_{\rm He}$ is the total density.} is estimated as

\begin{equation}\label{eq:deuterium_mpf}
\lambda_{\Dzero}^{\mathrm{mfp}} = \frac{1}{\sqrt{2}\,n\,\pi \sigma_{\Dzero}^2},
\end{equation}

from which the Knudsen number is defined in the usual way as

\begin{equation}\label{eq:knudsen_number}
\mathrm{Kn} = \frac{\lambda_{\Dzero}^{\mathrm{mfp}}}{d_h},
\end{equation}

where $d_h$ is the equivalent hydraulic diameter in equation (\ref{eq:equivalent_diameter}). 

\paragraph{Neutral-wall surface interaction model: CLL formulation.}\label{sec:methods__pop_phys__interactions__wall}

All plasma-facing walls are assumed as purely reflecting, and assigned a temperature of $1160~\mathrm{K}$ ($0.1~\mathrm{eV}$)---consistent with SOLPS-ITER defaults \cite{Zito_2025}.

Neutral--wall interactions are described using the Cercignani--Lampis--Lord (CLL) model \cite{Lord_1991, Lord_1995}. Compared with TRIM/SDTrimSP-based tabulated reflection models commonly employed in EIRENE \cite{ECKSTEIN1984550, Reiter2019EIRENE}, CLL is a lower-dimensional, phenomenological scattering kernel. This formulation offers a parametrisation via \cite{Varoutis2023DIVGASDEMO, Agrawal_2020}: the normal energy accommodation coefficient $\accN \in [0;1]$ (NEAC), representing the degree of gas--wall thermalisation; and the tangential momentum accommodation coefficient $\accT \in [0;1]$ (TMAC), quantifying the gas--wall friction. Experimental and numerical data for fusion-relevant conditions---namely rarefied, high-temperature neutrals impinging on tungsten---remain uncertain \cite{Agrawal_2008, Porodnov_1974, Shields_1982, Shields_1980}. Although a systematic mapping between TRIM and CLL parameters is in principle possible \cite{eirene_d_on_w_trim}, it is outside the scope of the present work.

A heuristic but physically-motivated choice of accommodation coefficients is adopted:
\begin{itemize}
    \item The baseline NEAC is set to $\accN=0.2$, so that atomic reflection does not fully temperature-equilibrate with the wall. This choice is coherent with the neutral-temperature fields reported by Subba \textit{et al.}~\cite{Subba_2021}, where neutral remain above the 0.1-eV wall temperature across most of the divertor volume. A similar value of $\accN=0.1$ is adopted by Varoutis \textit{et al.}~\cite{Varoutis2023DIVGASDEMO}, who instead model molecules\footnote{In EIRENE wall models \cite{Reiter2019EIRENE}, by default atomic recycling proceeds through either fast reflection or thermal re-emission, with only the latter branch being reset to the wall temperature. In contrast, molecular recycling (i.e. alike \cite{Varoutis2023DIVGASDEMO}) defaults to fully thermal re-emission at the wall. Since the present CLL model is intended to represent an effective atom--molecule mixture, rather than a purely molecular population (section \ref{sec:methods__pop_phys__interactions__wall}), a somewhat larger $\accN$ than the molecular value used in \cite{Varoutis2023DIVGASDEMO} is expected.}.

    \item The baseline TMAC is chosen as $\accT = 0.93$ \cite{Porodnov_1974}, representing a strong sink of directed momentum. Since the ITER simulations do not show pronounced gradients in the toroidal mean neutral velocity (contours in figure~\ref{fig:ITER_123013}), this assumption is interpreted as conservative. Yet, TMAC remains an important parameter to be scanned \cite{Agrawal_2008}.
\end{itemize}

\paragraph{Plasma-like boundary conditions along the separatrix divertor legs}\label{sec:methods__pop_phys__interactions__plasma}

The present section embodies the \textit{key departure from conventional DSMC approaches}, where neutrals are typically injected with \textit{stationary} Maxwellian distributions (zero bulk velocity) \cite{VAROUTIS2019120, Tantos_2022, Varoutis_2024, Tantos_2024}.

Instead, plasma-facing boundaries along the inner and outer separatrix divertor legs are here modelled with a non-zero toroidal velocity $\vTorBC$, which attains opposite signs along each one---consistently with section \ref{sec:evidence__simulations}.

Each leg is also further divided into two regions, visible in figure \ref{fig:pop_phys_domain_sketch}:
\begin{itemize}
    \item \textit{Detached zones.} Inner and outer detached zones act as a neutral source, extending from the strike point towards the X-point. The fictitious detachment fronts sit at one fifth of the leg length from the strike point.
    
    Neutrals impinging upon this surface are 100\% re-emitted using a CLL boundary with $\accN=0.1$ and $\accT=0.3$. This boundary condition, not representing a material surface, is intended to mimic charge exchange and recombination processes in detached plasma, where neutrals retain most of the incoming ion energy (section \ref{sec:evidence__theory}). Elastic collisions, which transfer less than 100\% of energy, are implicitly averaged into the chosen accommodation coefficients (hence raising $\alpha_t$ from 0.0, if only RC/CX, to 0.3).
    
    A source is then added in this region to mimic recycling, with neutrals sampled from a Maxwellian distribution at nominal temperature $\TplasmaBC$ and translated by the imposed toroidal velocity $\vTorBC$.

    \item \textit{Plasma zones.} The remaining four fifths of the inner and outer SEP divertor legs extend from the detachment front to the X-point, and are treated as a plasma-like neutral sink. From the branching ratio\footnote{Specifically, \texttt{ADAS ADF11 SCD} for EI and \texttt{ADAS ADF11 CCD} for CX \cite{Summers2001ADAS, Summers2006ADAS}. In pure deuterium, $n_{\Dzero^+} = n_{\mathrm{e}}$, and therefore the branching ratio as a function of reactivities $\langle \sigma v \rangle$ matches that of reaction rates $n_{\mathrm{e}} \, n_{\Dzero} \, \langle \sigma v \rangle$.} of electron impact ionisation (EI) and charge exchange (CX), a recycling probability is defined as
    \begin{equation}\label{eq:branching_ratio}
        \BEIZ = \frac{\langle \sigma v\rangle_{\mathrm{EI}}}{\langle \sigma v\rangle_{\mathrm{EI}} + \langle \sigma v\rangle_{\mathrm{CX}}}.
    \end{equation}
    This represent the fraction of the neutrals impinging on the plasma leg which is removed and subsequently recycled in the detached region. The remaining fraction $(1-\BEIZ)$ undergo CX-like reflection. This is modelled using the same CLL parameters as in the detached zones ($\accN=0.1$ and $\accT=0.3$), and neutrals are sampled from the same Maxwellian distribution at temperature $\TplasmaBC$ drifting at $\vTorBC$.
\end{itemize}

\subsubsection{PoP-Phys simulation database}\label{sec:methods__pop_phys__database}

\paragraph{Scanned parameters and baseline values}\label{sec:methods__pop_phys__database__parameters}

Key parameters are either intrinsically uncertain or strongly intertwined in coupled edge-plasma simulations. They cannot be self-consistently resolved within a stand-alone DSMC framework, and are hence scanned independently.

\begin{figure}
    \centering
    \subfloat[]{\includegraphics[width = 0.45\textwidth]{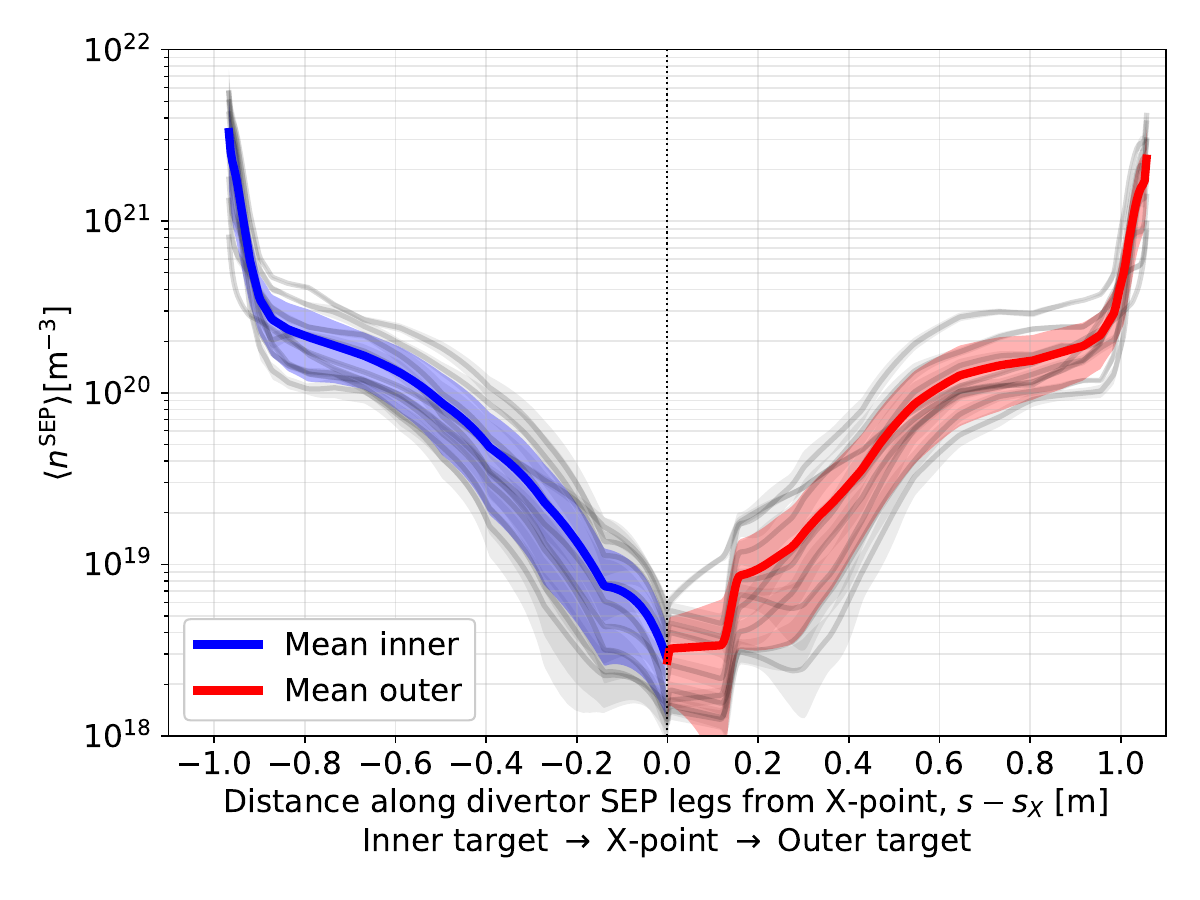}\label{fig:div_legs_density_Dplus2xD2}}\\
    \subfloat[]{\includegraphics[width = 0.45\textwidth]{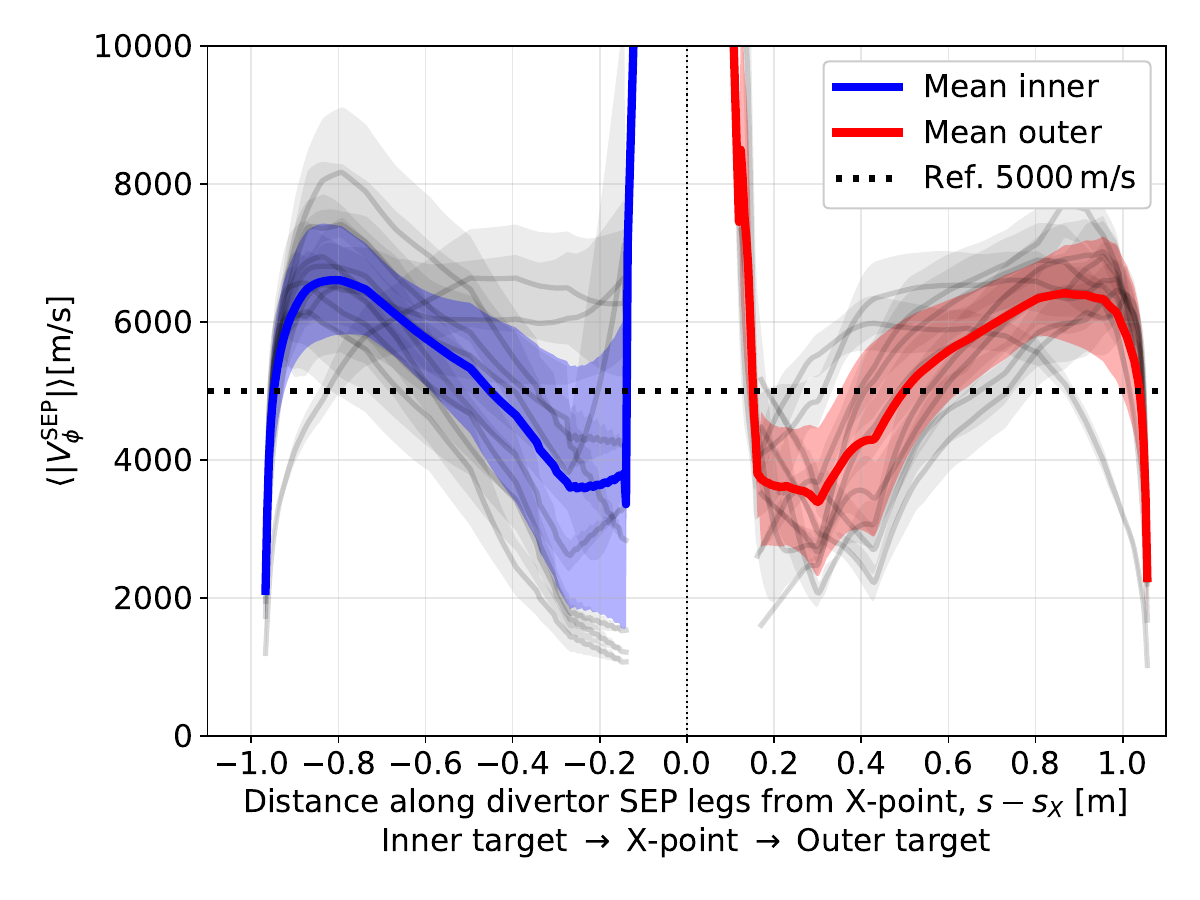}\label{fig:div_legs_velocity_z_Dplus2xD2}}\\
    \subfloat[]{\includegraphics[width = 0.45\textwidth]{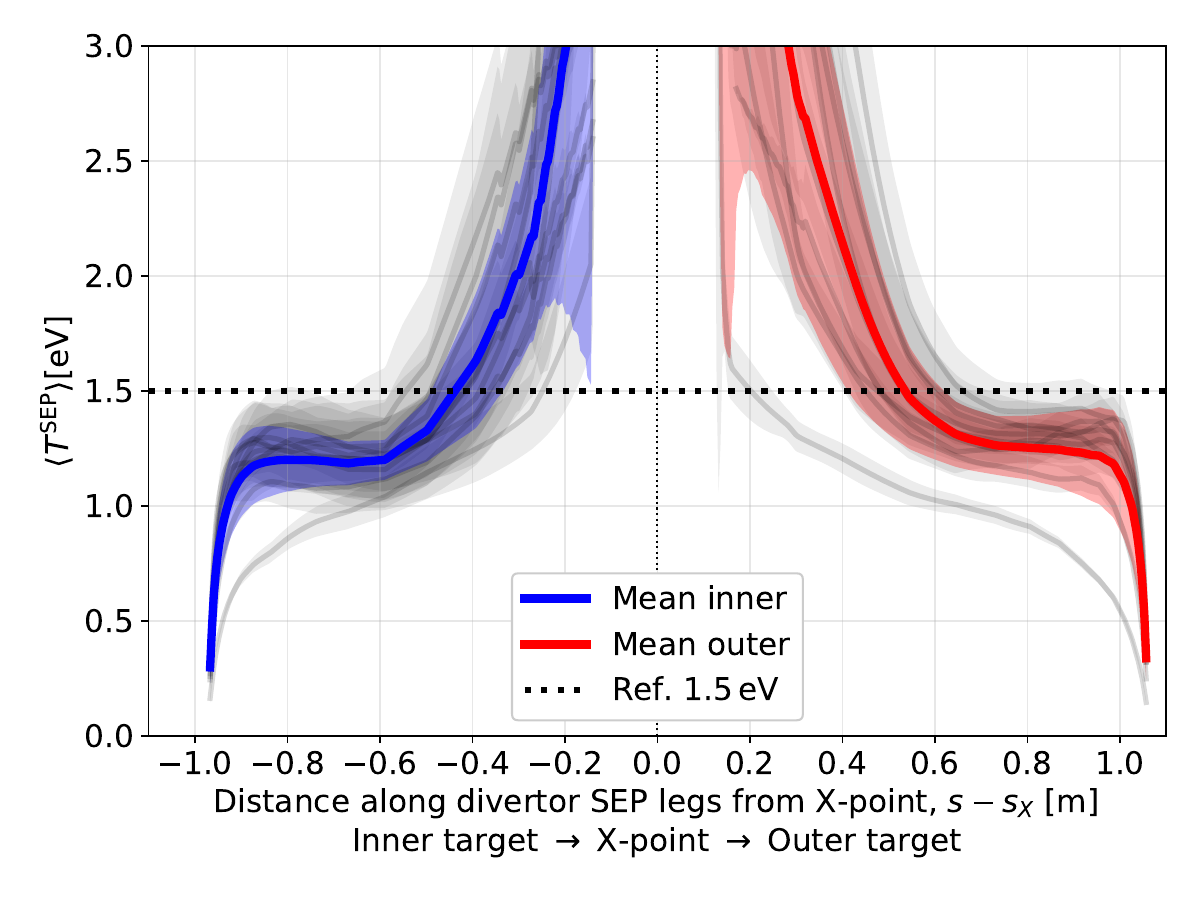}\label{fig:div_legs_temperature_Dplus2xD2}}
    \caption{Neutral density (a), magnitude of the toroidal velocity $|\vTor|$ (b), and temperature (c) averaged within the first $2~\mathrm{cm}$ into the private flux region (PFR) for atomic-equivalent deuterium. The underlying SOLPS-ITER simulations (grey curves) are those of Lore \textit{et al.} \cite{Lore_2022}. These include the triangle series (figure \ref{fig:ITER_123013_detachment_trend}) and diamond series (figure \ref{fig:ITER_123013_detachment_trend__d_ne_series}). The abscissa represents the distance along the legs, from the inner  to the outer targets, through the X-point. Dotted horizontal lines indicate the baseline values adopted in the present study.}
    \label{fig:div_legs}
\end{figure}

Representative figures are summarised below and collated in table \ref{tab:pop_phys_database}:

\begin{itemize}

    \item \textit{Geometry.} The radial and vertical extents of the PFR domain are varied in the ranges $L_r \in [0.20,\,0.80]~\mathrm{m}$ and $L_z \in [0.05,\,0.60]~\mathrm{m}$, spanning from compact COMPASS-U-/WEST-like configurations \cite{Komm_2024, Bucalossi_2022} up to divertor dimensions approaching ITER scale (figure \ref{fig:ITER_123013}). The corresponding hydraulic diameter attains values within $d_h \in [0.04, \, 0.29] \,\mathrm{m}$, with the baseline being $0.15 \,\mathrm{m}$.

    \item \textit{Neutral density.} A baseline average density of $\langle n \rangle = 10^{20}~\mathrm{m^{-3}}$ is adopted, representative of detached AUG conditions \cite{WU2023114023}. AUG measurements also report neutral densities stretching up to $10^{21}~\mathrm{m^{-3}}$ in high-density discharges \cite{Lang_2020}. The density is therefore scanned over the range $10^{19}$--$10^{21}~\mathrm{m^{-3}}$, spanning attached to deeply detached/fuelled regimes \cite{Kallenbach_2018}.
    
    \item \textit{Plasma-boundary temperature $\TplasmaBC$.} A baseline value $\TplasmaBC = 1.5~\mathrm{eV}$ is adopted, consistent with SOLPS-ITER averages near the divertor legs in figure \ref{fig:div_legs_temperature_Dplus2xD2}.

    Yet, the plasma-boundary temperature does not imply thermal equilibration with the plasma \cite{Tendler01031987}. This is consistent with SOLPS-ITER results showing that only a small fraction of high-energy, non-equilibrated neutrals populate the divertor legs (figure~6a of \cite{Moscheni_2022} and the discussion in section~5.2 therein, as well as figure~13a of \cite{Moscheni_2025}). This gives raise to the temperature spike in figure \ref{fig:div_legs_temperature_Dplus2xD2}, additionally plagued by reduced statistics. Although non-Maxwellian re-emission could be implemented, this remains outside the scope of the present work. Effective thermal equilibration instead occurs in the detachment zones, where neutral densities (figure \ref{fig:div_legs_density_Dplus2xD2}) and collision rates are high.
    
    The parameter $\TplasmaBC$ is then scanned in the range $1$--$50~\mathrm{eV}$.

    \item \textit{Imposed toroidal speed $\vTorBC$.} The imposed toroidal speed is the most influential control parameter. The baseline value $\vTorBC = \pm 5~\mathrm{km\,s^{-1}}$ is chosen based on experimental measurements (section \ref{sec:evidence__experiments}) and on ITER simulations \cite{Lore_2022} (figure \ref{fig:div_legs}). Its magnitude is then scanned over the range $1$--$20~\mathrm{km\,s^{-1}}$.

    \item \textit{Recycling probability.} The baseline recycling probability $\BEIZ$ in equation (\ref{eq:branching_ratio}) reads $0.50\pm0.05$ when computed as an average for plasma temperatures 1--200 eV and densities $10^{19}$--$10^{21} \,\mathrm{m}^{-3}$ \cite{Summers2001ADAS, Summers2006ADAS, Sciortino2021Aurora, Dux2004Habilitation}. This parameter is then scanned over $\BEIZ\in[0.50,\,0.95]$, conservatively accounting for differences between atomic and molecular deuterium channels, isotope effects, uncertainties in the local plasma conditions etc.

    \item \textit{Recycling flux.} In appendix \ref{apx:particle_balance}, the net injected flux $\GammaCirc$ is confirmed to be physically-realistic for detached conditions in the divertor sizes explored across the database.

\end{itemize}

The resulting Knudsen number, defined in equation~(\ref{eq:knudsen_number}), evaluates to 0.20 at the nominal density, but varies by nearly two orders of magnitude across the database, spanning the interval $[0.02,\,1.95]$. The flow therefore ranges from near-slip to clearly transitional conditions, justifying the adoption of an all-Knudsen-number kinetic (DSMC) approach.

The pressures instead spans $[0.3,\,70]~\mathrm{Pa}$. This broad stress-test envelope exceeds the highest values predicted in AUG \cite{Kallenbach_2018} (10 Pa) and ITER \cite{Lore_2022} (30 Pa), while approaching the realistic minimum of 0.1 Pa reported in AUG detached conditions \cite{Kallenbach_2018}.

Overall, the parameter space is sampled with at least four points along each control dimension, enabling the extraction of meaningful scaling trends.

\input{00_scaling_PoP_Phys_summary}

\paragraph{Vertical profile analysis: characteristic spatial relaxation and decay length}\label{sec:methods__pop_phys__decay}

\begin{figure}
    \centering
    \includegraphics[width = 0.475\textwidth]{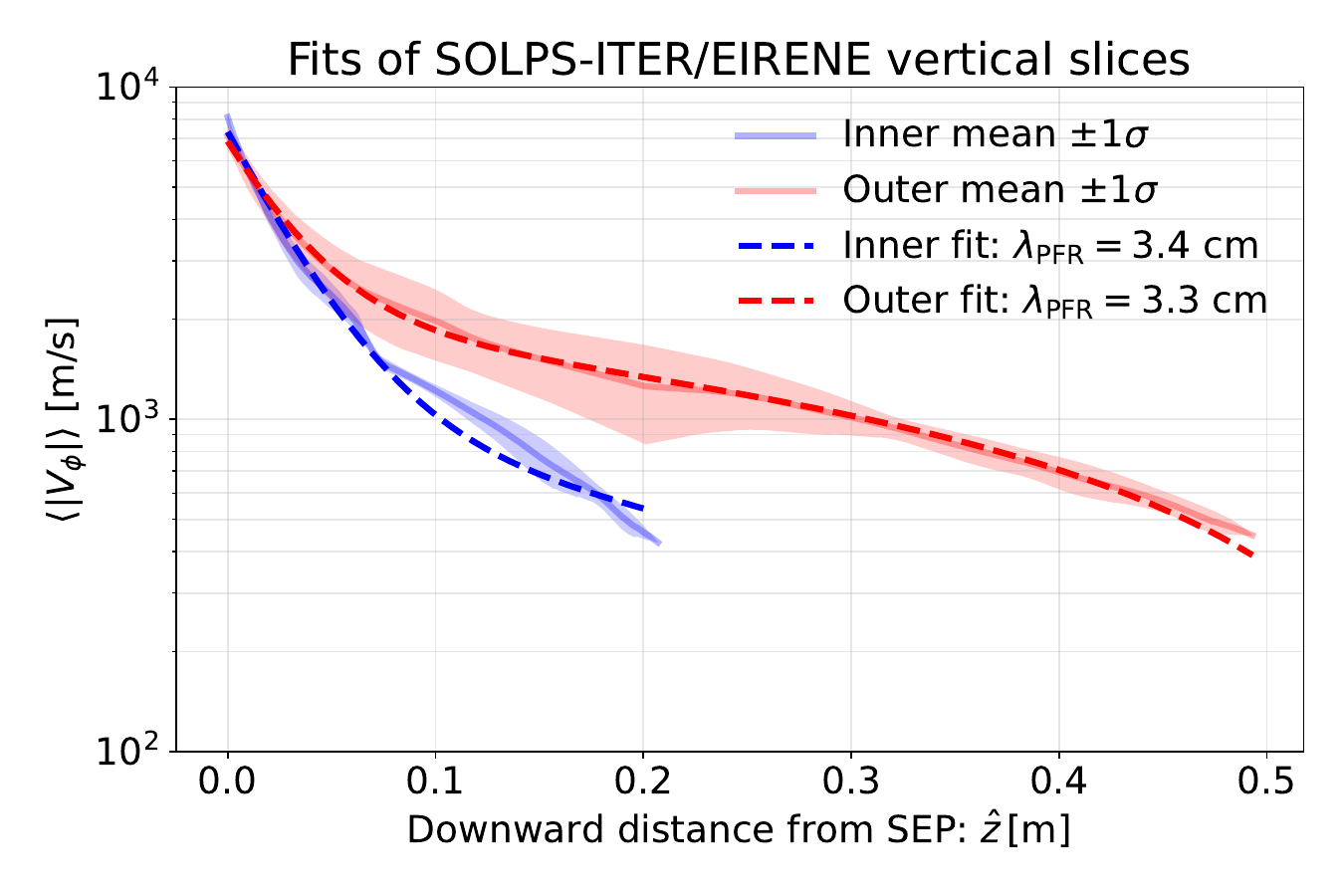}
    \caption{Vertical profiles of the toroidal neutral speed of atomic-equivalent deuterium in the SOLPS-ITER simulation of figure \ref{fig:ITER_123013_Dplus2_D2_velocity_z}. The spatial average is computed over $r\in[4.28,4.46]~\mathrm{m}$ and $r\in[5.27,5.52]~\mathrm{m}$ for the inner and outer target regions, respectively, i.e.\ above the divertor reflectors \cite{PITTS2019100696}. Different extents reflect the varying separatrix-to-wall distance. The fitting procedure used is described in appendix~\ref{apx:vphi_fitting}.}
    \label{fig:vphi_slice}
\end{figure}

Across all SPARTA and SOLPS-ITER (section \ref{sec:evidence__simulations__trends}) simulations exhibiting ordered toroidal neutral motion, a well-defined and reproducible cross-sectional relaxation pattern is observed. Figure \ref{fig:vphi_slice} provides a representative example. The magnitude of the toroidal neutral velocity decreases monotonically from the plasma-facing region (momentum-driving boundary) towards the wall, defining a characteristic sigmoid-like, two-exponential footprint. This is governed by a ``toroidal wind decay length'', denoted $\lambdaPfr$, which is a quantity of interest. Fundamentally, it represents the vertical spatial extent over which the toroidal wind exists in the PFR, not dissimilarly to plasma-related decay lengths \cite{Eich_2013, Brida_2025, stangeby2000plasma}. The full formulation of the newly-introduced fitting model to extract this parameter is provided in appendix~\ref{apx:vphi_fitting}.

Despite geometric variations and moderate slice-to-slice scatter, $\lambdaPfr \simeq 3.4 \,\mathrm{cm}$ is recovered across the inner and outer PFR domains of ITER, respectively, for the atomic-equivalent deuterium (figure \ref{fig:ITER_123013_Dplus2_D2_velocity_z}). When averaged over the full set of SOLPS-ITER-based simulations in figure \ref{fig:detachment_trend_Deq}, this procedure yields a characteristic $\lambdaPfr$ of order $3.8~\mathrm{cm}$.

\subsection{Exhaust proof-of-principle setup}\label{sec:methods__pop_exh}

\subsubsection{Geometry}\label{sec:methods__pop_exh__geometry}

The idealised PoP-Exh domain is displayed in figure \ref{fig:pop_exh_domain_sketch}. It is constructed in the toroidal plane $(r,\phi)$, where $\phi \, [\rm m]$ is the straightened toroidal coordinate (appendix \ref{apx:methods__pop_phys__geometry__simplification}). The domain represents the outer portion of a divertor PFR volume---up to a fictitious radial location \textit{before} the X-point (left), where the toroidal wind vanishes. This region therefore isolates the volume in which ordered toroidal motion is present.

Two simplified ducts are embedded in this domain to mimic figure \ref{fig:overview_wind}, and are separated sufficiently from one another to remain as mutually independent as possible. Control simulations omitting either duct confirm that differences are at the percent level and well below the intrinsic spatial variability of the fields.

The first duct has an inlet facing the PFR in the poloidal direction and develops radially outward along $r$. It is representative of a conventional, pressure-driven IPDP-like capture geometry (section \ref{sec:philosophies__ipdp}) and diagnostic approaches (e.g. figure 1 of \cite{SHAFER2019487}). The second duct is identical but oriented toroidally, with its inlet facing the direction of the ordered toroidal neutral wind---thus representing the proposed TFP strategy. This arrangement also remarks that the two approaches could be combined in a complementary particle-exhaust architecture (section \ref{sec:tfp__features}).

Both ducts feature a length of $35~\mathrm{cm}$ and an inlet width of $3~\mathrm{cm}$---computationally convenient, as this lowers the active volume. No noticeable change in the primary judgement metrics is observed when doubling the duct width. The $3~\mathrm{cm}$ width is therefore retained for the present study---although systematic geometric variations of the inlets would be worthwhile investigating.

    \input{00_PoP_Exh_domain}

\subsubsection{Neutral species}\label{sec:methods__pop_exh__neutrals}

In addition to atomic deuterium, helium is included as a proxy for minority impurity transport \cite{Reiter_1991, Zito_2025}. Its concentration is set to approximately $10\%$ to keep the DSMC problem computationally tractable, while retaining enough statistics. This value is deliberately higher than helium concentrations expected in ITER-relevant scenarios, of order $0.05\%$ (figure \ref{fig:detachment_trend_He}). However, it remains representative of helium-seeding experiments \cite{Wade_1995} and of divertor impurity concentrations reached in detached AUG conditions, for example for nitrogen seeding (figure~2b of \cite{Kallenbach_2018}). Future targeted studies should address concentration-related effects.

Both He--He and D--He collisions are treated using the same VSS framework of section \ref{sec:methods__pop_phys__interactions__neutrals}, with more details reported in appendix \ref{apx:numerics__collisions}.

    \subsubsection{Boundary conditions}\label{sec:methods__pop_exh__boundary_conditions}

    Identical boundary conditions are imposed for deuterium and helium, implicitly assuming flow entrainment and thermalisation---the latter being supported by SOLPS-ITER results in section \ref{sec:evidence__simulations__trends}. Any difference in the response of the two species can hence be attributed to relative concentration and mass-dependent transport effects.

A velocity boundary condition $\vTorBC<0$ is imposed at the top boundary of the PFR volume in figure \ref{fig:pop_exh_domain_sketch}, which acts as the inlet for the ordered toroidal neutral flow. The outlet boundaries around the entrance of the toroidal duct are assigned an effective capture coefficient in the range 0.35--0.75, the latter representing the baseline. This choice mimics periodic boundaries that cannot be imposed in the present 2D configuration. Relative variations of $\lesssim 6\%$ are achieved for the leading quantities in the baseline simulation throughout the PFR---confirming a satisfactory uniformity.

A symmetry, ``free'' boundary condition is applied on the left boundary, facing towards---but not sitting at---the X-point. This avoids imposing an artificial sink or source and allows the local neutral distribution to adjust self-consistently.

The plasma-facing outer target is assigned a temperature of $0.1~\mathrm{eV}$, while non-plasma-facing duct walls sit at $0.05~\mathrm{eV}$, representative of colder sub-divertor structures \cite{Zito_2025}.

The downstream ends of both ducts are treated as perfect sinks, with $100\%$ absorption. This mimics ideal particle removal and enables a direct comparison between the conventional poloidal duct and the toroidally-oriented TFP duct under identical exhaust conditions.

All remaining boundary and interaction models match those described in section \ref{sec:methods__pop_phys__interactions}.

\subsubsection{Output judgement metrics}\label{sec:methods__pop_exh__metrics}

The advantage of the toroidally-oriented duct relative to the conventional poloidal duct is quantified through the pressure gain
\begin{equation}\label{eq:pressure_gain}
    \pGain(\alpha) = \frac{\pTor(\alpha)}{\pPol(\alpha)} ,
\end{equation}
which enters equation (\ref{eq:throughput_gain}). Here $\pPol$ and $\pTor = \pPol + \Delta\pTor$ are the partial pressures of species $\alpha$ evaluated mid-way along the poloidal and toroidal ducts (magenta boxes in figure \ref{fig:pop_exh_domain_sketch}), respectively. In the toroidal duct, $\Delta \pTor > 0$ is the additional momentum-flux-related contribution. The results are insensitive, within the reported error bars, to moderate variations of the axial location of the sampling regions.

The back-flow fraction for a certain species can be evaluated as:
\begin{equation}\label{eq:backflow_fraction}
    f_{\rm back} = \frac{\Gamma_{\rm back}}{\Gamma_{\rm in}}
\end{equation}
for a pump inlet crossed by an incoming ($\Gamma_{\rm in}$) and a back-flowing ($\Gamma_{\rm back}$) particle flux. This metric therefore quantifies how many particles come back for every one entering, i.e. the effective albedo \cite{Tantos_2022, Tantos_2024}.

Finally, the characteristic decay length of $\vTor$ along the toroidal duct is $\lambdaDuct$, obtained by fitting the velocity profile with a standard exponential-decay function starting from the duct inlet.

    \subsubsection{PoP-Exh simulation database}\label{sec:methods__pop_exh__database}

The main control parameters are scanned to assess the robustness of the exhaust proof-of-principle. The resulting macroscopic quantities, averaged across the PFR volume upstream the ducts, are reported in table \ref{tab:pop_exh_database}. These values remain consistent with, and largely overlap to, those obtained in PoP-Phys, confirming that the PoP-Exh simulations operate within the same physically-relevant region of parameter space. Here, however, TMAC $\alpha_t$ is kept constant at a value of 0.80.

As an \textit{indicative} measure of rarefaction in the PFR volume, the Knudsen number is computed using the equivalent hydraulic diameter $d_h=0.15~\mathrm{m}$ of the baseline PoP-Phys poloidal geometry (section \ref{sec:methods__pop_phys__geometry}). This value is intended to approximately characterise the flow upstream of the duct entrances. Along the ducts themselves, the local Knudsen number can vary significantly owing to changes in density.

\input{00_scaling_PoP_Exh_summary}

\subsection{Numerics: DSMC good practices.}\label{sec:methods__dsmc_numerics}
\label{subsec:numerics_dsmc}

In all simulations, standard Bird-type best practices for rarefied gas flows are followed \cite{GLEASONGONZALEZ20141042, GLEASONGONZALEZ2016693, VAROUTIS2019120, Tantos_2022, Varoutis_2024, Tantos_2024, TANTOS2025115021}. Numerical parameters are chosen to accurately resolve collisional momentum transport over millimetre spatial scales, with particular emphasis on preserving bulk motion and ensuring small Monte Carlo statistical errors on velocity and pressure at steady-state. Spatial and temporal resolutions satisfy conventional DSMC constraints across all regimes considered.

Full details of the numerical setup are provided in appendix~\ref{apx:numerics}.

\section{Results}\label{sec:results}

\subsection{Physics proof-of-principle}\label{sec:results__pop_phys}

\subsubsection{Baseline}\label{sec:results__pop_phys__baseline}

The baseline DSMC simulation enforces the conditions described in section \ref{sec:methods__pop_phys}, featuring an imposed toroidal boundary velocity $\vTorBC = \pm 5~\mathrm{km\,s^{-1}}$. The resulting steady-state neutral population exhibits an average density of $(1.0 \pm 0.3)\times10^{20}~\mathrm{m^{-3}}$, corresponding to transitional conditions with $\mathrm{Kn} = 0.20 \pm 0.06$. With a temperature of $0.6 \pm 0.1~\mathrm{eV}$, alike \cite{Welch_2001}, the resulting pressure is $9 \pm 3~\mathrm{Pa}$. All macroscopic quantities remain spatially smooth across the domain.

\begin{figure*}
    \centering
    
    \subfloat[]{\includegraphics[width = 0.475\textwidth]{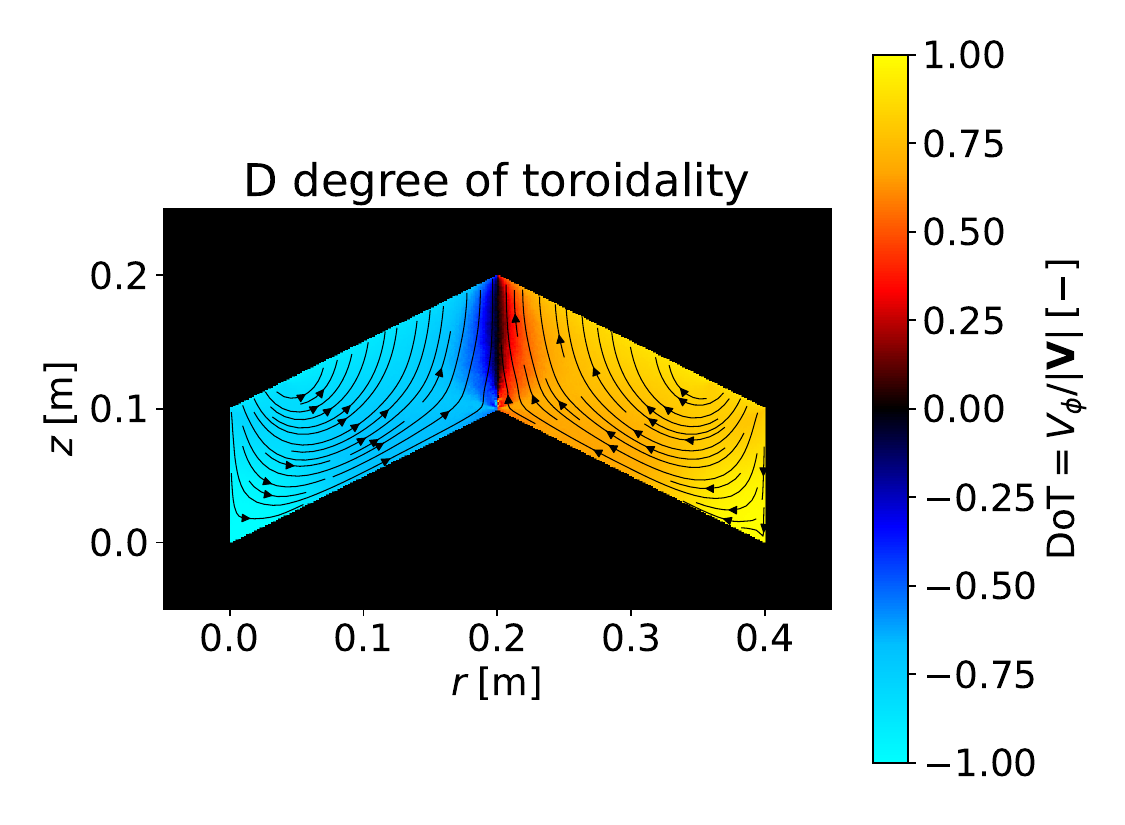}\label{fig:pop_phys_baseline__dot}}\hspace{0.05cm}
    \subfloat[]{\includegraphics[width = 0.475\textwidth]{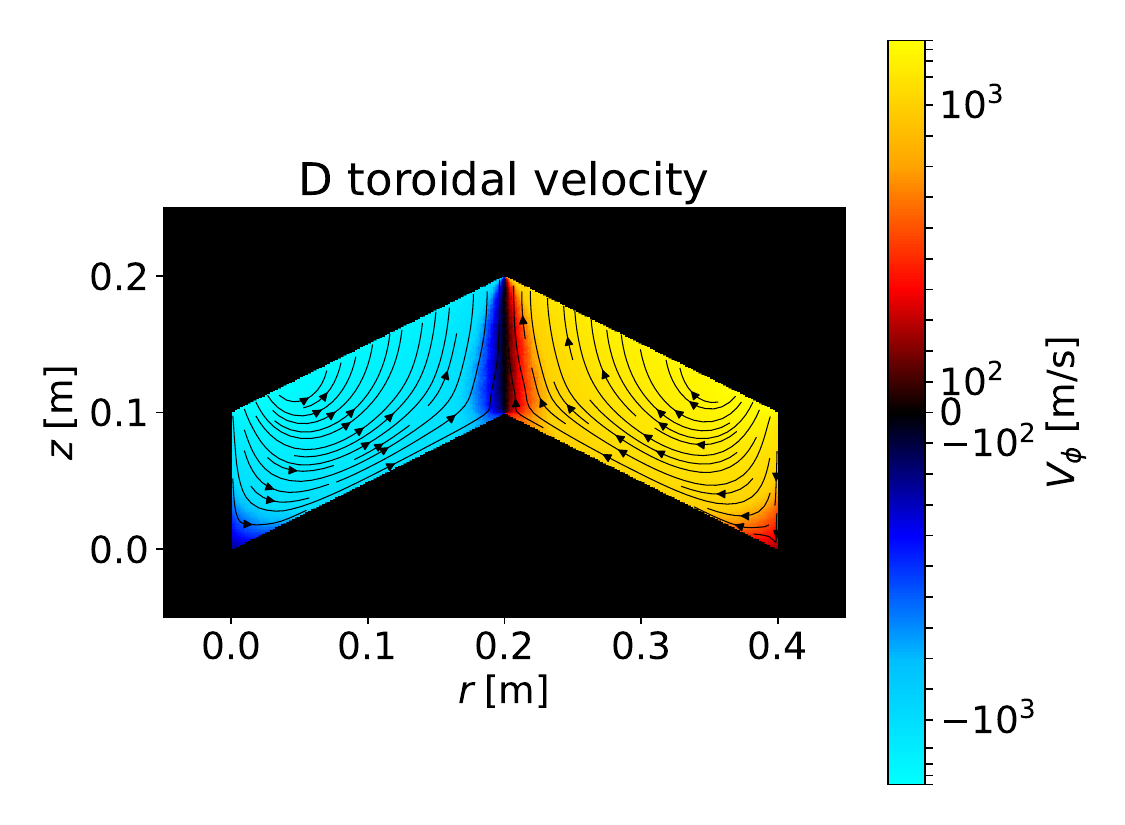}\label{fig:pop_phys_baseline__vphi}}
    
    \caption{SPARTA results: baseline simulation of physics proof-of-principle (section \ref{sec:methods__pop_phys}), with $\vTorBC = \pm 5~\mathrm{km\,s^{-1}}$, $\langle n\rangle \simeq 10^{20}~\mathrm{m^{-3}}$, and $\langle p\rangle \simeq 9~\mathrm{Pa}$. (a) Degree of toroidality $\DoT(r,z)$. (b) Toroidal velocity field $\vTor(r,z)$. Streamlines of the poloidal velocity field are shown in black.}
    \label{fig:pop_phys_baseline}
\end{figure*}

The main outcomes are summarised in figure \ref{fig:pop_phys_baseline}. Panels~(a) and~(b) show the spatial distribution of the degree of toroidality, $\mathrm{DoT}$, and the toroidal velocity field $\vTor$, respectively. Superimposed streamlines represent the poloidal velocity field $\vPolVec = [\vR, \vZ]$.

On a macroscopic level, both quantities attain large values throughout the regions of interest, with domain-averaged values $\absavg{\vTor} = 2.1 \pm 0.9~\mathrm{km\,s^{-1}}$ and $\absavg{\DoT} = 0.81 \pm 0.08$. Locally, the flow structure reveals the two expected counter-streaming toroidal wind currents (section \ref{sec:evidence__simulations}). In the present symmetric geometry, the $\vTor = 0$ surface lies midway between the two divertor legs, consistent with the imposed odd symmetry of the boundary conditions.

Taken together, the velocity fields describe a forced poloidal convection cell, with particle sources localised in the detached regions and sinks along the plasma-like boundaries. The presence of a finite toroidal component superimposed on this circulation implies two counter-rotating half-helicoidal convective motions of neutral particles through the divertor volume.

\begin{figure}
    \centering

    \subfloat[]{\includegraphics[width = 0.475\textwidth]{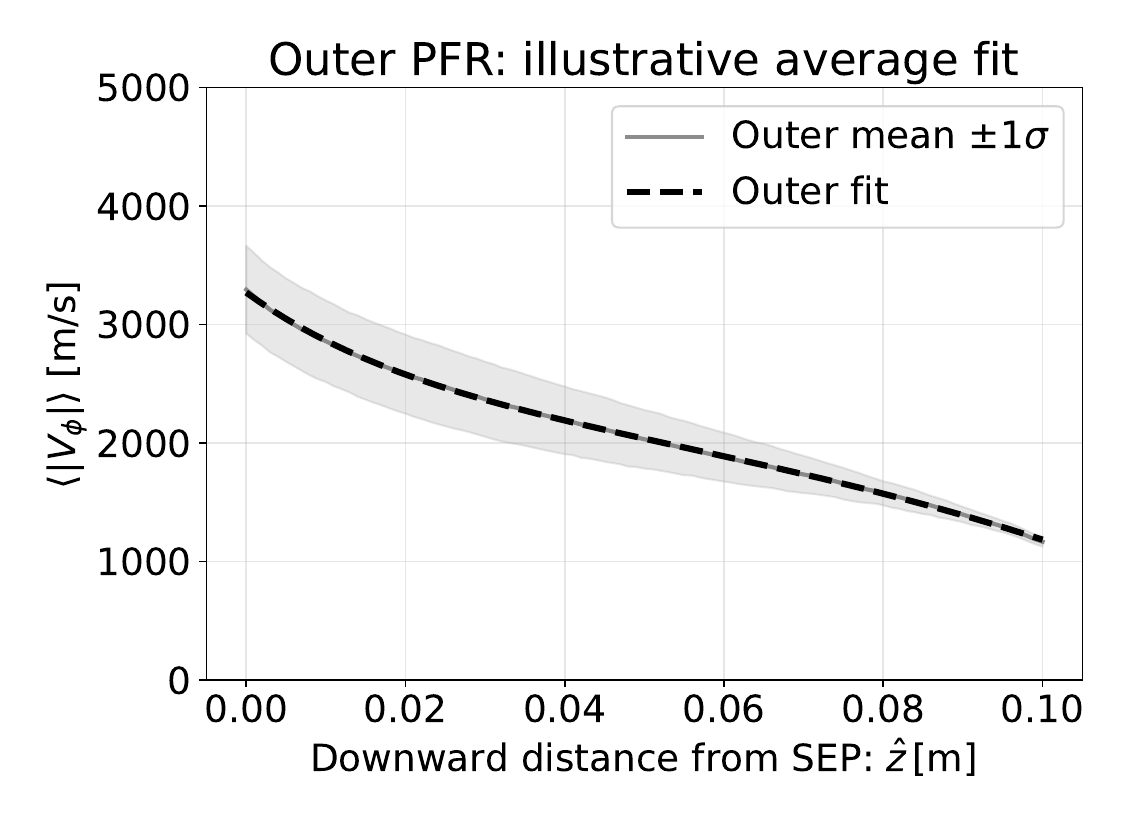}\label{fig:pop_phys_baseline__fit_slice}}\\
    \subfloat[]{\includegraphics[width = 0.475\textwidth]{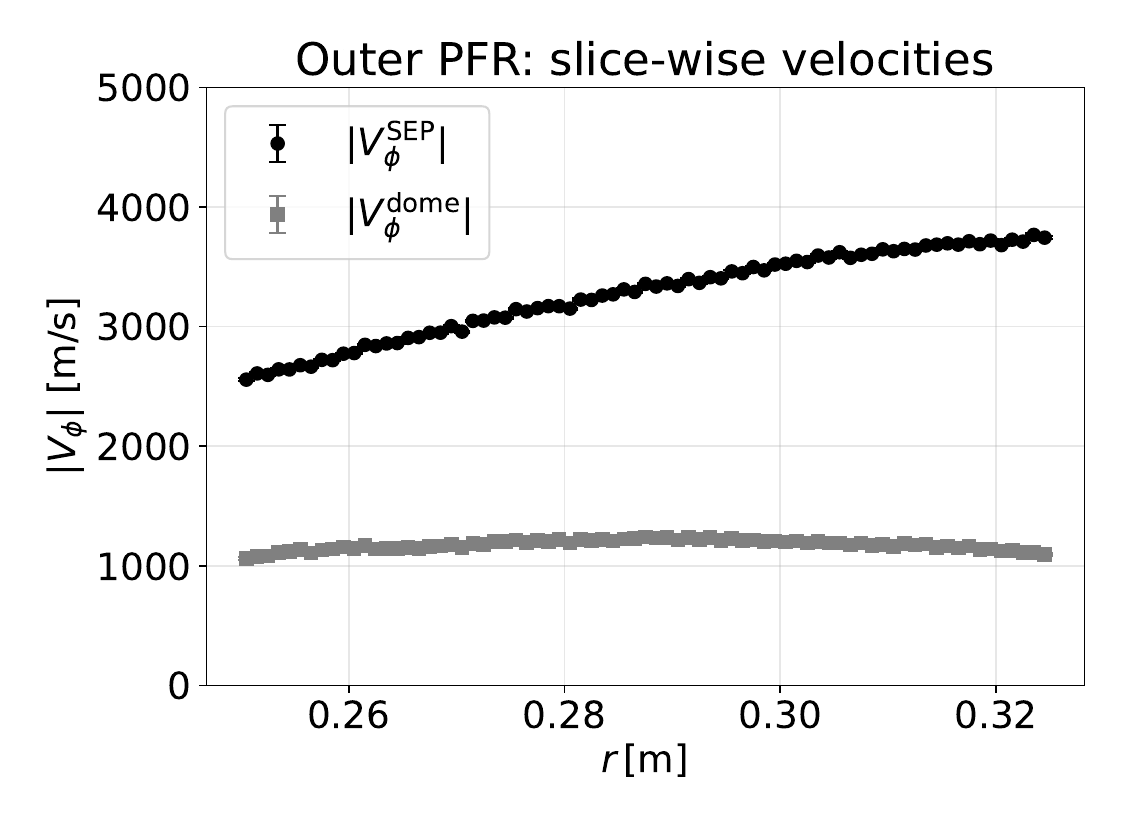}\label{fig:pop_phys_baseline__fit_vphi}}\\
    \subfloat[]{\includegraphics[width = 0.475\textwidth]{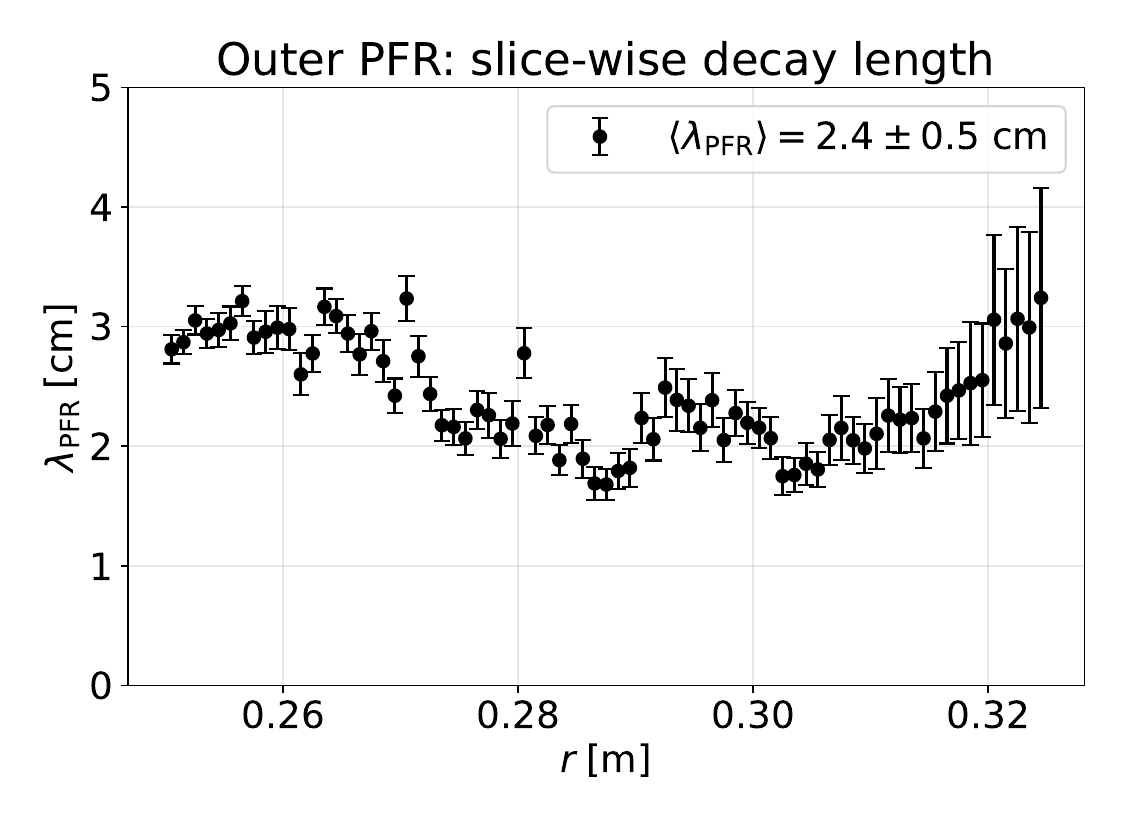}\label{fig:pop_phys_baseline__fit_lpfr}}
    
    \caption{SPARTA results from the same simulation in figure \ref{fig:pop_phys_baseline}. (a) Vertical velocity profiles averaged in the outer private-flux region and illustrative fit (appendix \ref{apx:vphi_fitting}). The shape is reminiscent of figures 1.10 of \cite{Agrawal_2020} and 7.4 of \cite{Roohi_2025couette} in transitional conditions. (b) Macroscopic toroidal velocities obtained via the actual slice-wise fit, as a function of $r \in [0.250,0.325]~\mathrm{m}$. (c) Corresponding toroidal wind decay length, $\lambdaPfr$, over the same radial interval.}
    \label{fig:pop_phys_baseline__fits}
\end{figure}

The quantitative characterisation of the toroidal flow is obtained by fitting $|\vTor|$ along vertical slices (section \ref{sec:methods__pop_phys__decay}). The average behaviour is illustrated in figure \ref{fig:pop_phys_baseline__fit_slice}. The slice-wise fitted profiles exhibit excellent agreement with $R^2 = 0.99$. The extracted effective toroidal velocities at the separatrix boundaries, $|\vTorSep|$, and at the dome, $|\vTorWall|$, are both in the kilometre-per-second range and vary smoothly with $r$ in figure \ref{fig:pop_phys_baseline__fit_vphi}. Representative values read $\absavg{\vTorSep} = 3.3 \pm 0.04~\mathrm{km\,s^{-1}}$ and $\absavg{\vTorWall} = 1.18 \pm 0.04~\mathrm{km\,s^{-1}}$.

The effective velocity at the plasma-facing boundary $\absavg{\vTorSep}$ is systematically lower than the imposed $|\vTorBC| = 5.0 \, \mathrm{km\,s^{-1}}$, likely owing to velocity slip \cite{Roohi_2025couette}, the TMAC/NEAC settings (section \ref{sec:methods__pop_phys__interactions__plasma}), collisional redistribution, thermal dispersion, and the superimposed poloidal convection. This confirms that the baseline choice remains conservative. The reduction of $\absavg{\vTorSep}$ from right to left is reminiscent of figure \ref{fig:div_legs_velocity_z_Dplus2xD2} (red) obtained by SOLPS-ITER---although other plasma-related effects/inhomogeneities might play a role there.

Notably, the near-dome velocity remains substantial, indicating that wall friction alone does not suppress the ordered toroidal motion.

The average characteristic decay length $\langle\lambdaPfr\rangle$ is shown in figure \ref{fig:pop_phys_baseline__fit_lpfr} and evaluates to $2.4 \pm 0.5~\mathrm{cm}$. Both the magnitude and the qualitative shape of the profiles are consistent with EIRENE simulations (figure \ref{fig:vphi_slice}), despite the fundamentally different numerical approaches and collision treatments. This agreement suggests that the present DSMC model captures the dominant physical mechanisms responsible for the ordered toroidal neutral motion---although more in depth comparisons should be carried out.

\subsubsection{Scaling laws and correlations}\label{sec:results__pop_phys__database}

\begin{figure*}
    \centering
    
    \subfloat[]{\includegraphics[width = 0.45\textwidth]{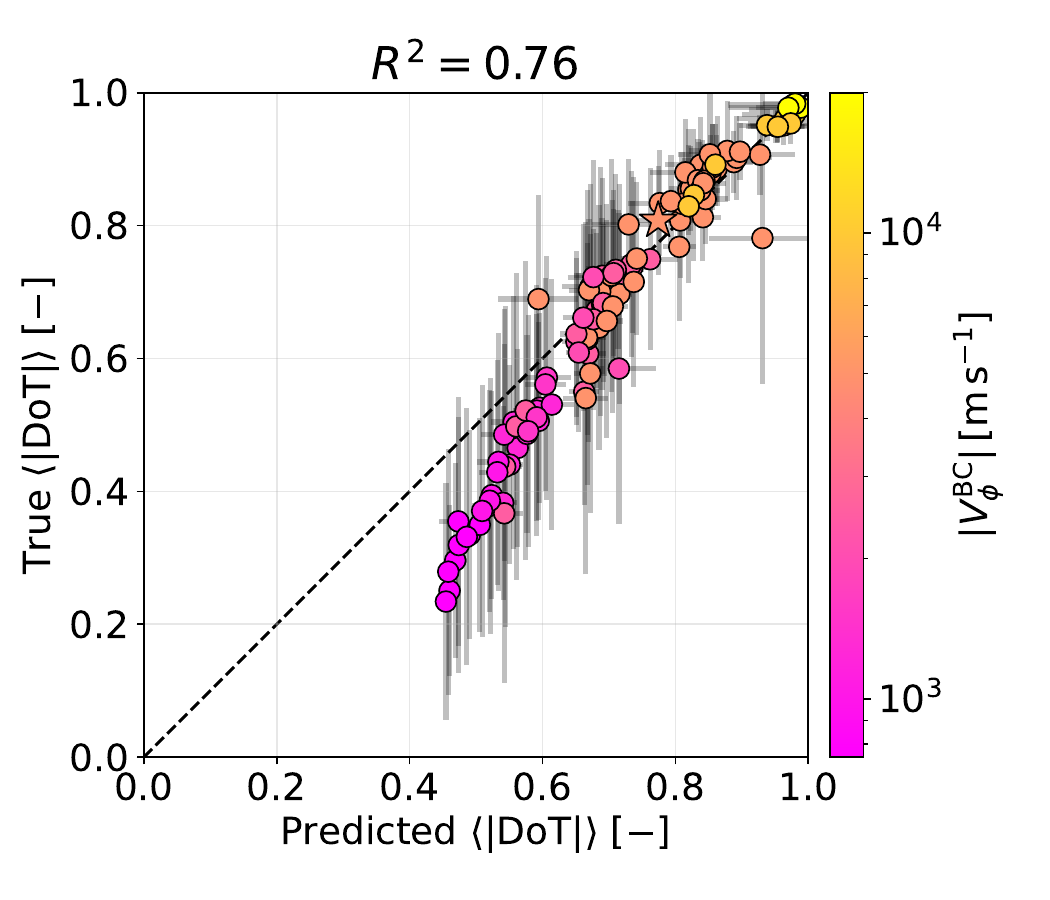}\label{fig:pop_phys_database__dot}}
    \hspace{0.05 cm}
    \subfloat[]{\includegraphics[width = 0.45\textwidth]{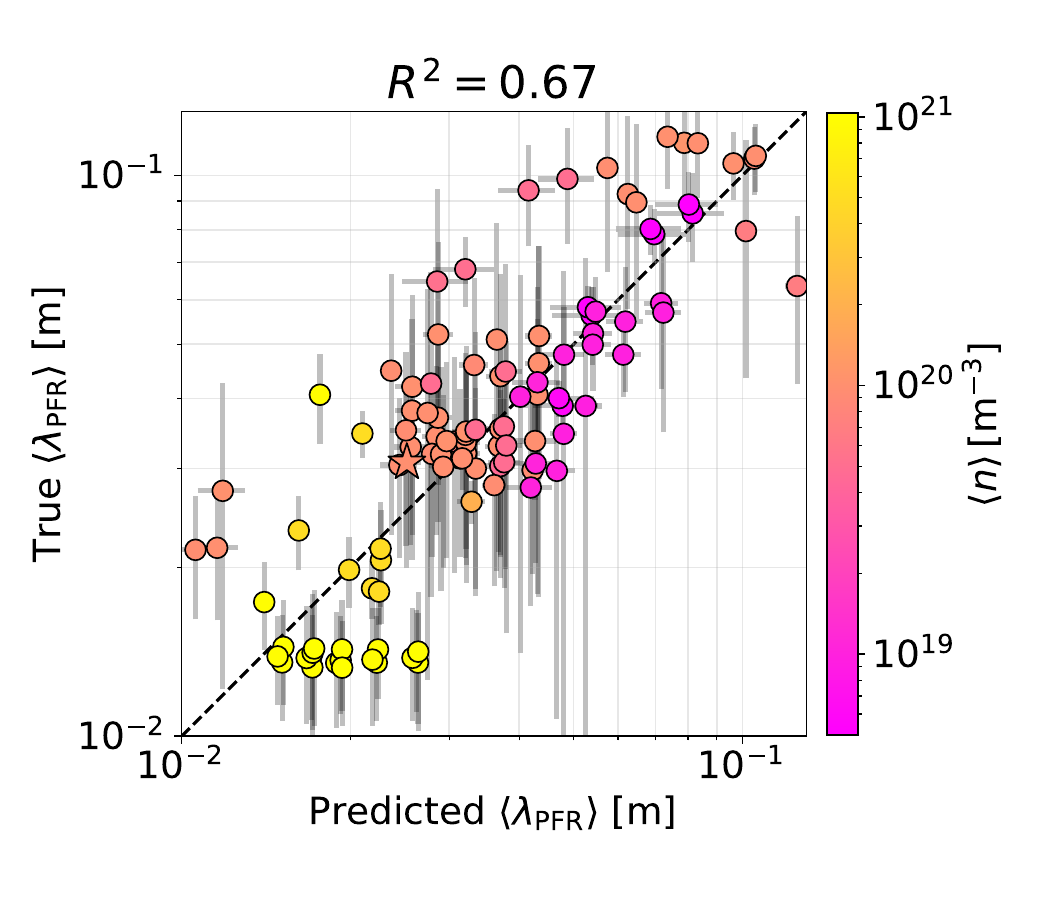}\label{fig:pop_phys_database__lpfr}}
    \\
    \subfloat[]{\includegraphics[width = 0.45\textwidth]{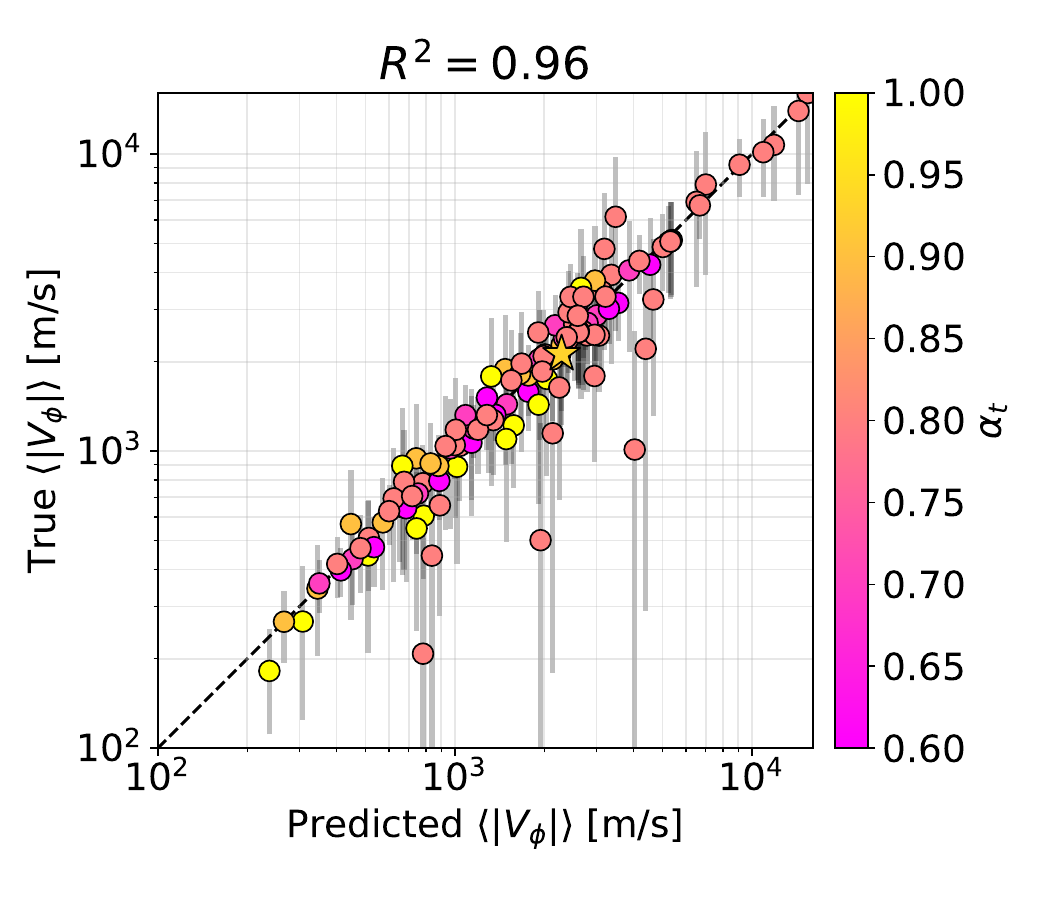}\label{fig:pop_phys_database__vphi}}
    \hspace{0.05 cm}
    \subfloat[]{\includegraphics[width = 0.45\textwidth]{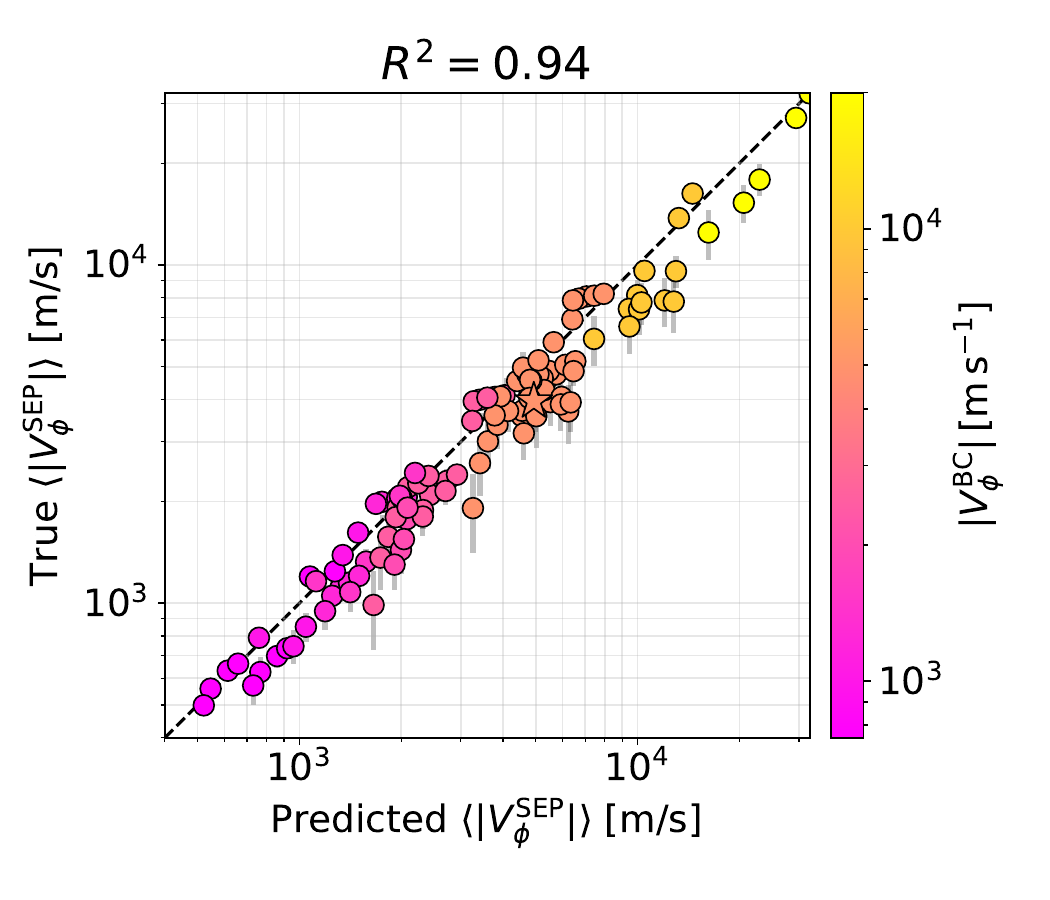}\label{fig:pop_phys_database__vphi_sep}}
    
\caption{Regression results across the physics proof-of-principle database for the main average quantities: (a) degree of toroidality, $|\DoT|$; (b) toroidal wind decay length in the private flux region, $\lambdaPfr$; (c) toroidal speed, $|\vTor|$; (d) toroidal speed along the divertor separatrix legs, $|\vTorSep|$. Vertical bars represent the standard deviation from the space-average procedure. Horizontal error bars quantify the uncertainty in the regressions. The baseline simulation of figure \ref{fig:pop_phys_baseline} is starred.}

    \label{fig:pop_phys_database}
\end{figure*}

Scanning parameters according to table \ref{tab:pop_phys_database} around the baseline operating point yields a database of 124 independent SPARTA simulations. The explored parameter space and the resulting trends are summarised in figure \ref{fig:pop_phys_database} with the corresponding goodness-of-fit ($R^2$), while the regressions are reported in the following. These bear SI units, $\langle T \rangle$ is in [eV], the rescaling constant implicitly embeds the dimensional adjustments and all the fitted parameters are displayed till the first uncertain digits. The relative $1\sigma$ uncertainty is 5.1\% for the exponents, on average, with a maximum of 22\%. For the prefactors, these read 16\% and 47\%, respectively.

Though not intended for detailed design studies nor extrapolation, the proposed regressions provide satisfactory agreement and enable meaningful sorting of the database with respect to the highlighted macroscopic variables.

\paragraph{Degree of toroidality.}\label{sec:results__pop_phys__database__dot}

Across the database, the high degree of toroidality is found to be robust:
\begin{equation}
\label{eq:pop_phys_database_scaling_dot}
\begin{aligned}
\absavg{\DoT} =\,&
0.014 \; \langle n\rangle^{0.021} \; \langle T\rangle^{-0.21}\\ & \, \times\, |\vTorBC|^{0.32} \; \alpha_t^{-0.28} \; \BEIZ^{-0.32}.
\end{aligned}
\end{equation}
In roughly half of the cases, the domain-averaged degree of toroidality satisfies $\absavg{\DoT} \geq 0.60$, with the fitted scaling reaching $R^2 = 0.76$ within this subset. All fitted exponents remain moderate in magnitude, and no exponent exceeds $0.32$ in absolute value. This weak-to-moderate parametric dependence indicates that, once established, toroidal ordering tends to persist robustly across the scanned operating conditions.

Notably, the scaling law for $\absavg{\DoT}$ requires $\langle n \rangle$ and $\langle T \rangle$ to enter independently, rather than through the combined average pressure $\langle p \rangle = \langle nT \rangle$. Physically, higher $\langle n \rangle$ enhances viscous transport and flow cohesion, whereas $\langle T \rangle$ increases the magnitude of thermal velocity dispersion, which blurs directed motion. Collapsing both effects into $\langle p \rangle$ would obscure this competition, instead clear from the fitted exponents, $0.021$ for density and $-0.21$ for temperature.

The positive exponent of $|\vTorBC|$ confirms that stronger toroidal speed promotes toroidal ordering. Conversely, the negative exponent of the momentum accommodation coefficient $\alpha_t$ indicates that increasing wall friction reduces the persistence of ordered toroidal motion. A degradation of toroidality is also observed at high recycling probability $\BEIZ$, which reflects the competition between toroidal ordering and ionisation-driven poloidal momentum drive.

The quality of the scaling deteriorates once $\absavg{\DoT}$ drops below approximately 0.60, and particularly below a threshold of approximately $|\vTorBC| \simeq 2~\mathrm{km\,s^{-1}}$. At low toroidal drive, the neutral population progressively approaches an isotropic state, and any correlation must vanish: a stagnant gas in a box features $\DoT = 0$ irrespective of any parameters.

Crucially, however, net toroidal motion is observed in the baseline case and in experiments (section~\ref{sec:evidence__experiments}). The collapse of $\absavg{\DoT}$ at low drive therefore implies that the mere \textit{presence} of toroidal speed is not sufficient to produce an organised toroidal neutral wind. The toroidal drive must also be \textit{strong enough} relative to thermal dispersion, wall losses, and poloidal convection to emerge as a measurable ordered component.

\paragraph{Toroidal velocities.}\label{sec:results__pop_phys__database__vphi}

The toroidal velocities $\absavg{\vTorSep}$ and $\absavg{\vTor}$ are, as expected, primarily controlled by the imposed boundary velocity $\vTorBC$ and scale approximately linearly with it:

\begin{equation}\label{eq:pop_phys_database_scaling_v}
\begin{aligned}
\absavg{\vTor} =
0.17 & \; \langle p\rangle^{0.110} \; |\vTorBC|^{0.99} \; \alpha_t^{-1.03} \;  d_h^{-0.39},\\
\absavg{\vTorSep} =
1.3 & \; \langle p\rangle^{0.143} \; |\vTorBC|^{0.96} \; \alpha_t^{-0.39} \; \BEIZ^{0.38}.
\end{aligned}
\end{equation}

However, $\absavg{\vTorSep}$, which characterises the \textit{effective} toroidal velocity at the plasma-like boundary, systematically satisfies $|\absavg{\vTorSep}| < |\vTorBC|$ and exhibits measurable dependence on additional parameters.

Conversely to the degree of toroidality, both $\absavg{\vTorSep}$ and $\absavg{\vTor}$ are well described when expressed in terms of the average pressure $\langle p \rangle$. Higher neutral divertor pressure---commonly a synonym of higher density---is here systematically associated to stronger average toroidal flows. This behaviour is consistent with enhanced collisional coupling at higher density, which promotes collective ordering (section \ref{sec:evidence__simulations__trends}).

The dataset in figure \ref{fig:pop_phys_database__vphi} is coloured by $\alpha_t$. While the smallest values of $\absavg{\vTor}$ are expectedly associated with the largest $\alpha_t$ (i.e. largest friction), velocities of comparable magnitude are also observed at the lowest $\alpha_t$. This indicates that wall friction contributes to the reduction of toroidal flow, but is not the dominant control parameter across the database.

When the remaining parameters are consider, the scaling of the two velocities begin to differ. The volume-averaged toroidal velocity $\absavg{\vTor}$ shows no dependence on $\BEIZ$, indicating decoupling between toroidal momentum transport and poloidal convection driven by ionisation losses---at least in a domain-averaged sense. Additionally, the two also differs markedly in a geometrical sense. The near-plasma $\absavg{\vTorSep}$ intuitively appears insensitive to the extent of the domain, i.e. hydraulic diameter $d_h$. Instead, a larger volume includes extended regions remote from the toroidal momentum source, thereby reducing the volume-averaged toroidal velocity $\absavg{\vTor}$.

\paragraph{Toroidal flow decay length.}\label{sec:results__pop_phys__database__lpfr}

The decay length $\lambdaPfr$ displays noticeable scatter but nevertheless follows a consistent scaling:
\begin{equation}
\label{eq:pop_phys_database_scaling_lpfr}
\langle \lambdaPfr \rangle =\,
1.4 \; \langle p\rangle^{-0.22} \; \alpha_t^{-1.1} \; d_h^{1.9}.
\end{equation}
Thirteen cases are excluded due to lack of fit convergence: in domains with large vertical extent $L_z$, secondary flow structures develop near the bottom wall---expected in the absence of dome clearly separating inner and outer volumes \cite{IAEA_TECDOC_2049_2024}. These locally invalidate the assumed monotonic decay shape of appendix~\ref{apx:vphi_fitting}.

Across the 111 well-behaving cases, $\lambdaPfr$ remains in the centimetre to decimetre range and shows a dominant dependence on the hydraulic diameter $d_h$---which increases the distance between the plasma boundary source and the wall friction sink. \textit{Ordered toroidal transport therefore systematically persists over macroscopic distances in all scanned configurations}, in contrast to characteristic plasma decay lengths in the SOL \cite{Eich_2013} and PFR \cite{Brida_2025}---in the order of sub-centimetre-to-centimetre. Hence, an inlet positioned centimetres away from the plasma legs would intercept a finite directed neutral flux and a vanishing plasma flux---an advantage further discussed in section \ref{sec:discussion__integration}.

Unlike the velocity regressions, $\lambdaPfr$ features a negative exponent in $\langle p \rangle$. As a result, higher pressure is associated to higher near-plasma toroidal speed while shortening the characteristic decay length---a steeper but elevated velocity decay profile.

\paragraph{Synthesis.}\label{sec:results__pop_phys__database__synthesis}

Taken together, the database reveals a coherent but trade-off-driven picture. Ordered toroidal neutral motion remains present across the explored parameter space, including ITER-sized divertor geometries. Larger hydraulic diameter extends the spatial persistence of the toroidal, but reduces the volume-averaged toroidal speed. Conversely, higher divertor pressure is associated with stronger toroidal velocities, but a shorter decay length.

Even without self-consistently accounting for the non-linearities of a detached edge plasma (section \ref{sec:evidence__simulations__trends}), these compensating mechanisms indicate the absence of a singularly optimal configuration. 

\subsection{Exhaust proof-of-principle}\label{sec:results__pop_exh}

This section investigates the exhaust proof-of-principle simulations---designed to isolate the inlet-orientation mechanism in an idealised, controlled scenario.

\subsubsection{Baseline}\label{sec:results__pop_exh__baseline}

\begin{figure*}
            \centering
            
            \subfloat[]{\includegraphics[width=0.475\textwidth]{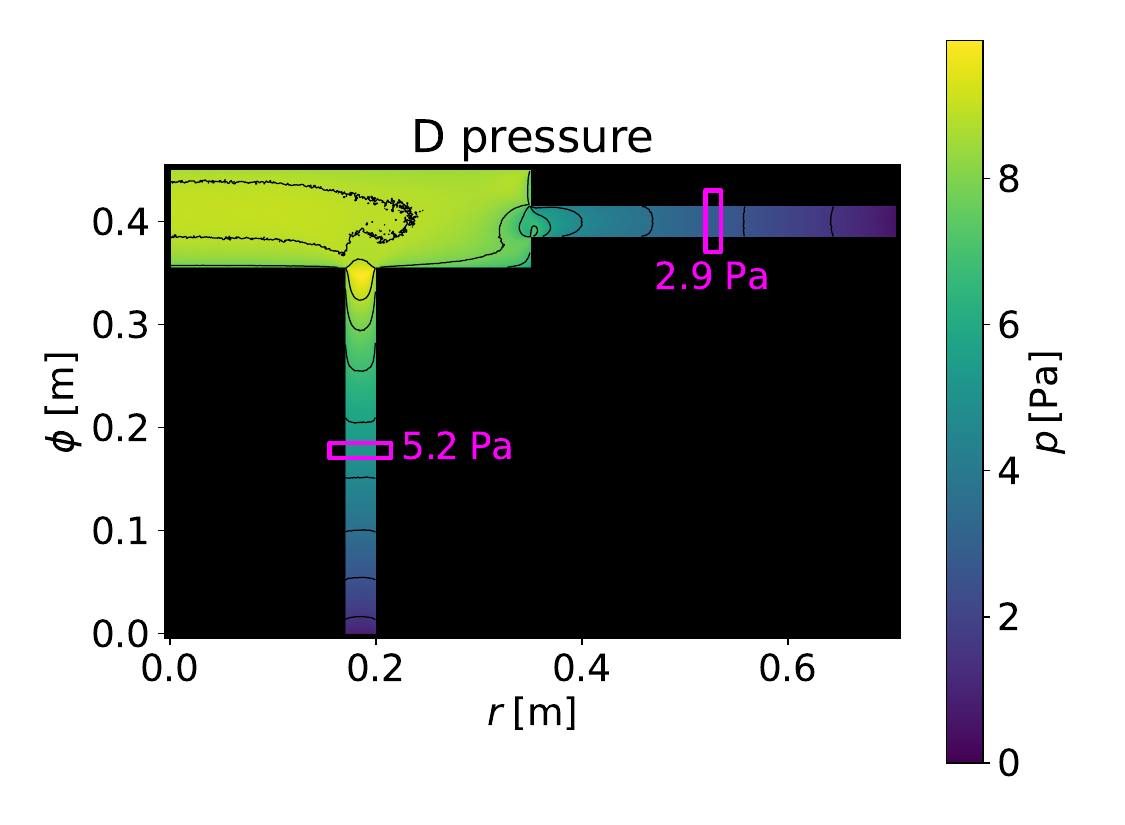}\label{fig:pop_exh_baseline__pressure_d}}
            \hspace{0.075cm}
            \subfloat[]{\includegraphics[width=0.475\textwidth]{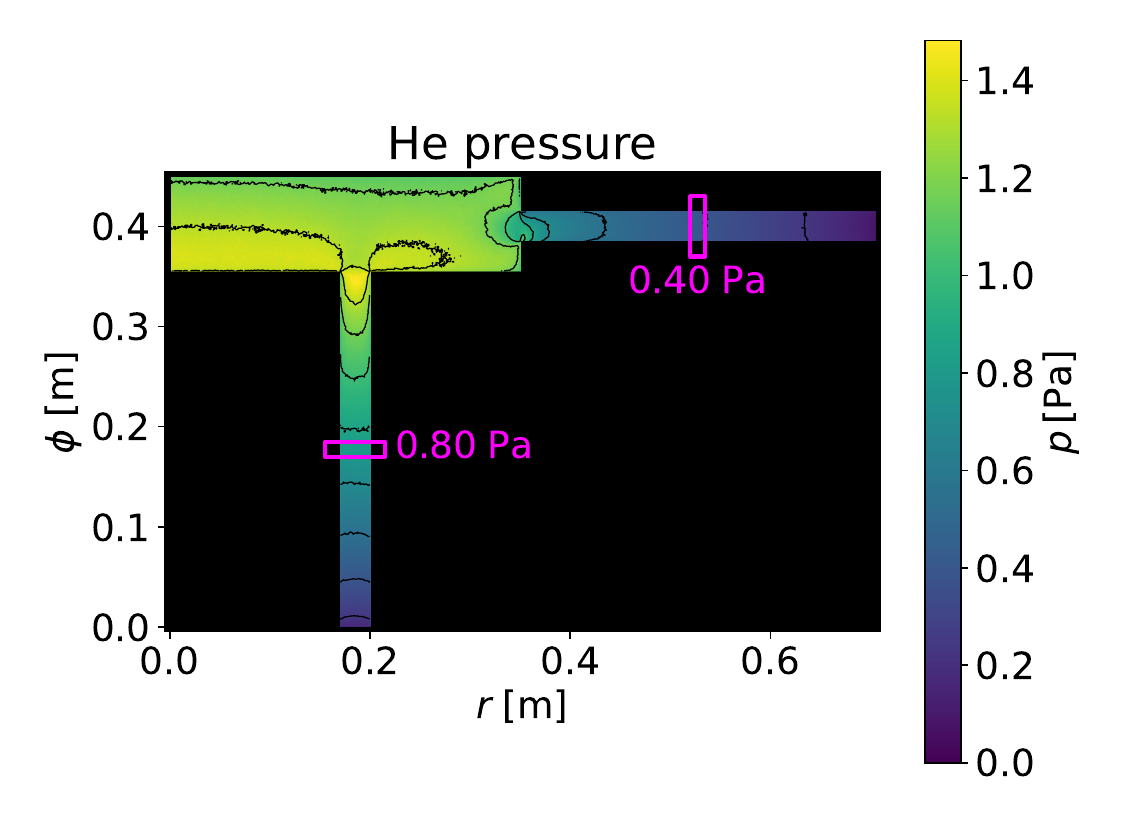}\label{fig:pop_exh_baseline__pressure_he}}\\
            
            \subfloat[]{\includegraphics[width=0.475\textwidth]{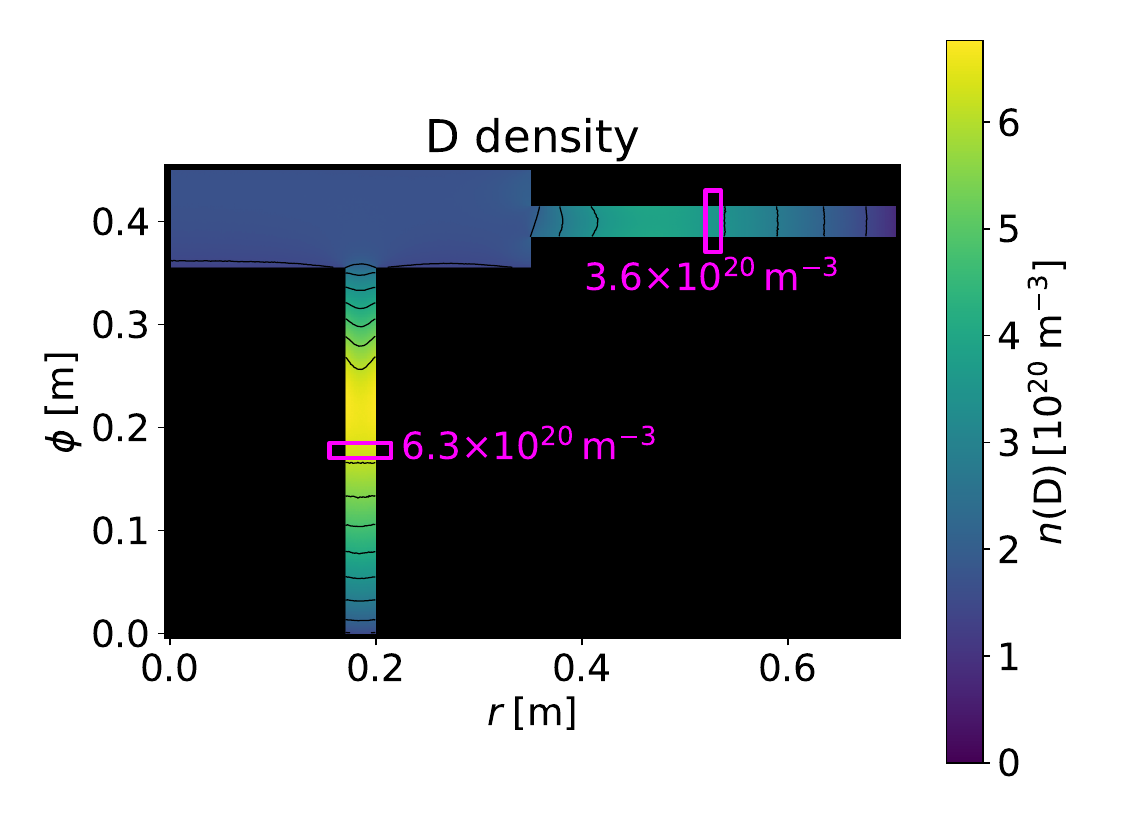}\label{fig:pop_exh_baseline__density_d}}
            \hspace{0.075cm}
            \subfloat[]{\includegraphics[width=0.475\textwidth]{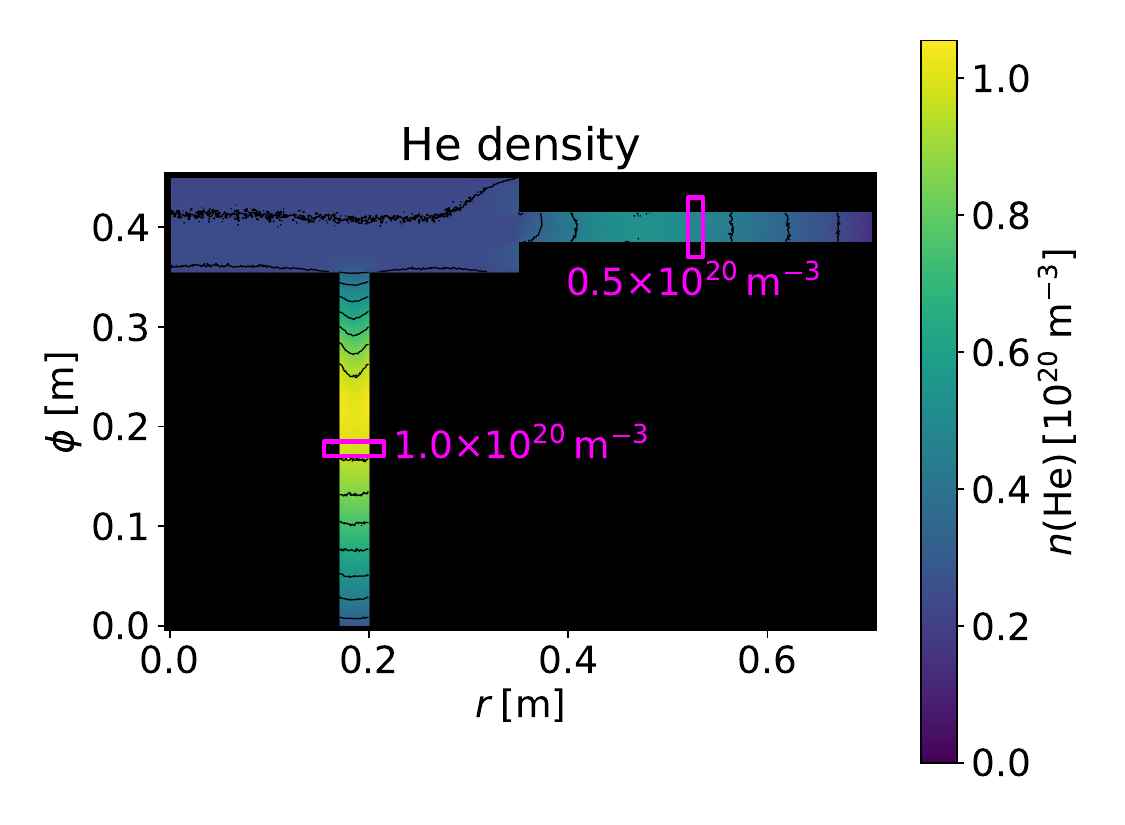}\label{fig:pop_exh_baseline__density_he}}\\
            
            \subfloat[]{\includegraphics[width=0.475\textwidth]{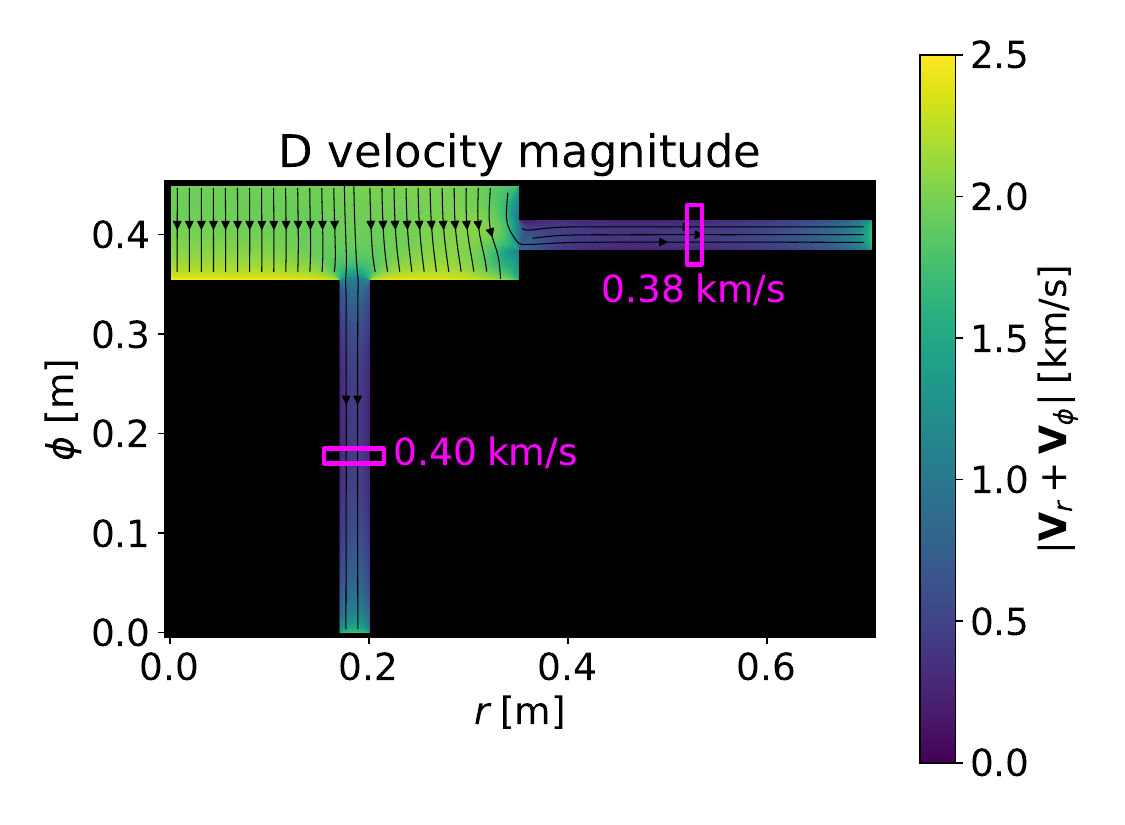}\label{fig:pop_exh_baseline__vmag_d}}
            \hspace{0.075cm}
            \subfloat[]{\includegraphics[width=0.475\textwidth]{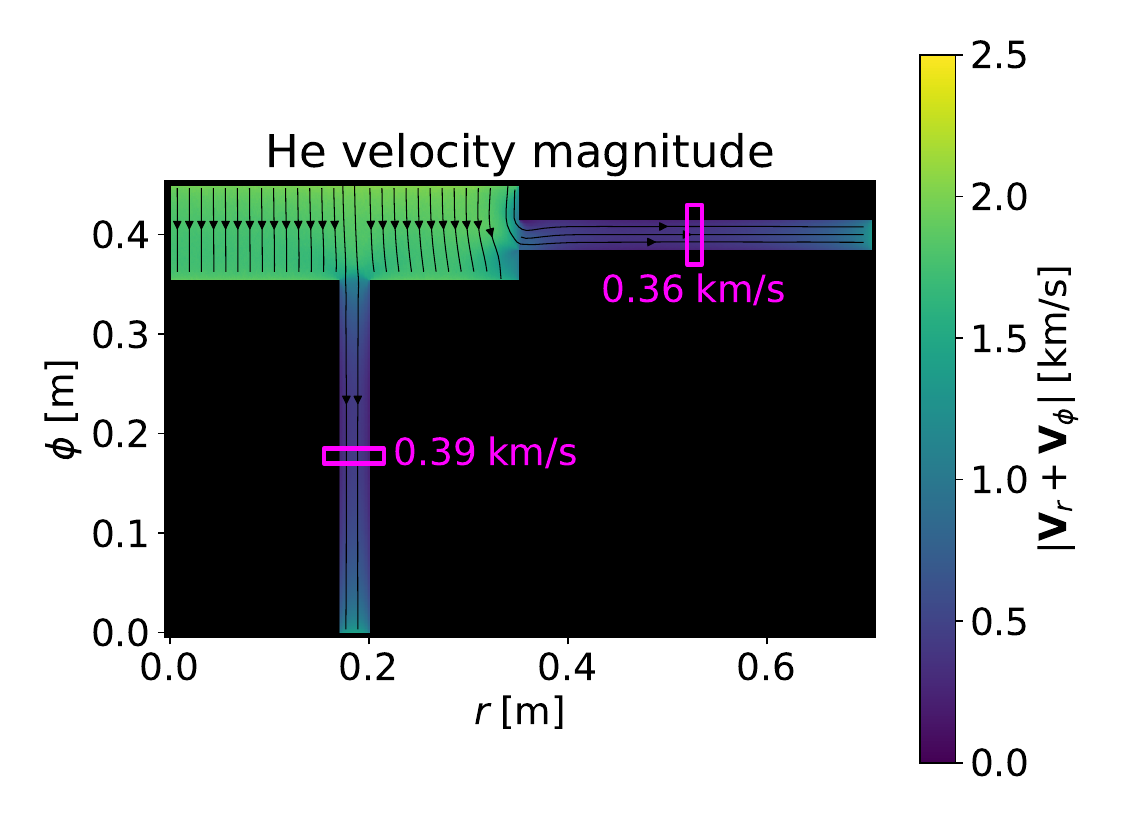}\label{fig:pop_exh_baseline__vmag_he}}

            \caption{SPARTA results: baseline simulation of exhaust proof-of-principle (section \ref{sec:methods__pop_exh}) in the toroidal plane, with $\phi$ being the straightened toroidal coordinate. Upstream divertor-average are approximately: $9~\mathrm{Pa}$, $1.6 \times 10^{20}~\mathrm{m^{-3}}$, and $2.0~\mathrm{km\,s^{-1}}$ for deuterium pressure, density, and velocity magnitude (left); and $1.3~\mathrm{Pa}$, $0.2\times10^{20}~\mathrm{m^{-3}}$, and $1.8~\mathrm{km\,s^{-1}}$ for helium at $10\%$ concentration (right). Iso-lines are overlaid to pressure and density maps and stream-lines to velocity magnitude maps. Values at duct mid-lengths are in magenta.}
            \label{fig:pop_exh_baseline}
        \end{figure*}

The result of the baseline exhaust proof-of-principle simulation in the toroidal plane is shown in figure \ref{fig:pop_exh_baseline}. The velocity magnitude $\left|\vRVec + \vTorVec\right|$, which is predominantly toroidal, is approximately uniform throughout the divertor PFR volume, as are the density and pressure fields. In the region upstream of the ducts, these quantities evaluate to $2.0 \pm 0.1~\mathrm{km\,s^{-1}}$, $(1.64 \pm 0.09)\times10^{20}~\mathrm{m^{-3}}$, and $8.8 \pm 0.3~\mathrm{Pa}$ for deuterium, and to $1.80 \pm 0.07~\mathrm{km\,s^{-1}}$, $(0.234 \pm 0.008)\times10^{20}~\mathrm{m^{-3}}$, and $1.29 \pm 0.08~\mathrm{Pa}$ for helium at approximately $10\%$ concentration. Compared with the nominal physics proof-of-principle case at $|\vTorBC| = 5~\mathrm{km\,s^{-1}}$ (figure \ref{fig:pop_phys_baseline__vphi}), these smaller velocities are representative of bulk-to-near-wall locations, approximately $10~\mathrm{cm}$ away from the plasma-like boundary. This corresponds to a realistic placement of the TFP inlet (section \ref{sec:discussion__integration}).

The back-flowing fraction is evaluated according to equation~(\ref{eq:backflow_fraction}). At the poloidal duct inlet, the back-flowing fractions are 0.53 for deuterium and 0.41 for helium. At the toroidal duct inlet, these values decrease to 0.43 and 0.29, respectively. The toroidal configuration therefore reduces back-flow by 20\% and 33\% for the two species, directly contributing to a higher particle flux through the toroidal exhaust path---and to the observed over-density and over-pressure, relative to the poloidal counterpart. The corresponding pressure gains, defined according to equation~(\ref{eq:pressure_gain}), are
\begin{equation}\label{eq:pop_exh_baseline__pressure_gains}
    \begin{aligned}
        \pGain(\mathrm{D})  &= 1.78 \pm 0.04 ,\\
        \pGain(\mathrm{He}) &= 2.00 \pm 0.05 .
    \end{aligned}
\end{equation}
This demonstrates---at proof-of-principle level---that a tangible pressure amplification may be obtained by exploiting the ordered toroidal neutral flow. The helium response follows that of deuterium, even featuring marginally bigger gains.

Notably, the ordered toroidal velocity entering through the toroidal inlet decays fast, over $\lambdaDuct = 3.8\pm0.04~\mathrm{cm}$. This decay is consistent with the conversion of directed toroidal momentum to increased local density and static pressure, while also reducing inlet back-flow. In the baseline case, the conversion is nearly one-to-one: in the upstream divertor volume, the toroidally-directed contribution of the deuterium--helium mixture is $\langle m n\vTor^2\rangle = 2.74~\mathrm{Pa}$. This closely matches the static-pressure excess of the toroidal duct over the poloidal duct, $\Delta\pTor=\pTor-\pPol=2.67~\mathrm{Pa}$. For helium alone the agreement is less exact, within approximately $20\%$.

This near-equality should be interpreted as a case-specific consistency check, not as a general one-to-one conversion law. Indeed, helium already indicates that additional parameters can modify the effective conversion between upstream directed momentum and downstream static pressure. The extent to which this relation persists across operating conditions is assessed in the following.

    \subsubsection{Scaling laws and correlations}\label{sec:results__pop_exh__database}

Scanning the parameters in table \ref{tab:pop_exh_database} yields 90 simulations populating the PoP-Exh database. SI units are used apart from $[T] = \rm eV$. Density, temperature and speed figuring in the regressions are the species-specific ones. The fitted parameters are reported till their first uncertain digit. The relative $1\sigma$-uncertainties for the exponents average 9\%, with a maximum of 36\%. For the prefactors, the average uncertainty is 29\%, with a maximum of 42\%.

The empirical laws are intended to identify trends and quantify the relative influence of the scanned parameters within the PoP-Exh database. They should therefore not be interpreted as design correlations nor extrapolated outside the explored parameter range.

The trends remain remarkably similar for deuterium and helium. Unless otherwise stated, the following interpretation therefore applies to both species.

\paragraph{Pressures.}\label{sec:results__pop_exh__pressure}

Duct-pressure regressions, $\pPol$ and $\pTor$, feature $R^2 > 0.99$. For deuterium:
\begin{equation}\label{eq:pop_exh_database__regressions_d_p}
\begin{split}
\pPol(\mathrm{D}) = 1.7 \times 10^{-22} & \, \langle n \rangle^{1.128} \, \langle T \rangle^{0.68} \, \langle |V_{\phi}| \rangle^{-0.07},\\
\pTor(\mathrm{D}) = 4 \times 10^{-22} & \, \langle n \rangle^{1.049} \, \langle T \rangle^{0.48} \, \langle |V_{\phi}| \rangle^{0.33},
\end{split}
\end{equation}

and for helium:
\begin{equation}\label{eq:pop_exh_database__regressions_he_p}
\begin{split}
\pPol(\mathrm{He}) = 7 \times 10^{-22} & \, \langle n \rangle^{1.108} \, \langle T \rangle^{0.65} \, \langle |V_{\phi}| \rangle^{-0.11} ,\\
\pTor(\mathrm{He}) = 3.2 \times 10^{-21} & \, \langle n \rangle^{1.003} \, \langle T \rangle^{0.44} \, \langle |V_{\phi}| \rangle^{0.35} .
\end{split}
\end{equation}

They primarily scale with the species-specific neutral density, roughly in a linear fashion. The temperature exponents are also positive, but smaller than the density's. While a stagnant neutral gas would suggest a linear dependence on $p=nT$, the observed density-temperature dependence likely encodes the additional effects of directed flow, finite conductance, and back-flow at the duct entrance.

In addition, both $\pPol$ and $\pTor$ exhibit a measurable dependence on the average toroidal wind speed $\langle |V_{\phi}| \rangle$, with opposite signs for the two duct orientations. The dependence is weakly negative for $\pPol$ and clearly positive for $\pTor$.

The positive dependence of $\pTor$ on $\langle |V_{\phi}| \rangle$ represents the over-pressure produced by capturing a directed toroidal neutral flow. Consistently with this interpretation, the static over-pressure is directly correlated with the upstream toroidal momentum reservoir. Across the database, $\Delta\pTor=\pTor-\pPol$ scales as $\langle m n \vTor^2\rangle^{0.83}$ for deuterium and as $\langle m n \vTor^2\rangle^{0.76}$ for helium, with $R^2=0.93$ and $0.95$, respectively. The sub-linear exponents indicate that the conversion efficiency is not one-to-one over the scan, likely because kinetic effects such as varying duct conductance, rarefaction level, and back-flow do play a role. Nevertheless, the high correlation confirms that \textit{the static over-pressure in the toroidal duct is governed primarily by the available directed toroidal momentum flux}.

Conversely, the weak negative dependence of $\pPol$ on $\langle |V_{\phi}| \rangle$ may be interpreted as a rarefied, Venturi-like cross-flow effect. Unlike a classical continuum Venturi effect \cite{Gallitto_2021}, this is not here connected to Bernoulli acceleration through a constriction.

Let us consider a \textit{localised}, non-toroidally-symmetric poloidal duct inlet which directly faces the divertor volume (e.g.\footnote{By contrast, designs like \cite{VAROUTIS2019120, Tantos_2024} feature a continuous toroidally-symmetric plasma-facing inlet connecting the divertor to the sub-divertor. The toroidally-localised duct only branches out from the sub-divertor itself, where the toroidal wind has already vanished (figure \ref{fig:ITER_123013}, centre).} in CFETR \cite{YU2024101826} or in probing channels \cite{SHAFER2019487}). The toroidal wind is tangential to the entrance and therefore does not directly feed the poloidal path. Hence, increasing toroidal wind speed reduces the particle residence time near the inlet and sweeps past the aperture a fraction of neutrals that could otherwise supply the duct. This may lower the effective capture probability into the poloidal exhaust path and may manifest macroscopically as a weak reduction of the inferred duct pressure---and a negative exponent on $\absavg{\vTor}$. Whether this effect appreciably influences diagnostic measurements remains to be assessed.

\paragraph{Pressure gains.}\label{sec:results__pop_exh__gain}

Pressure-gain regressions are affected by uncertainty ($R^2\simeq0.66$) but remain nonetheless well-defined for most of the dataset:

\begin{equation}\label{eq:pop_exh_database__regressions_g}
\begin{split}
\pGain(\mathrm{D}) & = 3 \, \langle n \rangle^{-0.081} \, \langle T \rangle^{-0.20} \, \langle |V_{\phi}| \rangle^{0.40} ,\\
\pGain(\mathrm{He}) & = 5 \, \langle n \rangle^{-0.106} \, \langle T \rangle^{-0.21} \, \langle |V_{\phi}| \rangle^{0.47} .
\end{split}
\end{equation}

\begin{figure}
\centering
\subfloat[]{\includegraphics[width=0.45\textwidth]{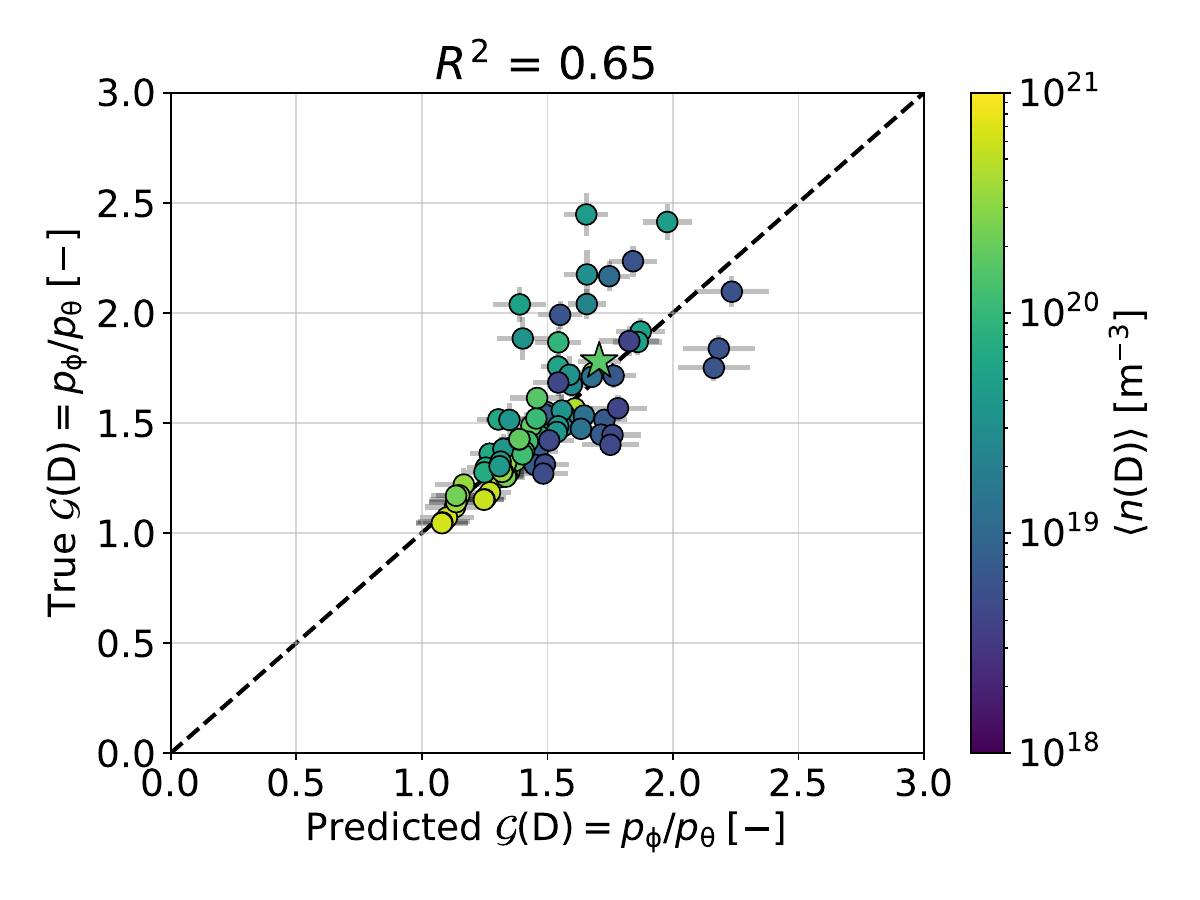}\label{fig:pop_exh_database__gain_d}}\\
\subfloat[]{\includegraphics[width=0.45\textwidth]{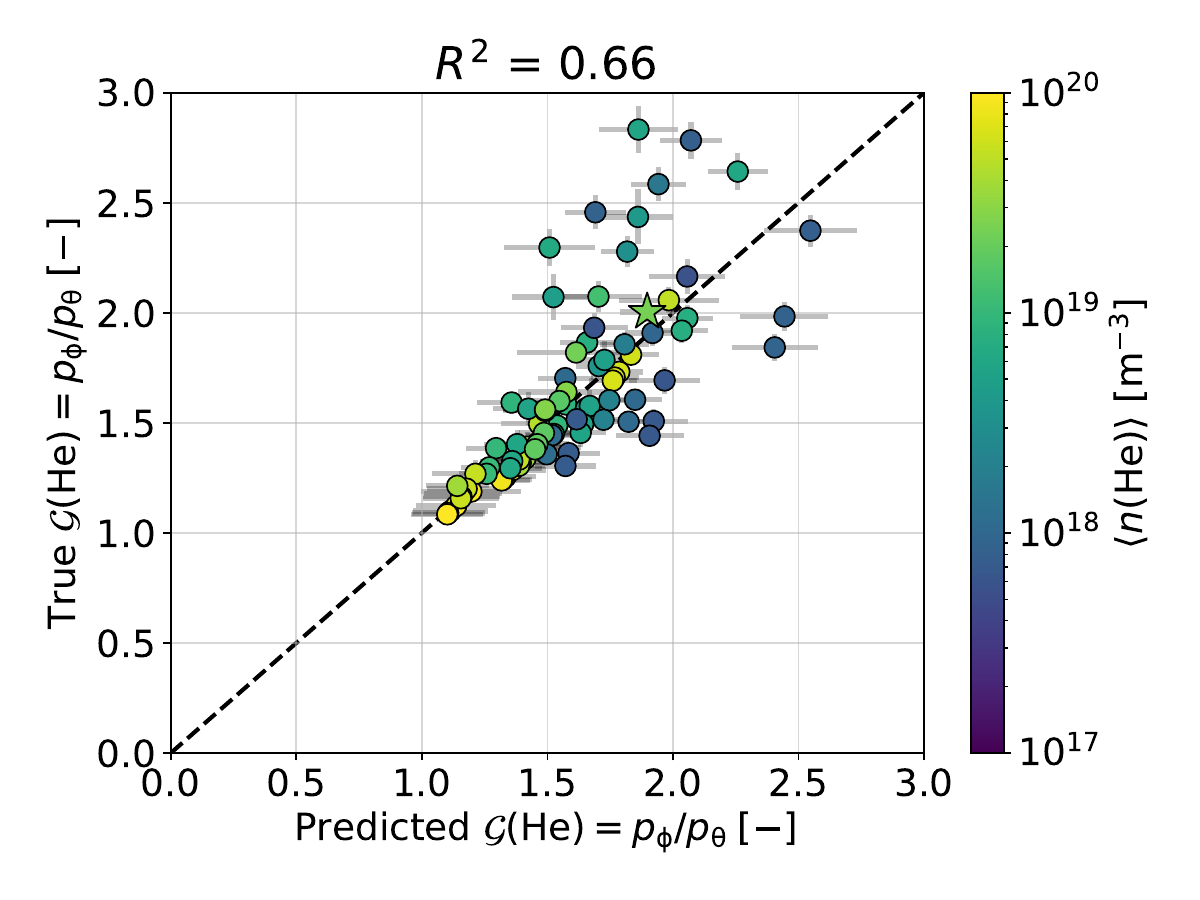}\label{fig:pop_exh_database__gain_he}}\\
\caption{Pressure-gain regressions for (a) deuterium and (b) helium across the exhaust proof-of-principle database. The colouring represents the species-specific average density throughout the divertor domain, upstream of the inlets. The baseline simulation in figure~\ref{fig:pop_exh_baseline} is starred.}
\label{fig:pop_exh_database}
\end{figure}

They are pictured in figure \ref{fig:pop_exh_database}. The weak negative density and temperature exponents should be interpreted with care. They do not imply that higher neutral density and temperature (i.e. pressure) are detrimental to the toroidal approach in absolute terms. Increasing divertor pressure raises both duct pressures---but comparatively favours the poloidal exhaust path. This is consistent with the principle of indirect pressure-driven pumping (IPDP; section~\ref{sec:philosophies__ipdp}), where increased pressure improves the capture of pseudo-random thermal motion \cite{Pitcher_1997}. The negative exponents remain small in absolute value compared with the dominant positive dependence on $\langle |V_{\phi}| \rangle$, which instead captures the operating principle of the TFP concept.

More generally, the moderate magnitude of the fitted exponents is encouraging, as it indicates robustness of the toroidal advantage against variations in divertor conditions.

Figure~\ref{fig:pop_exh_database} further shows that a significant fraction of the database occupies a region of practically-relevant pressure gains, with $\pGain \gtrsim 1.5$.

The increased scatter observed in the high-gain region, $\pGain \gtrsim 2$, tends to be associated with low-density deuterium cases at $\langle n \rangle \lesssim 10^{19}~\mathrm{m^{-3}}$. These cases indicatively correspond to elevated Knudsen numbers, $\mathrm{Kn} \gtrsim 1$, for which average thermodynamic quantities and macroscopic flow speed may no longer be sufficient descriptors of the exhaust dynamics. Dedicated analysis of this rarefied corner of parameter space is therefore warranted. While such conditions are characteristic of present-day stellarator sub-divertors \cite{Wenzel_2022,Haak_2023,Varoutis_2024,Litovoli_2026,Winters_2021,Boeyaert_2023,Wenzel_2026}, they are also relevant to tokamak operation in open divertor configurations \cite{FEVRIER2021100977} or at low fuelling \cite{Kallenbach_2018}, where divertor neutral pressures can likewise remain in the order of $\sim 0.1$ Pa \cite{FEVRIER2021100977, Kallenbach_2018}.

Finally, the deuterium and helium regressions must be interpreted consistently. In the present database, the helium density and boundary conditions co-vary with the deuterium's by construction. The regressions above should therefore be used simultaneously within the parameter range explored here.

Isolating the separate impact of species concentration and, more comprehensively, of density and temperature requires dedicated scans and is left to future work.

\paragraph{Toroidal speed decay length in the duct.}\label{sec:results__pop_exh__decay_duct}

The longitudinal decay length of the toroidal flow speed inside the duct, $\lambdaDuct$, is well described across the database: 
\begin{equation}\label{eq:pop_exh_database__regressions_l}
\begin{split}
\lambdaDuct(\mathrm{D}) = 3.1 \times 10^{-10} & \, \langle n \rangle^{0.390} \, \langle T \rangle^{0.23} \, \langle |V_{\phi}| \rangle^{0.11},\\
\lambdaDuct(\mathrm{He}) = 1.0 \times 10^{-9} & \, \langle n \rangle^{0.387} \, \langle T \rangle^{0.26} \, \langle |V_{\phi}| \rangle^{0.07}.
\end{split}
\end{equation}

The regression quality is $R^2 \simeq 0.98$ and values range from approximately $1$ to $10~\mathrm{cm}$ over the explored conditions.

The regressions indicate a primary dependence on the upstream neutral pressure. Beyond the variables scanned in the present database, the precise conversion length is expected to depend on several geometric and kinetic factors. These include the tangential momentum accommodation coefficient at the duct walls, the duct aspect ratio, and its absolute dimensions. The latter dependence is sub-linear in an illustrative test case: doubling the duct width increases $\lambdaDuct$ from $4.05 \pm 0.07~\mathrm{cm}$ only to $6.55 \pm 0.04~\mathrm{cm}$. Additional dependencies are also expected on the outlet boundary condition, the dimensionality of the model---which is presently two-dimensional---and, for helium, on the minority concentration. These effects require dedicated assessment.

Nevertheless, the key result is that conversion of directed toroidal motion into duct pressure occurs over a short spatial scale. This is practically favourable, since it avoids the need for long straight toroidal duct sections, which would complicate engineering integration.

\paragraph{Back-flow fraction.}\label{sec:results__pop_exh__backflow}

A minimal comparison is performed to provide a preliminary assessment of back-flow mitigation (section \ref{sec:philosophies__ipdp}). Since the back-flow diagnostic is not yet available for the full PoP-Exh database, the analysis is restricted to the four representative cases  in table \ref{tab:backflow_pressure_species}. The average toroidal speed is kept approximately constant at $2.2~\mathrm{km\,s^{-1}}$, so that the comparison isolates, to first order, the effect of neutral pressure.

In all cases the toroidal duct exhibits a lower back-flowing fraction than the poloidal duct. The reduction is significant throughout the scan, ranging from $12\%$ to $20\%$ for deuterium and from $20\%$ to $33\%$ for helium. The largest relative improvement is obtained at the reference pressure, suggesting the existence of an intermediate-collisionality optimum.

The systematically larger reduction of helium compared with deuterium is consistent with the lower helium density and partial pressure at fixed concentration. This places the minority species in a more rarefied regime where inlet orientation and directed-flow capture seemingly remain more influential.

Overall, these results confirm that the toroidal configuration reduces back-flow, directly contributing to the over-pressure reported above. On general grounds, mitigating back-flow targets a potentially important loss channel: back-flow fractions as high as $70\%$ have been reported in rarefied-flow configurations \cite{Varoutis_2024} (section \ref{sec:philosophies__ipdp}). A more systematic characterisation of back-flow across geometry, velocities, species concentration, and Knudsen number is therefore an important direction for future work. Consequences of a \textit{prescribed} constant back-flow/albedo are commented in appendix \ref{apx:pressure_dependent_capture}.

\begin{table}[t]
    \centering
    \caption{Back-flow fractions for poloidal and toroidal ducts, $f_{\mathrm{back}}(\theta)$ and $f_{\mathrm{back}}(\phi)$, respectively, at varying divertor pressure, with approximately constant average toroidal speed $\sim -2.2~\mathrm{km\,s^{-1}}$. The relative variation is defined as $\Delta f_{\mathrm{back}} = [f_{\mathrm{back}}(\phi) - f_{\mathrm{back}}(\theta)]/f_{\mathrm{back}}(\theta)$ and is reported in percent, so that negative values indicate a beneficial, reduced back-flow from the toroidal duct.}
    \label{tab:backflow_pressure_species}
    \begin{tabular}{rcccc}
        \toprule
        $\langle p \rangle$
        & Species
        & $f_{\mathrm{back}}({\theta})$ 
        & $f_{\mathrm{back}}({\phi})$ 
        & $\Delta f_{\mathrm{back}}$ \\
        \midrule
        $0.1$ Pa  & D  & $0.66$ & $0.55$ & $-17\%$ \\
        $1$ Pa    & D  & $0.63$ & $0.53$ & $-16\%$ \\
        $10$ Pa   & D  & $0.51$ & $0.41$ & $-20\%$ \\
        $100$ Pa  & D  & $0.25$ & $0.22$ & $-12\%$ \\
        \midrule
        $0.01$ Pa & He & $0.67$ & $0.51$ & $-24\%$ \\
        $0.1$ Pa  & He & $0.62$ & $0.45$ & $-27\%$ \\
        $1$ Pa    & He & $0.40$ & $0.27$ & $-33\%$ \\
        $10$ Pa   & He & $0.15$ & $0.12$ & $-20\%$ \\
        \bottomrule
    \end{tabular}
\end{table}

\section{Discussion}\label{sec:discussion}

\subsection{Assessment of plasma--neutral interaction effects with SOLPS-ITER and implications for stellarators}\label{sec:discussion__interactions}

\begin{figure*}
    \centering
    
    \hspace{0.075cm}
    \subfloat[]{\includegraphics[width = 0.32\textwidth]{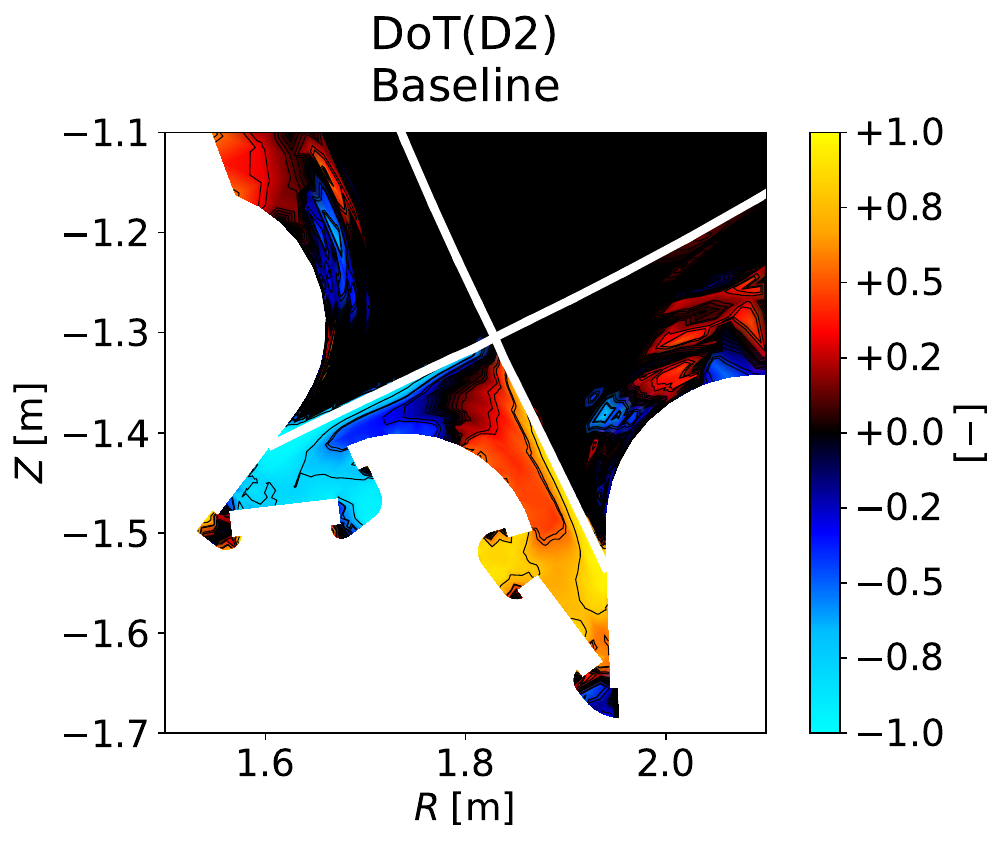}\label{fig:dtt__dot_d2__full}}
    \hspace{0.075cm}
    \subfloat[]{\includegraphics[width = 0.32\textwidth]{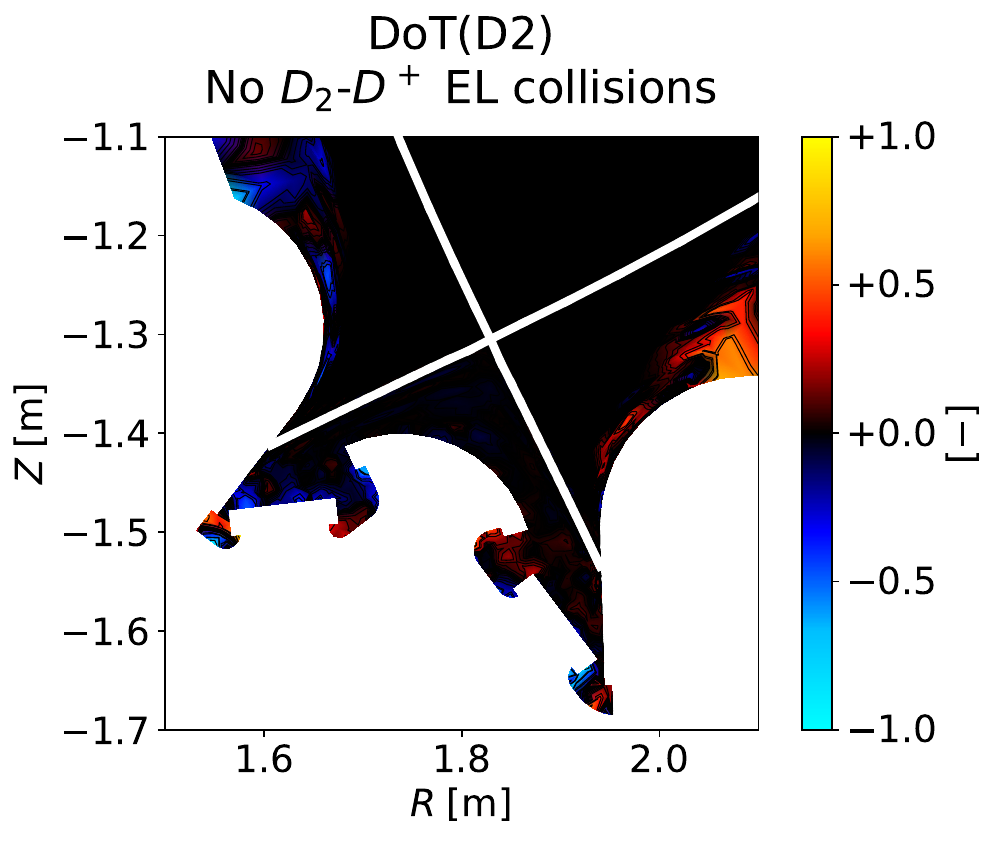}\label{fig:dtt__dot_d2__reduced}}
    \hspace{0.075cm}
    \subfloat[]{\includegraphics[width = 0.32\textwidth]{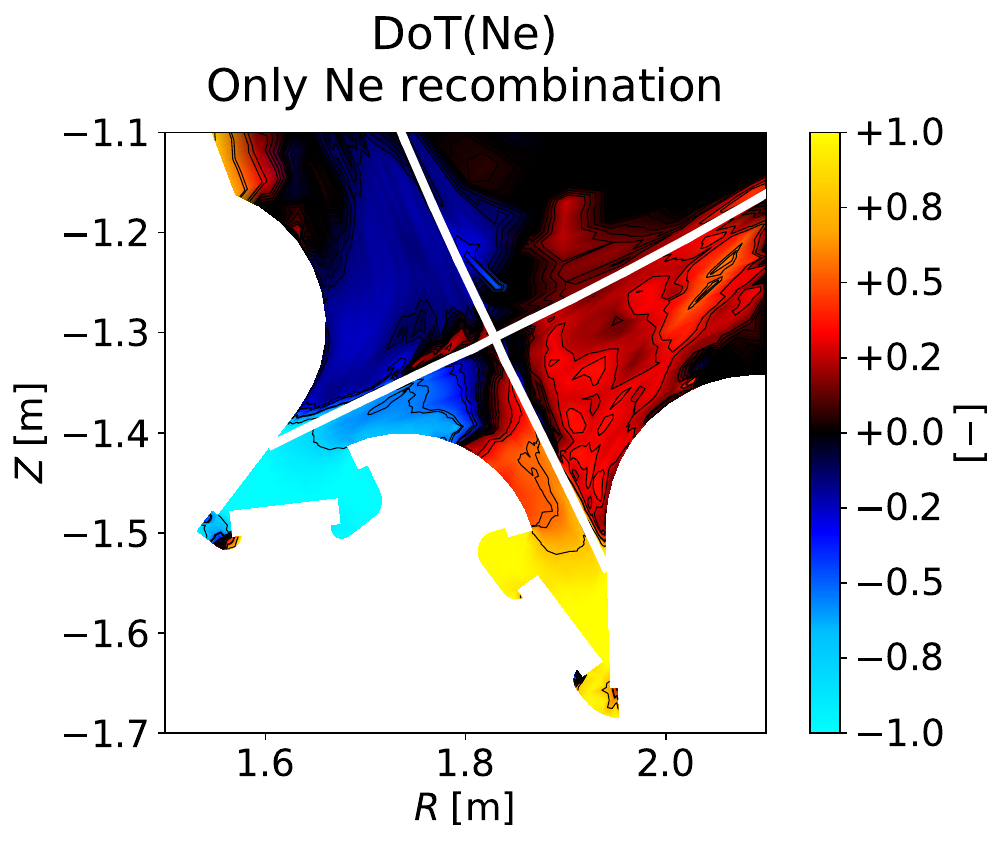}\label{fig:dtt__dot_ne}}
    
    \caption{SOLPS-ITER simulations of Moscheni \textit{et al.} \cite{Moscheni_2025} in DTT. Left: degree of toroidality of molecular deuterium with baseline plasma-neutral interaction set (no neutral-neutral collisions). Centre: same as left, but without elastic collisions between deuterium ions and molecules. Right: neon degree of toroidality, generated only by neon recombination and fast surface reflection.}
    \label{fig:dtt}
\end{figure*}

The role of plasma--neutral reactions in establishing toroidal neutral motion is illustrated in figure \ref{fig:dtt} for the DTT tokamak \cite{Martone_2019_DTT}. The comparison is based on a new post-processing of the original SOLPS-ITER simulations with different reaction sets enabled, taken from~\cite{Moscheni_2025}. In both cases, neutral--neutral BGK collisions are \textit{not} included.

The left panel corresponds to the full molecular reaction model, where a widespread degree of toroidal ordering is observed for $\DzeroMol$. The central panel shows the same simulation in which elastic scattering between molecules and plasma ions is suppressed, according to table 1 of \cite{Moscheni_2022}. In this reduced description, molecules can only be ionised or dissociated by the plasma. Since there is no direct ion-to-molecule recombination channel and no neutral--neutral collisional coupling, the degree of toroidality collapses toward zero.

In this limit, the plasma acts exclusively as a particle sink for molecular neutrals, rather than as an effective momentum source. This highlights the importance of plasma-driven momentum transfer channels for the neutral populations. Proper inclusion of these effects is hence paramount for capturing such dynamics in simulations.

In the same setup as the left panel, the right panel of figure \ref{fig:dtt__dot_ne} shows the $\DoT$ for neon---approaching unity in widespread regions of the divertor. Since neutral--neutral collisions are absent, \textit{the strong neon ordering is driven entirely by electronic recombination at the detachment front and by fast reflection at the surface}. The creation of non-collided/primary neutrals (section \ref{sec:evidence__theory}) with a strongly directed velocity distribution is therefore sufficient, by itself, to establish a toroidal neutral-neon wind with characteristic velocities of order $1\,\mathrm{km\,s^{-1}}$ near the divertor legs.

This argument suggests that \textit{toroidal neutral-flow generation need not be intrinsically limited to axisymmetric tokamaks}~\cite{Loarte_2001, Feng_2011}. The necessary condition is the presence of ordered \textit{plasma} motion capable of imprinting directionality on newly created neutrals. Such a condition also arises in stellarators with localised island divertors~\cite{Grigull_2001, Renner_2002, Wolf_2019, Feng_2021}---which would most benefit from pumping improvements \cite{Barbarino_2020, Wenzel_2024}---or other non-axisymmetric exhaust configurations~\cite{Ohyabu_1994, Bader_2017, Bader_2018, Garcia_2023, Garcia_2025}, where ordered toroidal plasma flows have been diagnosed~\cite{Effenberg_2019, Perseo_2021, Kriete_2023}. However, the extension to stellarator geometries would deserve dedicated assessment. The entrainment of collided/secondary neutrals, the finite toroidal extent of the momentum source, and the local magnetic and wall geometry all require dedicated three-dimensional assessment. In particular, toroidally-localised wetted areas~\cite{Hammond_2019} may lead to spatially-localised injections of directed neutral momentum---more akin to neutral ``gusts''---rather than a continuous toroidal neutral wind.

\subsection{Synergy with gas dynamics community: rarefied Couette flows}\label{sec:discussion__synergy_gas}

The physics proof-of-principle model allows to build a solid bridge with established scenarios investigated within the rarefied gas dynamics community. Results in section \ref{sec:results__pop_phys__baseline} can indeed be interpreted within a broader class of rarefied Couette-flow problems with heat transfer, which remain an active area of research \cite{Gorelov2015couette, Roohi_2025couette, Nejad_2021couette, Chernyak2010couette, Torrilhon_2015, Taheri_2009, marques2000couette, misdanitis2008couette}.

In each divertor half, the plasma-like boundary imposes a finite tangential velocity, $|\vTorBC|>0$, while the dome wall acts as a stationary momentum sink with finite slip and partial accommodation. This configuration is therefore Couette-like, although the moving boundary is not a material wall but an effective plasma-mediated source of toroidal momentum. This justifies the strong similarity between figure \ref{fig:pop_phys_baseline__fit_slice} herein and figures 1.10 of Agrawal \textit{et al.} \cite{Agrawal_2020} and 7.4 of Roohi \textit{et al.} \cite{Roohi_2025couette} in transitional conditions.

The complete geometry may then be viewed as a double rarefied-Couette system with two counter-streaming toroidal flows. By odd symmetry, $\vTor$ vanishes at an internal shear-reversal surface near the radial location of the X-point. This surface is no-slip-like only in the kinematic sense that $\vTor=0$ there, not a material no-slip boundary.

This Couette-like structure is superimposed on additional transport mechanisms. Neutral recycling and ionisation losses drive a forced poloidal convection cell. The temperature difference between the plasma-like boundary and the wall introduces a Fourier-flow component in the sense of rarefied Couette--Fourier configurations (section~7.9 of \cite{Roohi_2025couette}). In the present divertor configuration, however, the temperature gradient is \textit{unfavourable} with respect to the forced convective circulation. Classically, the heat source is located at the bottom of the convective cell, so that thermal forcing reinforces the upward branch of the circulation. Here, by contrast, the hottest boundary is the plasma-like ionising region, which lies at the sink side of the forced neutral circulation. The imposed temperature structure therefore does not act as a buoyant driver of the cell. This distinguishes the present system from classical buoyancy-driven Rayleigh--Bénard-type convection and related rarefied thermal-convection studies with a net velocity driver \cite{Dadzie_2016, mizzi2007effects}. The resulting neutral dynamics should therefore not be interpreted as a single textbook limit, but as the superposition of different known phenomena.

Additionally, using atomic deuterium for the baseline case in table~\ref{tab:pop_phys_database} gives a characteristic Mach number $\mathrm{Ma}\simeq0.7$ and a Reynolds number of $\mathrm{Re}\sim5$--$6$ for the toroidal flow \cite{Roohi_2025couette}. As a whole, the proof-of-principle database spans from near-continuum/slip conditions to strongly transitional or nearly free-molecular regimes, with compressibility becoming non-negligible whenever $\mathrm{Ma}\gtrsim0.3$~\cite{Roohi_2025couette}. This estimate is species dependent. For the same baseline conditions (section \ref{sec:evidence__simulations__fuel}), atomic deuterium gives $\mathrm{Ma}\simeq0.7$, whereas molecular deuterium gives $\mathrm{Ma}\simeq1.1$ because of its larger mass and lower specific heat ratio \cite{Roohi_2025couette}. The baseline regime is therefore mildly compressible for D and potentially near-sonic for $\mathrm{D_2}$. For molecules, aperture transmission, stagnation, and wake recovery may be sensitive to compressibility-like effects---which would require dedicated assessments.

Existing work has already demonstrated the value of cross-fertilisation between magnetic-fusion neutral exhaust and rarefied-gas dynamics, notably \cite{HAUER2009903, GLEASONGONZALEZ20141042, GLEASONGONZALEZ2016693, VAROUTIS2019120, Tantos_2022, Tantos_2024, Varoutis_2024, TANTOS2025115021, Litovoli_2026}. The present results suggest that this interface can be pushed further. Re-framing divertor neutral transport as a fusion-specific superposition of well-studied rarefied-flow mechanisms may provide a more systematic route to understanding and optimising neutral control. Whether the individual mechanisms superimpose linearly, interact weakly, or produce genuinely coupled kinetic effects remains an open question and motivates dedicated future work.

\subsection{Device integration of toroidally-facing pumping inlets}\label{sec:discussion__integration}

A key requirement for the TFP concept, and at the same time one of its main practical advantages, is the persistence of toroidal neutral-flow anisotropy over macroscopic distances. Consider the SOLPS-ITER simulations discussed in section \ref{sec:evidence__simulations__trends}, and figure \ref{fig:decay_pfr_iter} below. These indicate that, in ITER-relevant conditions, substantial toroidal neutral speeds and particle fluxes persist up to distances of order $10 \, \mathrm{cm}$ from the strike point, deep into the private flux region.

This property opens a practical placement window for a toroidally oriented pumping inlet, as illustrated by the grey band in figure \ref{fig:decay_pfr_iter}. Excepting off-normal operation \cite{MAVIGLIA2022113067, Scarpari_2024}, a TFP inlet can be positioned sufficiently far from the strike point to avoid direct plasma exposure, while remaining close enough to intercept a significant toroidal neutral flux. For example, in the private flux region of ASDEX Upgrade plasma decay lengths do not exceed approximately $1\,\mathrm{cm}$ once magnetic flux expansion is accounted for, according to Brida \textit{et al.} \cite{Brida_2025}. In larger devices and closer to reactor-relevancy, characteristic plasma decay lengths are expected to decrease further according to established empirical scalings~\cite{Eich_2013}. The plasma-exclusion zone required for TFP placement may therefore remain limited to only a few centimetres, even accounting for a degree of broadening \cite{Eich_2020}.

A distributed array of toroidally-facing apertures could then be used to provide the cumulative capture area required for steady-state particle balance. The flux estimates discussed in section \ref{sec:evidence__simulations__fuel} suggest that a conservative cumulative inlet area of order $1\,\mathrm{m^2}$ would satisfy the necessary particle-exhaust condition in an ITER-scale device. In turn, the results of section \ref{sec:results__pop_exh} indicate that the condition can become sufficient when the inlet exploits the directed toroidal velocity component, thereby reducing backflow and increasing the attainable over-pressure.

For example, the inner and outer divertor regions in ITER correspond to a toroidal length of $2\times2\pi R\simeq 60\,\mathrm{m}$. Thus, an $\Delta r \times \Delta z = 20 \times 7 \,\mathrm{cm^2}$ toroidally-facing area per inlet (alike the sketch in figure \ref{fig:ITER_123013}, right) would correspond to approximately one inlet per metre of toroidal circumference, summing to $1 \,\rm m^{2}$. This spacing appears geometrically plausible according to appendix \ref{apx:flow_past_obstacle}---which preliminarily shows flow recovery lengths of only a few tens of centimetres downstream a localised perturbation (e.g. a TFP inlet) \cite{Varade_Agrawal_Pradeep_2014}. The constraint could be softened by staggering the inlets in the radial--poloidal direction, thereby reducing direct wake overlap. However, dedicated assessments of inlet spacing and local flow recovery remains necessary. 

The detailed shape of each inlet in the array should then be subject to further optimisation. The transmission probability of drifting Maxwellian distributions through apertures remains an active topic of both fundamental rarefied-gas research~\cite{Binder_2016, SATO201960, YUAN2024113366, Sepahi_2018, Manela_Gibelli_2020} and engineering optimisation~\cite{ROMANO2021225, ZHENG2023223, ZHANG202451, YAKUNCHIKOV2025102, Almeida2021A, Ruetten_2013, Miansari_2020}. Similar considerations would apply to a TFP inlet, where aperture orientation, aspect ratio, and surrounding lip geometry could all influence the capture of directed neutrals, the suppression of backflow, and the achievable pressure gain. Divertor manipulators in ASDEX Upgrade \cite{HERRMANN20151496, Herrmann_2015} and W7-X \cite{HUBENY201977, HUBENY2021112297} would provide possible experimental placements, as RFMEA-like structures in MAST-U~\cite{DAMIZIA2025101912} and WEST~\cite{Gunn_2026_PSI} would---although any direct plasma exposure entails engineering complications.

One alternative implementation is a submerged/recessed inlet located below the nominal plasma-facing divertor outline \cite{Almeida2021A, Miansari_2020}, e.g. in a NACA-duct-like configuration~\cite{Mossman_1948, Holzhauser_1948, Li_2020}. Such an inlet would remain optically-shadowed from direct plasma impingement \cite{PITTS201760, Looby02012022}, while still being exposed to the neutral wind. Despite an expectedly lower transmission probability, this could allow the inlet to be placed closer to the strike point---where the toroidal neutral speed and flux are maximal---without imposing a corresponding increase in direct plasma loading.

A submerged configuration could also extend the accessible design space beyond the private-flux-region, and to the SOL side of the separatrix. Concrete examples are provided by the lower outer divertors of DIII-D and EAST, which have already been the subject of pumping-related investigations in~\cite{Maingi_1999, Unterberg_2010} and~\cite{Lore_2019}, respectively. In such geometries, a toroidally-oriented inlet could be integrated within the channel connecting the divertor volume to the pumping plenum (e.g., figures~1 and 2 of \cite{Maingi_1999} in DIII-D). 

The trade-off between plasma shielding, toroidal velocity budget, neutral transmission probability, cumulative capture area, inlet spacing, and manufacturability should therefore be treated as a dedicated rarefied-flow optimisation problem. Although the above suggests that the TFP concept could in principle constitute a stand-alone exhaust pathway, the same mechanism could also be used to enhance conventional pressure-driven poloidal pumping.

\begin{figure}
    \centering
    \includegraphics[width=0.475\textwidth]{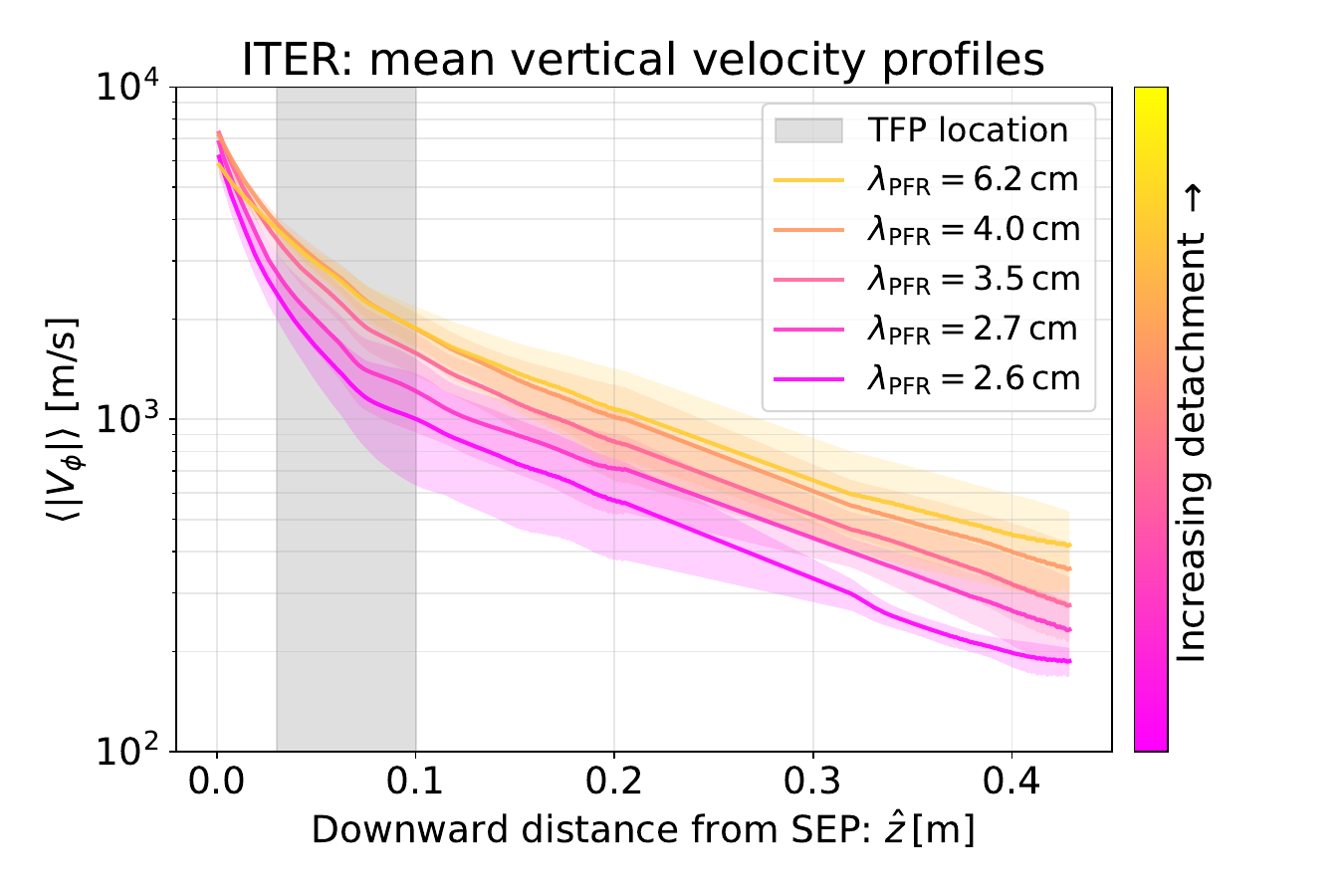}
    \caption{Illustrative placement (grey) of the TFP in an ITER-like environment. The top-most inlet surface lies $3 \, \rm cm$ below the strike point ($\hat{z} = 0$), above the divertor target reflectors \cite{PITTS2019100696}. Overlaid are the average vertical velocity profiles of the atomic-equivalent deuterium at increasing degree of detachment for the simulations in figure \ref{fig:ITER_123013_detachment_trend}.}
    \label{fig:decay_pfr_iter}
\end{figure}

\subsection{Aerodynamic considerations for divertor dynamics:
toroidally-localised non-plasma-contacting structures}\label{sec:discussion__aerodynamics}

Design and optimisation of divertor baffles, reflectors, and domes \cite{FEVRIER2021100977, Reimerdes_2022, PITTS2019100696, Loarte_2001, Cowley_2026} have traditionally focused on macroscopic shaping within an effectively two-dimensional, toroidally-symmetric framework---although these structure do not directly intercept the plasma strike-points. 

In addition to recognised three-dimensionalities \cite{Lore_2015, LORE2015515, Xie_2025}, the observation of an organised toroidal neutral wind further motivates a complementary perspective. While divertor targets remain constrained by plasma contact and heat-flux handling \cite{PITTS201760, Looby02012022}, surrounding structures---including localised apertures---may be shaped not only to control poloidal neutral pathways, but also to interact ``aerodynamically'' with the \textit{toroidal} component of the neutral flow. A tailored design may guide, separate, or redistribute neutral streams in three dimensions. For instance, rather than extracting particles, a suitably oriented non-plasma-facing structure could redirect part of the toroidal neutral wind toward the near-strike-point SOL. Such a concept would aim to increase the local neutral source where it is most effective for detachment.

These examples are proposed to simply illustrate that an organised toroidal neutral wind may enlarge the divertor design space. Particle exhaust through a toroidally-facing inlet is only one direct application---and other uses of directed neutral motion should be explored.

\section{Conclusions and outlook}\label{sec:conclusion}

This work shows that divertor neutral populations possess a usable form of order that is normally left unexploited. Plasma--neutral momentum transfer generates macroscopic toroidal neutral ``winds''. Here we have assessed whether it can be converted into a practical advantage for particle exhaust.

Previous theoretical arguments, experimental observations and SOLPS-ITER simulations indicate that ordered toroidal neutral motion is a recurring feature of tokamak divertors, with measured and predicted mean velocities reaching several $\mathrm{km\,s^{-1}}$. The numerical trends further suggest that---before extremely deep detachment reduces the available plasma momentum---the spatial organisation of this flow strengthens across a realistic, reactor-relevant detachment window. The essential prerequisite appears to be an ordered plasma flow---rather than tokamak axisymmetry itself. Analogous phenomena may therefore manifest in stellarators, which would require dedicated assessment and would most benefit to particle-exhaust improvements.

Our physics proof-of-principle simulations isolate the underlying mechanism with SPARTA direct simulation Monte Carlo calculations and plasma-like boundary conditions. The resulting Couette-like neutral dynamics form two counter-streaming toroidal wind currents in the private-flux region, superimposed on ionisation-driven poloidal convection. Even when wall friction and an unfavourable thermal gradient act to blur the ordered motion, the simulations retain high degree of toroidality, kilometre-per-second velocities and spatial decay lengths up to ten centimetres. Dedicated regression analyses across the assembled database confirm that these quantities remain robust across variations in the key parameters.

The exhaust proof-of-principle then assesses a possible practical exploitation in a controlled idealised scenario. A toroidally-oriented inlet aligned with the neutral wind reduces inlet back-flow by up to $20\%$ for deuterium and up to $33\%$ for helium compared to an otherwise equivalent poloidal pressure-driven inlet. In the baseline case, partial pressures in the toroidal duct are enhanced by a factor $\simeq 1.78$ for deuterium and $\simeq 2.00$ for helium at a nominal $10\%$ concentration. Across the database, this pressure gain is controlled primarily by the toroidal wind velocity, with the largest gains expected at lower pressures---while a clear advantage remains present under nominal mid-range conditions.

These results constitute a detailed proof-of-principle exercise---not a complete reactor pumping design. The simulations use simplified two-dimensional geometries, prescribed boundary conditions, and a helium concentration not reproducing reactor ash concentrations. Enlargement of the database with inclusion of three-dimensional cases and realistic helium fractions would therefore require future attention.

Yet, these results simultaneously support the toroidal flow pump (TFP) as a distinct particle-exhaust philosophy to be explored---one which deliberately captures the toroidal winds. For fixed throughput and otherwise unchanged pumping characteristics, an increased inlet pressure suggests a reduction in the required effective pumping speed. In reactor exhaust systems, duct conductance, neutron streaming, shielding and integration constraints are tightly coupled---and have recently been recognised as even more crucial than anticipated~\cite{Meschini_Moscheni_2026}. A practical gain in this sense would therefore not be a mere gas-dynamic advantage: it would be a direct engineering lever.

More generally, the present work hints at a wider design space for neutral control based on non-continuum divertor aerodynamics. Inlet transmission, submerged NACA-like geometries, and three-dimensional shaping of non-plasma-contacting divertor structures constitute optimisation directions worth exploring. The present exhaust implementation may therefore be viewed as a first demonstration of a broader principle: toroidally-directed neutrals can be exploited---rather than solely viewed through the lens of pseudo-random two-dimensional motion commonly depicted in poloidal cross-sections.

The results also have modelling implications. Interfaces between edge-plasma simulations and stand-alone vacuum-duct models require particular care: density and pressure are not sufficient if momentum anisotropy and source--sink recycling balance are not consistently transferred. Experimentally, existing toroidal-neutral-flow measurements remain sparse and limited to atomic deuterium. An expanded diagnosis would provide additional anchor points for validation.

To conclude with---toroidal neutral winds sit at the boundary between edge plasma physics, divertor engineering and rarefied gas dynamics. The central message of this work is that this boundary is not only a modelling complication, but also an opportunity. A long-neglected feature of divertor neutral transport may be turned into a practical tool for enhanced reactor-relevant particle exhaust.

\section{Acknowledgements}
    
The authors sincerely thank the ITER Organization (IO), the SOLPS-ITER community, and X. Bonnin for making the SOLPS-ITER code openly available \cite{SOLPSITERGitHub, SOLPSZenodoCommunity}. The resulting ability to share and reuse simulation data was an important enabler of the present study.

The authors acknowledge the use of ChatGPT for language refinement, visualisation support, and scientific brainstorming.

The authors gratefully acknowledge the SPARTA development team for creating and maintaining a capable, well-documented, and user-friendly simulation code \cite{SPARTA}. The Qarnot HPC team \cite{QarnotCloudHPC} is thanked for providing all the necessary computational resources. H. Wu is thanked for providing the original DTT data reported in~\cite{Moscheni_2025}, F. Subba for the EU-DEMO data in~\cite{Subba_2021}, and A. Zito for the ASDEX Upgrade data in~\cite{Zito_2025}; these datasets provided useful supporting insight throughout the study.

The authors are grateful to A. Kallenbach, T. P\"utterich, A. Zito, and the ASDEX Upgrade Team at large for useful discussions---and in particular to H. Zohm for the exciting conversations.

Finally, the authors wish to acknowledge the late Peter Christian Stangeby, whose work has been a continuing source of inspiration. We hope that consolidating and building upon a physical effect discussed in his papers is a fitting tribute to his lasting influence on edge plasma and divertor physics.

\section{Data Availability}

The SPARTA DSMC input files defining all simulation cases analysed in this study, as well as the database output used for regression are openly available on Zenodo~\cite{zenodo_Moscheni_2026}.

\appendix

\numberwithin{equation}{section}
\counterwithin{figure}{section}
\counterwithin{table}{section}

\renewcommand{\theequation}{\Alph{section}.\arabic{equation}}
\renewcommand{\thefigure}{\Alph{section}.\arabic{figure}}
\renewcommand{\thetable}{\Alph{section}.\arabic{table}}

\section*{Appendix}


\section{Two-dimensional space-averaging procedure in SOLPS-ITER}\label{apx:solps_average}

\begin{figure}
    \centering
    \subfloat[]{\includegraphics[width = 0.475\textwidth]{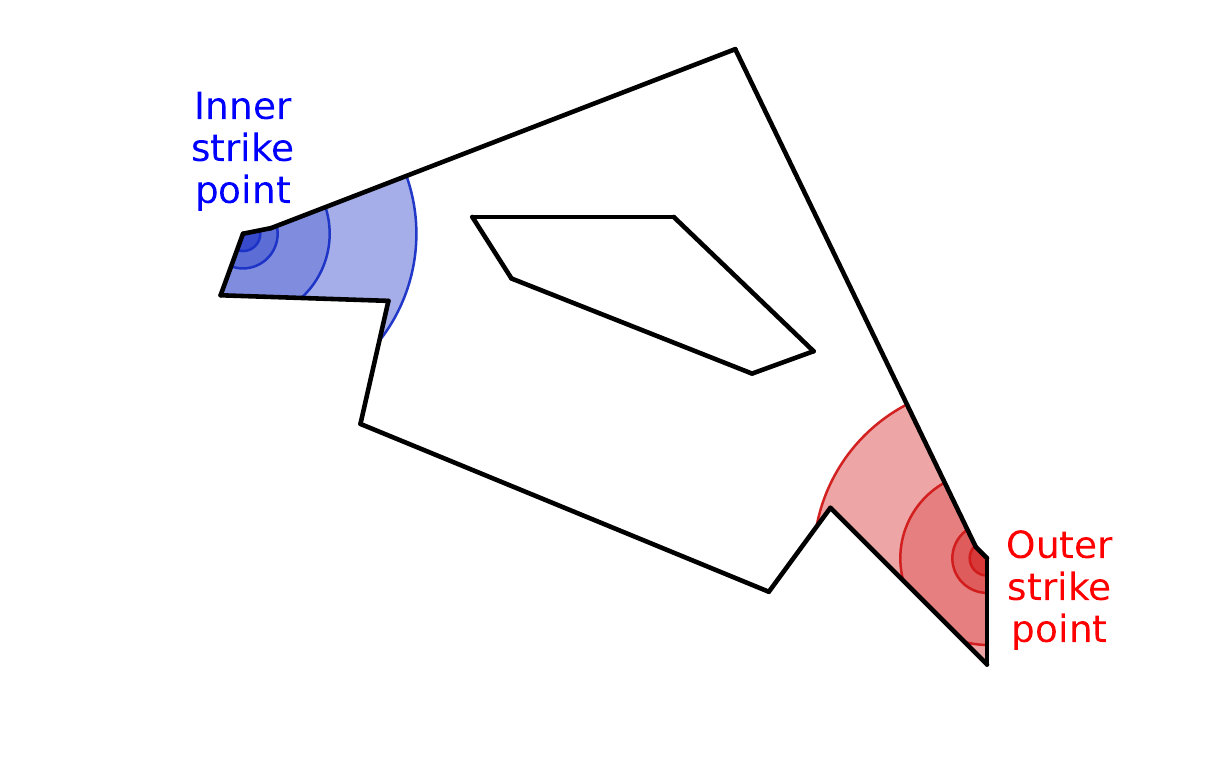}\label{fig:solps_average_annuli}}\\
    \hspace{0.075cm}
    \subfloat[]{\includegraphics[width=0.475\textwidth]{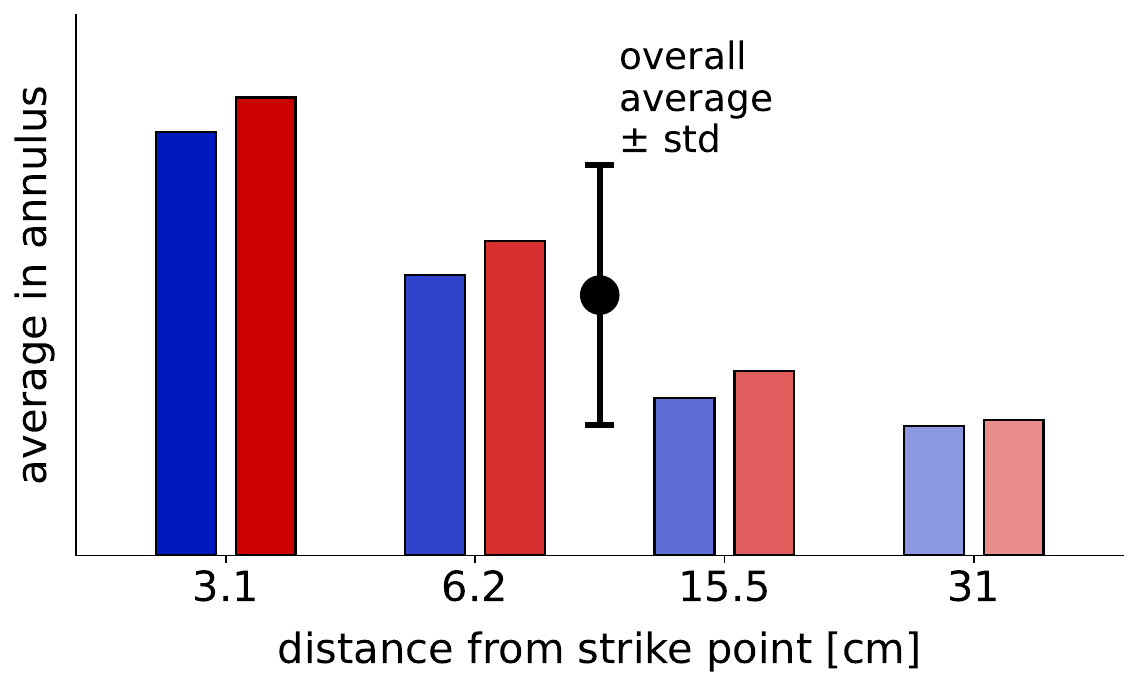}\label{fig:solps_average_bars}}\\
    \caption{Spatial-averaging procedure applied to SOLPS-ITER two-dimensional output data. (a) Annular domains at increasing distance from strike points within the ITER private-flux region. (b) Resulting histogram of average values, and the overall average with variation range.}
    \label{fig:solps_average}
\end{figure}

To compare SOLPS-ITER two-dimensional maps across different cases, a dedicated spatial-averaging procedure is introduced. For each simulation, circular regions are defined in the poloidal plane, centred on the inner and outer strike points. The selected radii are
\begin{equation}
    r_i \in \{3.1,\,6.2,\,15.5,\,31.0\}\,\mathrm{cm},
\end{equation}
Spatial averages are then computed over annular regions satisfying
\begin{equation}
    r_i \leq r < r_{i+1},
\end{equation}
while retaining only the portion of each annulus located inside the private flux region, as shown in figure \ref{fig:solps_average_annuli}. As a whole, this domain roughly corresponds to the private-flux region in between divertor reflectors and separatrix legs (see figure 1 of \cite{PITTS2019100696}).

The resulting values are ordered by alternating inner- and outer-target averages at increasing distance from the corresponding strike point, as per figure \ref{fig:solps_average_bars}. An overall mean across all selected regions is also reported, together with a variability estimate given by the standard deviation of the sample (black marker).

\section{Toroidal neutral-wind trends with increasing detachment: Lore's diamond series}\label{apx:ITER_123013_detachment_trend__d_ne_series}

\begin{figure*}
    \centering
    \subfloat[]{\includegraphics[width = 0.975\textwidth]{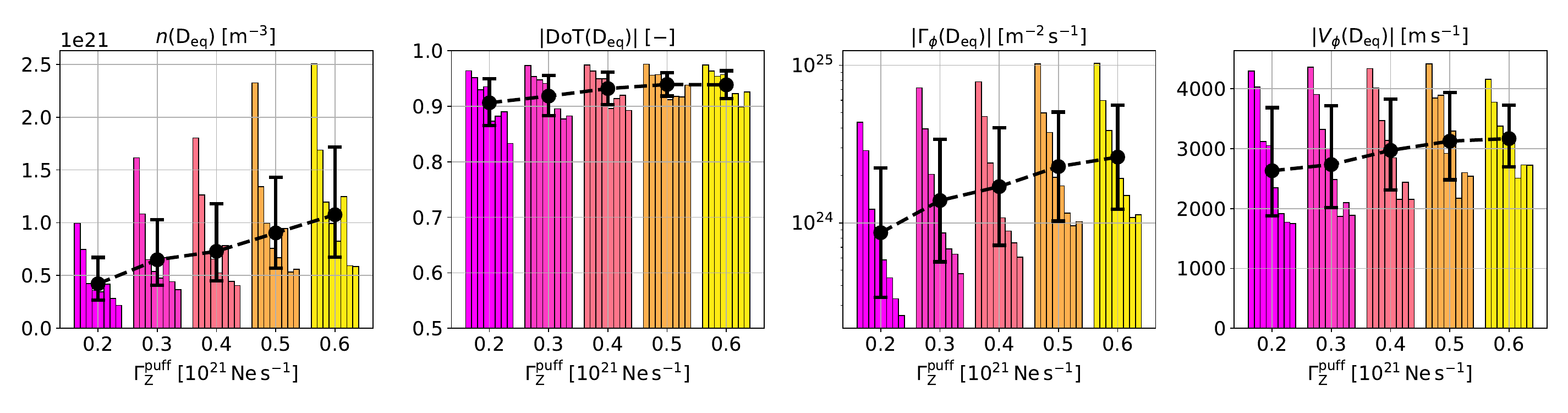}}\\
    \subfloat[]{\includegraphics[width = 0.975\textwidth]{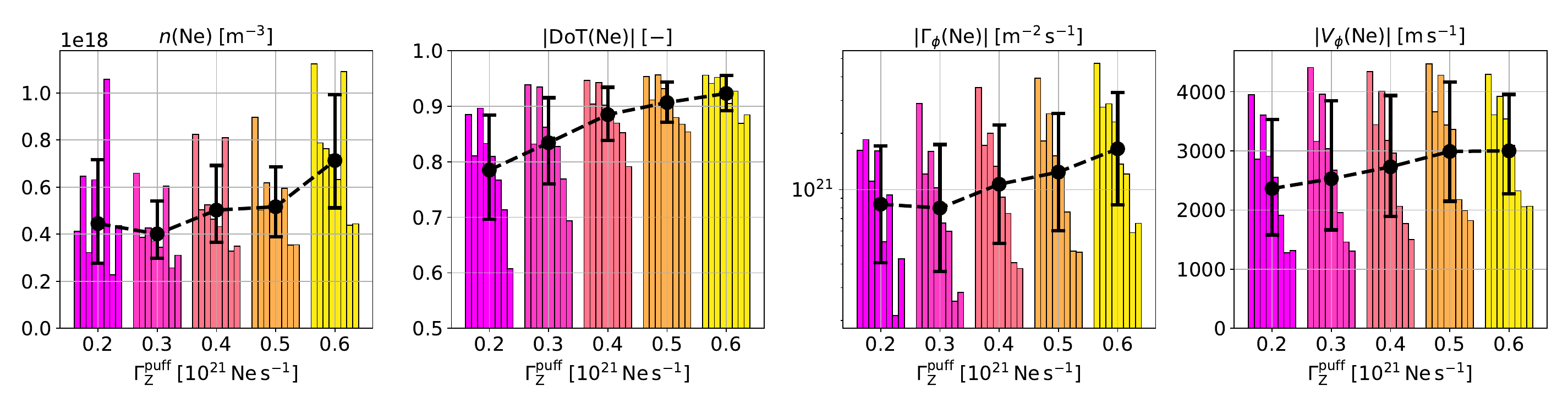}}\\
    \subfloat[]{\includegraphics[width = 0.975\textwidth]{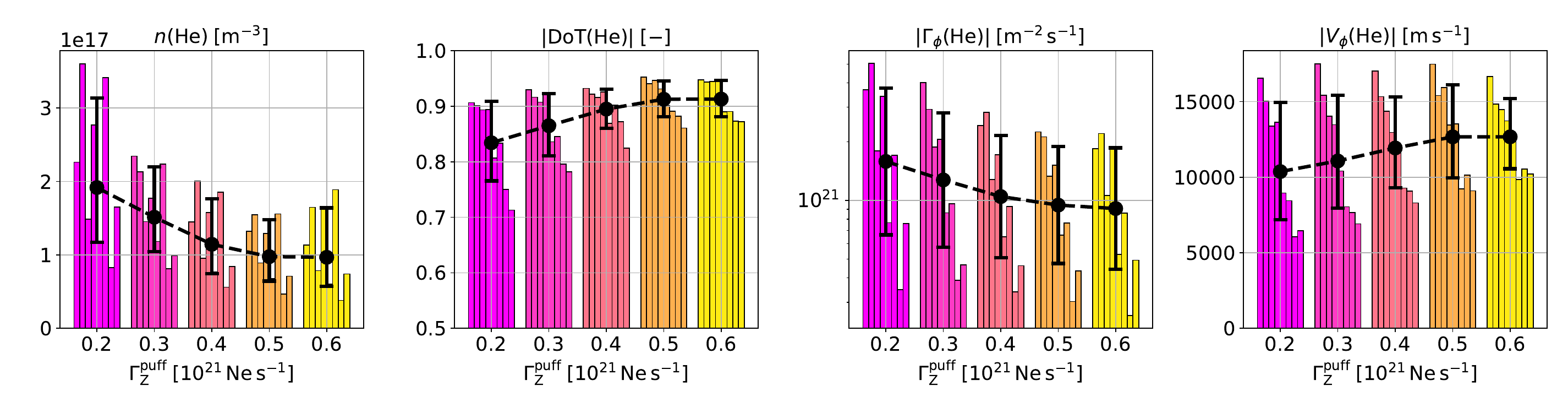}}
    \caption{Trends of toroidal neutral-wind quantities in the ITER private-flux region. Deuterium and neon puffing rates increase at constant ratio. The underlying SOLPS-ITER simulations are those of Lore \textit{et al.} \cite{Lore_2022} (diamond series).}
    \label{fig:ITER_123013_detachment_trend__d_ne_series}
\end{figure*}

The diamond series of Lore \textit{et al.} \cite{Lore_2022} (figure 4b) leads to the same qualitative conclusion as the triangle series of section \ref{sec:evidence__simulations__trends}. Figure \ref{fig:ITER_123013_detachment_trend__d_ne_series} confirms that, as fuel puffing and impurity seeding are increased together at fixed ratio in SOLPS-ITER, the macroscopic cohesion of the toroidal neutral wind increases with puffing and deeper detachment. This demonstrates that the favourable trend is not specific to the particular---and to some extent arbitrary---fuelling/seeding recipe used to access detachment. Rather, over the explored operating window, increasing detachment consistently promotes the formation of a more spatially-organised toroidal neutral wind.

A notable difference from the triangle series is that the maximum neon seeding rate in the diamond scan is only $\Gamma_{\mathrm{Ne}}\simeq0.6\times10^{21}~\mathrm{Ne}\,\mathrm{s}^{-1}$, compared with $4.0\times10^{21}~\mathrm{Ne}\,\mathrm{s}^{-1}$ in figure \ref{fig:ITER_123013}. The corresponding degree of detachment is therefore lower (figure 5 of \cite{Lore_2022}), and the trend inversion observed at the highest seeding level of the triangle series---where the toroidal neutral speed begins to degrade---is not reached here.

\section{Supporting evidence: limitations and caveats}\label{apx:evidence_limitations}

\subsection{Experimental evidence}\label{apx:evidence_limitations__experiments}

The existing experimental evidence of section \ref{sec:evidence__experiments} exhibits several limitations when assessed specifically in the context of toroidal neutral winds:
\begin{itemize}
    \item Measurements to date rely on Doppler spectroscopy, which provides line-integrated velocities in an environment characterised by strong poloidal gradients.
    \item Only atomic deuterium neutrals have been diagnosed, with no direct information on molecular species or neutral impurities. The molecular population should attain a similar velocity---whereby any speed differential between majority species in presence of effective collisional channels would tend to be suppressed. Impurity-deuterium flow entrainment would lead to the same conclusion. However, this has not been measured.
    \item Observations are then generally limited to individual discharges rather than systematic scans across plasma regimes and degrees of detachment, unlike comparable studies of ion flows \cite{Silburn_2014, Effenberg_2019, Perseo_2021}. Although Goetz \textit{et al.} \cite{Goetz_1999} show a tendency of the toroidal neutral speed to increase as detachment is established and maintained, this remains an isolated case.
    \item Measurements focus on the parallel ($\sim$toroidal) velocity component alone, without direct comparison to the poloidal or isotropic velocity components \cite{Yadava_2019}. Still, the detectability of a Doppler shift despite substantial thermal broadening already suggests a significant degree of ordering.
    \item Finally, neutral density is not measured \cite{AIMETTA2023113513}, precluding direct estimates of neutral particle fluxes---equally-relevant for exhaust.
\end{itemize}

\subsection{Numerical evidence}\label{apx:evidence_limitations__simulations}

Several limitations of the numerical evidence of section \ref{sec:evidence__simulations} should be noted:

\begin{itemize}
    \item Conventional divertor pumping surfaces can locally increase $\DoT$ by preferentially removing poloidally-directed neutrals. However, these tend to be located tens of centimetres deep inside the PFR, and the modest absorption probabilities involved likely render this effect secondary: $0.72\%$ in ITER \cite{Lore_2022,PITTS2019100696} and, for comparison, of order $2\%$ in DTT \cite{Moscheni_2025} and below $1\%$ in EU-DEMO \cite{Subba_2021}.

    \item Several early studies \cite{Knoll_1996,Knoll_1998} relied on fluid-neutral models, such as those embedded in UEDGE \cite{ROGNLIEN1992347}. Their applicability is restricted to highly collisional regimes, and the associated limitations are discussed in detail in \cite{Moscheni_2022,Moscheni_2025}.

    \item Even in kinetic-neutral frameworks, additional caveats apply. Convergence and mesh-independence of EIRENE solutions are not automatically guaranteed \cite{Ghoos_2019}, and this requirement may be more stringent when addressing momentum transport rather than density alone.
    
    \item Neutral--neutral collisions in EIRENE (and other similar codes such as DEGAS2 \cite{Stotler_1994, Wilkie_2025}) are treated through a simplified linearised BGK approximation \cite{Chernyak2010couette, Torrilhon_2015, Pfeiffer2018} with constant effective cross sections. This introduces a model error that is presently not quantified. The limitation may partly explain the significant speed differences observed between species, in particular the ordering $\vTor(\mathrm{Ne}) \lesssim \vTor(\DzeroEq) < \vTor(\mathrm{He})$ in figure \ref{fig:ITER_123013_detachment_trend}. With fully-resolved inter-species neutral collisions, minority-species speeds would be expected to relax towards the dominant hydrogenic neutral flow, i.e. towards $\vTor(\DzeroEq)$ \cite{GLEASONGONZALEZ20141042, GLEASONGONZALEZ2016693}. Part of the apparent separation may also arise from the \textit{plasma}-boundary conditions adopted in SOLPS-ITER, and/or the averaging procedure described in appendix~\ref{apx:solps_average}---every averaging domain, even if far away from the strike points, necessarily includes near-plasma regions where velocities may be locally high. The hydrogenic neutral speed nevertheless remains large and consistent with available experimental evidence.

    \item Backscattering of neutrals and ions impinging on rough tungsten surfaces is not modelled in the EIRENE---only specular (fast) and diffusive (thermal) reflections are \cite{Reiter2019EIRENE}. However, backscattering may not necessarily be zero and can be modelled with $\alpha_t = 2$, at least in DIVGAS \cite{Varoutis2023DIVGASDEMO}. This may primarily relevant in attached regimes \cite{Gradic_2018} where plasma particles directly impinge on material surfaces.

    \item The SOLPS-ITER simulations considered neglect ELMs \cite{Zohm_1996, Horacek_2023}. ELMs act as periodic power and particle bursts that can temporarily burn through the detached front \cite{Flanagan_2025}, displacing the plasma--neutral interaction region and modifying neutral observables such as divertor pressure \cite{Federici_2024, Zito_2026_PSI}. At the same time, the associated increase in power and particle fluxes to the divertor produces transient rises in target observables \cite{Zito_2026_PSI}: many-fold increases in electron density and temperature have been reported \cite{Flanagan_2025}, alongside experimentally-diagnosed increases in ion temperature \cite{DAMIZIA2025101912}. Such bursts would periodically enhance the plasma momentum flux towards the target, scaling approximately as $n_i c_s \sim n_i\sqrt{(T_e+T_i)/m_i}$. Although the plasma sheath might simultaneously be impacted \cite{Vasileska_2018, Vasileska_2019, Vasileska_2020, Tskhakaya_2010, Tskhakaya_2017}, these energetic bursts are expected to increase the time-averaged momentum available to drive toroidal neutral flows relative to conventional inter-ELM edge-plasma simulations.

    \item Finally, the simulations analysed in-depth here for ITER are \textit{predictive} models. These are known to face persistent challenges in simultaneously reproducing plasma and neutral observables when subjected to validation \cite{Zito_2025, Lore_2022, Wigram_2026, Wilcox_2026}. However, \textit{validated} cases\footnote{Specifically, \texttt{50701022\_2\_11MW\_D\_puff=1.7e22\_N\_puff=1e17} and \texttt{50701011\_2\_11MW\_D\_puff=1.7e22\_Ne\_puff=1e17} in \cite{kaveeva_2026_20325809}.} of Kaveeva \textit{et al.} \cite{KAVEEVA2021101030, kaveeva_2026_20325809} suggest that a quantitatively similar dynamics manifests in JET---where, however, no direct diagnosis of the toroidal wind was accomplished.
\end{itemize}

\section{Removal efficiency}\label{sec:philosophies__generalities_capture_efficiency}

The overall effectiveness of a given pumping setup can be quantified through the removal efficiency \cite{TEXTORTEAM1989115, Wenzel_2022, Wenzel_2024, Bader, Overskei_1981, Ohyabu_1992},
\begin{equation}\label{eq:etaCapture}
    \etaCapture =
    \frac{\Gamma^{\mathrm{pump}}}{I^{\mathrm{sat}}} ,
\end{equation}
where $I^{\mathrm{sat}}$~[s$^{-1}$] is the cumulative ion saturation current collected at the divertor targets, interpreted as a proxy for the total neutral source generated by recycling in the divertor. The quantity $\Gamma^{\mathrm{pump}}$~[s$^{-1}$] is the \textit{net} neutral particle flux crossing the pump inlet toward the pumping system, i.e. $\Gamma^{\mathrm{pump}} = Q / (k_{\mathrm{B}} T)$ where $T$~[K] is the neutral gas temperature in the vacuum duct.

This definition of $\etaCapture$ has limitations. Since $I^{\mathrm{sat}}$ varies with the degree of detachment, comparing cases at different divertor conditions does not necessarily provide a like-for-like assessment of exhaust performance. In the limit of complete detachment, $I^{\mathrm{sat}} \rightarrow 0$ would even imply $\etaCapture \rightarrow \infty$. This singular behaviour can be mitigated by alternative normalisations, for example by using $\Gamma^{\mathrm{pump}} + I^{\mathrm{sat}}$ in the denominator, or the definition in \cite{Loarer_1995}.

Nevertheless, such metrics still do not fully distinguish between increased pumping driven by higher source rates and genuine improvements in capture performance. For instance, increasing the deuterium puffing rate in steady-state may simultaneously increase $\Gamma^{\mathrm{pump}}$ and reduce $I^{\mathrm{sat}}$ through stronger detachment, thereby increasing $\etaCapture$ without necessarily indicating a more efficient exhaust mechanism---which, e.g., may still be plagued by high back-flow (section \ref{sec:methods__pop_exh__metrics}).

Despite the above, $\etaCapture$ remains useful for two reasons: it is experimentally measurable, and hence reported in several studies (section \ref{sec:philosophies__dfdp}); and it provides a practical basis for distinguishing pumping approaches when efficiency differs significantly between them.

Crucially, a high removal efficiency is desirable to avoid bottlenecks in the global particle balance, but it cannot be maximised in isolation. Particle exhaust must be traded against other equally-essential divertor functions, including power exhaust and erosion mitigation. The attainable removal efficiency is therefore a central but constrained figure of merit, whose value depends strongly on the underlying pumping philosophy (section \ref{sec:philosophies__synthesis}).


\section{Physical setup: additional settings}\label{apx:setup}

\subsection{Regions of interest}\label{apx:methods__pop_phys__geometry__rois}

Regions of interest (ROIs) are defined in the PoP-Phys domain (section \ref{sec:methods__pop_phys__geometry}) to extract meaningful macroscopic quantities while avoiding geometrical or symmetry-induced artefacts. In the PoP-Phys domain (figure \ref{fig:pop_phys_domain_sketch}):
\begin{itemize}
    \item ROIs for signed quantities, such as the degree of toroidality (DoT) and the toroidal velocity $\vTor$, are defined over the radial intervals $0 \leq r \leq 0.4\,L_r$ for the inner divertor volume and $0.6\,L_r \leq r \leq L_r$ for the outer divertor volume. In both cases, the vertical coordinate spans the full domain, $0 \leq z \leq L_z$. These ROIs exclude regions where $\vTor$ and DoT necessarily vanish and would otherwise bias the averaged quantities $\absavg{\DoT}$ and $\absavg{\vTor}$;
    
    \item Separate radial intervals, $0.15\,L_r \leq r \leq 0.35\,L_r$ and $0.65\,L_r \leq r \leq 0.85\,L_r$, are used to analyse transverse velocity profiles $\vTor(z)$ in regions where the flow is well behaved and approximately quasi-one-dimensional (appendix~\ref{apx:vphi_fitting});
    
    \item Positive-definite and smoothly varying scalar quantities, such as density and pressure, are instead averaged over the entire accessible domain unless otherwise specified.
\end{itemize}

For the PoP-Exh cases, ROIs are defined to enable direct comparison between the two capture strategies. The first ROI corresponds to the divertor PFR volume. In addition, rectangular ROIs are placed at the mid-points of the poloidal and toroidal ducts, as indicated by dashed boxes in figure \ref{fig:pop_exh_domain_sketch}. These allow a one-to-one comparison of averaged quantities at the same distance along each duct.

\subsection{Absence of traditional pumping}\label{apx:methods__pop_phys__geometry__no_pump}

No conventional pumping/absorbing boundary condition is applied on the wall across PoP-Phys simulations. This choice avoids preferential removal of poloidally-directed particles, which could otherwise artificially bias the results in favour of toroidal transport (section \ref{apx:evidence_limitations__simulations}). Moreover, introducing explicit pumping would add additional free parameters that would require systematic scanning \cite{Tantos_2024}, without being essential for the present objective.

Importantly, the dominant mechanism for poloidal neutral removal and recirculation is already captured through the plasma-sink-like boundary condition described in section \ref{sec:methods__pop_phys__interactions__plasma}. This mechanism removes a substantial fraction of impinging neutrals (of order $50\%$) and extends over the entire plasma-facing region of the divertor legs, whereas typical pump capture coefficients in EIRENE are of order of a few percent at most, and spatially-localised (section \ref{apx:evidence_limitations__simulations}). As a result, the primary competition between poloidal recirculation and toroidal transport is already represented in the present framework, and the omission of explicit pumping does not exclude any fundamental physical process relevant to the proof-of-principle investigation.

In practice, the resulting configuration can be interpreted as a plasma discharge after pre-fill in a metallic machine----with inactive fuelling and pumping \cite{Wigram_2026}.

\subsection{Domain simplification}\label{apx:methods__pop_phys__geometry__simplification}

To retain computational tractability, the SPARTA simulations presented in the PoP-Phys adopt a non-axisymmetric Cartesian formulation, i.e.\ a ``straightened torus'' approximation. In practice, this treats the divertor as toroidally-straight and infinite, rather than explicitly retaining toroidal curvature.

This approximation is justified by the moderate geometric curvature of the domain considered. The baseline simulated region emulates a divertor annulus of width $L_r \simeq 0.4~\mathrm{m}$ centred at a major radius $R_{\mathrm{div}} \simeq 1.45~\mathrm{m}$, representative of AUG-like conditions and corresponding to a relative curvature $L_r/R_{\mathrm{div}} \simeq 0.27$. Moreover, the neutral mean free path tends to be short compared with the radial extent of the domain for most simulations: for the nominal density, $\lambda \simeq 2.7~\mathrm{cm}$. At the lowest scanned densities, where the mean free path becomes longer, curvature effects may become more pronounced and should be assessed in future dedicated studies.

Overall, this approximation is therefore not expected to alter the central conclusions of the present study. Additionally, the SOLPS-ITER results discussed in section~\ref{sec:evidence__simulations} and in \cite{kaveeva_2026_20325809} include axisymmetric geometry and still exhibit strong toroidal neutral ordering, indicating that toroidal curvature does not suppress the effect. A dedicated quantitative assessment of axisymmetry effects is left to future work.

\subsection{A-posteriori realism check: global particle balance}\label{apx:particle_balance}

In the nominal PoP-physics geometry (section \ref{sec:methods__pop_phys__geometry}), an AUG-sized toroidal circumference is taken as $2\pi R_{\mathrm{div}}$, with $R_{\mathrm{div}}\simeq 1.45~\mathrm{m}$ \cite{zenodo_repo, Moscheni_2026}. The baseline configuration hence corresponds to a circulating, recycling-like particle flux $\GammaCirc = 3.3\times10^{23}\,\Dzero\,\mathrm{s^{-1}}$, removed in the plasma-like regions and re-emitted in the detached regions to sustain a volume-averaged neutral density of $10^{20}\,\mathrm{m^{-3}}$. Across the full PoP-physics database, $\GammaCirc$ spans $3.0\times10^{22}$--$5.8\times10^{24}\,\Dzero\,\mathrm{s^{-1}}$, reflecting both the two orders of magnitude variation in scanned neutral density and the increase in divertor volume up to near ITER-relevant dimensions.

In the present framework, where no active pumping is imposed, $\GammaCirc$ represents the effective recycling flux from plasma-facing components rather than the external fuelling rate alone. This distinction is important because, in detached conditions, recycling fluxes are expected to exceed gas puffing rates by up to approximately one order of magnitude \cite{PITTS2019100696, KAVEEVA2023101424}. This is consistent with AUG detached-regime fuelling levels, where puffing rates average $\Gamma^{\mathrm{puff}}=(2\pm1)\times10^{22}\,\mathrm{D/s}$, with detachment-onset values of $(6\pm2)\times10^{21}\,\mathrm{D/s}$ \cite{Moscheni_2026, Meschini_Moscheni_2026, zenodo_repo}, and reported maxima of approximately $5\times10^{22}\,\mathrm{D/s}$ \cite{Kallenbach_2015}. These values correctly lie roughly one order of magnitude below the baseline recycling flux $\GammaCirc$.

The upper end of the scanned database is likewise consistent with ITER-relevant conditions. ITER detached-scenario fuelling rates average $(1.3\pm1.0)\times10^{23}\,\mathrm{D/s}$ \cite{zenodo_repo, Moscheni_2026} and reach values up to $6\times10^{23}\,\mathrm{D/s}$ \cite{Lore_2022}. Applying the same order-of-magnitude separation between external puffing and recycling fluxes \cite{PITTS2019100696, KAVEEVA2023101424}, the maximum database value $\GammaCirc = 5.8\times10^{24}\,\Dzero\,\mathrm{s^{-1}}$ remains physically reasonable.

This consistency across device scales indicates that the imposed DSMC source terms correspond to realistic divertor particle throughputs, rather than artificially-inflated recycling levels.


\section{Numerics: DSMC methods and good practices}\label{apx:numerics}

\subsection{Particle-to-particle collision model}\label{apx:numerics__collisions}

In all simulations, DSMC numerical parameters were selected following standard Bird-type constraints to ensure accurate resolution of collisional momentum transport. Neutral--neutral collisions are modelled using the variable soft sphere (VSS) formulation \cite{Koura_1991}, with species-dependent parameters taken from \cite{Tantos_2024}, although variations of these parameters exist in the literature \cite{Tantos_2022, Varoutis_2024}. For atomic deuterium, a reference diameter $\sigma_\mathrm{D} = 0.281~\mathrm{nm}$ and viscosity index $\omega_\mathrm{D} = 0.74$ are used; for helium, $\sigma_{\mathrm{He}} = 0.260~\mathrm{nm}$ and $\omega_{\mathrm{He}} = 0.70$, all specified at a reference temperature of $1000~\mathrm{K}$. For D--He cross-collisions, the effective diameter and viscosity index are taken as the arithmetic averages of the corresponding single-species values.

\subsection{Numerical setup}\label{apx:numerics__setup}

Two-dimensional simulations are performed with a uniform spatial resolution of $\Delta x = \Delta y = 1~\mathrm{mm}$, following a hierarchical grid philosophy similar to that employed in DIVGAS \cite{Varoutis_2024}. This resolution corresponds to approximately one third of the minimum mean free path in the densest cases considered, $\lambda \simeq 2.7~\mathrm{mm}$ at $n = 10^{21}~\mathrm{m^{-3}}$ for D--D collisions, thereby satisfying the requirement $\Delta x \lesssim \lambda$ throughout the regions of greatest physical interest.

The time-step is fixed to $\Delta t = 10^{-8}~\mathrm{s}$, ensuring that: (i) the maximum particle displacement per time-step remains small compared to the cell size, with $v_{\max}\Delta t / \Delta x \lesssim 0.1$ for macroscopic velocities up to $15~\mathrm{km\,s^{-1}}$ and temperatures up to $5~\mathrm{eV}$; and (ii) that the time-step remains well below the local mean collision time, $\Delta t \ll \tau_c$, in all simulated regimes.

An average of approximately $50$ simulation particles per cell is targeted within the accessible flow volume. This choice ensures convergence of first-order velocity moments and allows accurate capture of momentum diffusion from imposed velocity boundary conditions. This criterion is satisfied throughout the domain in all simulations.

steady-state is identified from the stationarity of both the particle inventory and the source terms. Once this condition is reached, macroscopic quantities are averaged over $5\times10^{4}$ time-steps, excluding the initial transient. The resulting time traces of the quantities of interest, in particular $\vTor$ and pressure in representative regions, are required to remain stationary within relative fluctuations below $1\%$.

\section{Pressure dependence of the effective capture coefficient}
\label{apx:pressure_dependent_capture}

While appendix~\ref{sec:philosophies__generalities_capture_efficiency} discusses the global removal efficiency $\etaCapture$, the present section focuses instead on a local, boundary-condition-like capture coefficient defined from the gross atomic-neutral traffic through the PoP-Exh duct entrances. This can be inferred from the DSMC diagnostics of section~\ref{sec:results__pop_exh__backflow}, and a more comprehensive treatment is accomplished by Bonelli \textit{et al.} \cite{Bonelli_2017}.

The back-flow fraction, i.e.\ the effective duct albedo \cite{Tantos_2024}, is defined in equation~\eqref{eq:backflow_fraction}. The effective capture coefficient for atomic species $\alpha$ hence becomes
\begin{equation}
\xi(\alpha) = 1-f_{\rm back}(\alpha) .
\label{eq:capture_coefficient_from_backflow}
\end{equation}

For the two-dimensional duct scans of section~\ref{sec:results__pop_exh__database}, and table~\ref{tab:backflow_pressure_species} specifically, the atomic deuterium capture coefficients of the poloidal and toroidal inlets scale over $p=0.1$--$100~\mathrm{Pa}$ as
\begin{align}
\xi_{\theta}(\rm D) &\simeq 0.41 \, p^{0.12},
\\
\xi_{\phi}(\rm D) &\simeq 0.51 \, p^{0.08},
\end{align}
respectively, where $p$ is expressed in Pa. For atomic helium, over $p=0.01$--$10~\mathrm{Pa}$,
\begin{align}
\xi_{\theta}(\rm He) &\simeq 0.59 \, p^{0.14},
\\
\xi_{\phi}(\rm He) &\simeq 0.71 \, p^{0.09}.
\end{align}

The main implication is not the precise numerical value of the exponent, but the fact that \textit{the effective capture coefficient is not constant with pressure}. A single \textit{prescribed}---rather than DSMC-\textit{deduced} \cite{VAROUTIS201713}---pump capture coefficient therefore cannot simultaneously represent low- and high-pressure operation. This is precisely the type of dependence that dedicated sub-divertor DSMC calculations are designed to recover \cite{Tantos_2022,Tantos_2024}.

This observation is particularly relevant when the neutral or vacuum duct domain is truncated close to the divertor-facing region and replaced by a prescribed pump albedo, as is commonly done in predictive edge-code modelling \cite{Moscheni_2022,Moscheni_2025}. In such reduced representations, the feedback of fuel-puffing or divertor-pressure scans on the true particle return probability of the duct is absent: the boundary condition is held fixed. Retaining the sub-divertor region, or estimating its response through dedicated auxiliary simulations, would at least avoid this ambiguity \cite{Tantos_2022,Tantos_2024,VAROUTIS2019120,KAVEEVA2021101030,DEKEYSER2017899,Park_2018,Gao_2021,WU2023114023,Varoutis_2024, Zito_2025}.

Several limitations should be stressed. The coefficients above are specific to the present two-dimensional geometry and boundary conditions. They are expected to depend on the duct aspect ratio, wall accommodation model, helium concentration, and downstream boundary condition. The absolute values of $\xi$ should therefore \textit{not} be interpreted as universal pump coefficients. Rather, the power laws illustrate the magnitude of the pressure dependence that can be hidden by a fixed-albedo boundary condition.

\section{Vertical velocity profiles and two-exponential fitting model}
\label{apx:vphi_fitting}

To quantify how ordered toroidal motion relaxes across the private-flux-region cross-section, one-dimensional profiles of the toroidal velocity magnitude, $\left|\vTor\right|$, are extracted along vertical chords connecting the separatrix-side boundary to the dome-side boundary (as in figure \ref{fig:pop_phys_baseline__fit_slice}). For each chord, a local coordinate $\hat{z}$ is defined such that $\hat{z}=0$ lies on the separatrix side and $\hat{z}=H$ lies on the dome side, where $H$ is the chord length. In the PoP-Phys geometry, $H$ is fixed by construction, whereas in realistic divertor geometries it may vary from slice to slice, as illustrated in figure \ref{fig:vphi_slice}.

The profiles are fitted only in the radial regions of interest defined in appendix~\ref{apx:methods__pop_phys__geometry__rois}, where the vertical relaxation is sufficiently regular to be represented by a quasi-one-dimensional model. Although various functions may describe a rarefied Couette-like relaxation profile near the wall \cite{Roohi_2025couette}, we introduce here a compact empirical parametrisation of the entire profile convenient for our application. This is in the same spirit as profile-based characterisations routinely used in divertor plasma analyses \cite{Eich_2013, Brida_2025}.

\subsection{Fit ansatz}
\label{apx:vphi_fitting__ansatz}

Each profile is fitted over its available interval $0\leq\hat{z}\leq H$ using a superposition of two exponentials plus a constant offset,
\begin{equation}
\label{eq:twoexp_model_ABC}
\begin{split}
\left|\vTor(\hat{z};r)\right|
& = + A(r) \exp\!\left[-\frac{\hat{z}}{\lambdaPfr(r)} \right] \\
&\quad
- B(r) \exp\!\left[ -\frac{H(r)-\hat{z}}{\lambdaWall(r)} \right]\\
&\quad + \vTorInfty(r) .
\end{split}
\end{equation}
Here $\lambdaPfr$ is the characteristic decay length associated with the plasma-side drive, while $\lambdaWall$ characterises the influence of the dome-side momentum sink. The parameter $\vTorInfty$ is the interior offset, i.e. the fitted bulk toroidal velocity magnitude away from the immediate influence of either boundary.

A two-exponential form is used because the plasma-side drive and dome-side damping both affect the finite-width profile. A single-exponential fit would therefore generally return an effective decay length contaminated by the opposite boundary, rather than the separatrix-side influence length alone.

\subsection{Boundary-constrained parametrisation}
\label{apx:vphi_fitting__parameters}

Rather than fitting the amplitudes $A$ and $B$ independently, they are determined from physically-interpretable boundary values,
\begin{equation}
\label{eq:twoexp_boundary_values}
\begin{split}
& \left|\vTor(0;r)\right| = \vTorSep(r),\\    
& \left|\vTor(H;r)\right| = \vTorWall(r),\\
& |\vTorInfty(r)| = \text{bulk offset}.
\end{split}
\end{equation}
The independent fit parameters are therefore
\begin{equation}
\label{eq:twoexp_fit_parameters}
\left\{
\vTorSep,\,
\vTorWall,\,
\vTorInfty,\,
\lambdaPfr,\,
\lambdaWall
\right\}.
\end{equation}

Substituting equation~\eqref{eq:twoexp_boundary_values} into equation~\eqref{eq:twoexp_model_ABC} yields the following expressions
\begin{equation}
\label{eq:B_expression}
\begin{aligned}
B(r) = & \Bigl\{ \left[\vTorInfty-\vTorWall\right] + \\
& \left[\vTorSep-\vTorInfty\right] \exp\!\left[-H/\lambdaPfr\right] \Bigr\} \times \\
& \Bigl\{ 1 - \exp\!\Bigl[ -H \Bigl( \frac{1}{\lambdaPfr} + \frac{1}{\lambdaWall} \Bigr) \Bigr] \Bigr\}^{-1} \, ,
\end{aligned}
\end{equation}
and
\begin{equation}
\begin{aligned}
A(r) = & \left[\vTorSep-\vTorInfty\right] + \\
     &B(r) \exp\!\left[-H/\lambdaWall\right],
\end{aligned}
\end{equation}
with all parameter depending on $r$. Thus, $A$ and $B$ are derived quantities, while the fit itself is expressed in terms of boundary velocities and influence lengths.

This parametrisation reduces degeneracy between amplitudes and offset, enforces the boundary values by construction, and keeps the fitted quantities directly interpretable. 

In practice, the fit is performed on $\left|\vTor\right|$, and the ordering
\begin{equation}
\label{eq:twoexp_ordering}
\begin{split}
   & |\vTorSep| \geq |\vTorInfty| \geq |\vTorWall| \geq 0,\\
    & \lambdaPfr>0 \;\; \mathrm{and} \;\; \lambdaWall>0
\end{split}
\end{equation}
is imposed. Fits that violate these constraints, or for which the covariance matrix indicates poorly constrained parameters, are discarded.

\section{Preliminary assessment of rarefied flow past a TFP-like obstacle}
\label{apx:flow_past_obstacle}

A preliminary SPARTA calculation was performed to assess the downstream perturbation induced by a toroidally-localised obstacle representative of a TFP inlet. The calculation is carried out in the $(\phi,z)$ plane and is intended to estimate the flow reattachment length behind the obstacle. This is a relevant design quantity for ram-scoop-like intake concepts, where inlet wake and downstream recovery must be assessed explicitly~\cite{Almeida2021A}.

\begin{figure}
    \centering
    \subfloat[]{%
        \includegraphics[width=0.475\textwidth]{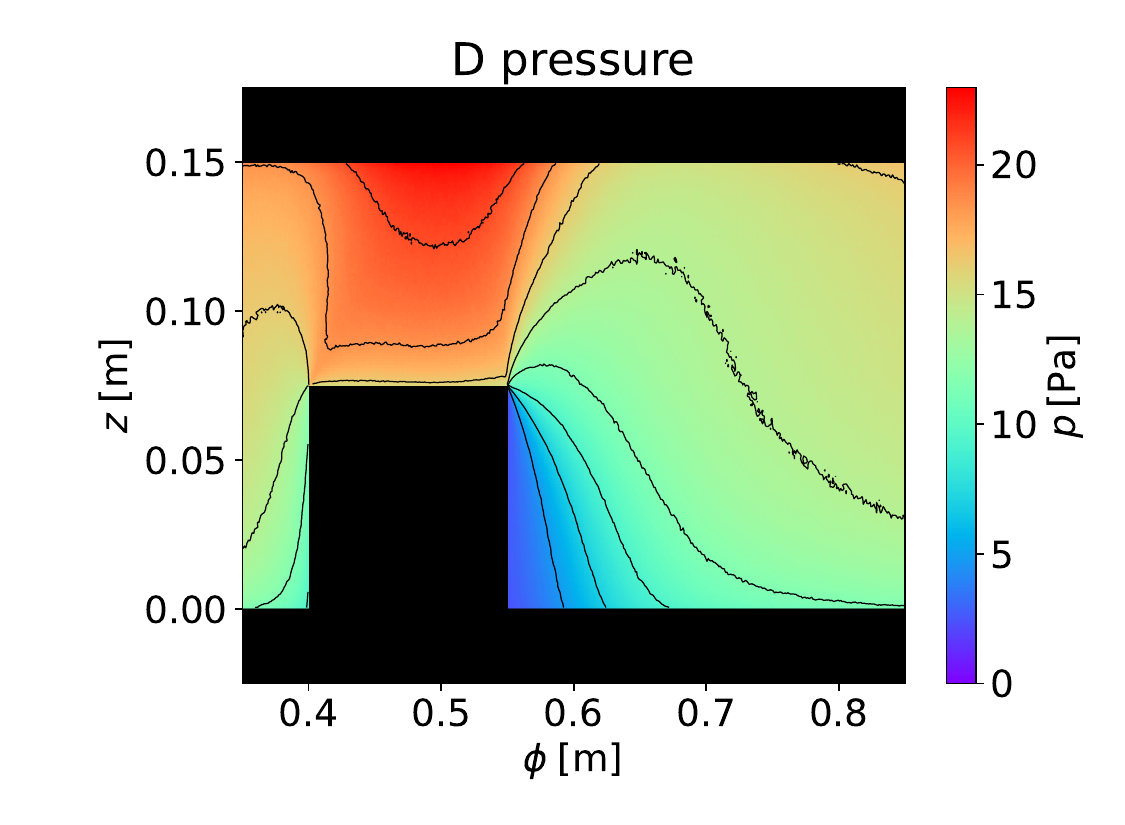}%
        \label{fig:flow_past_obstacle_p}}
    \\
    \subfloat[]{%
        \includegraphics[width=0.475\textwidth]{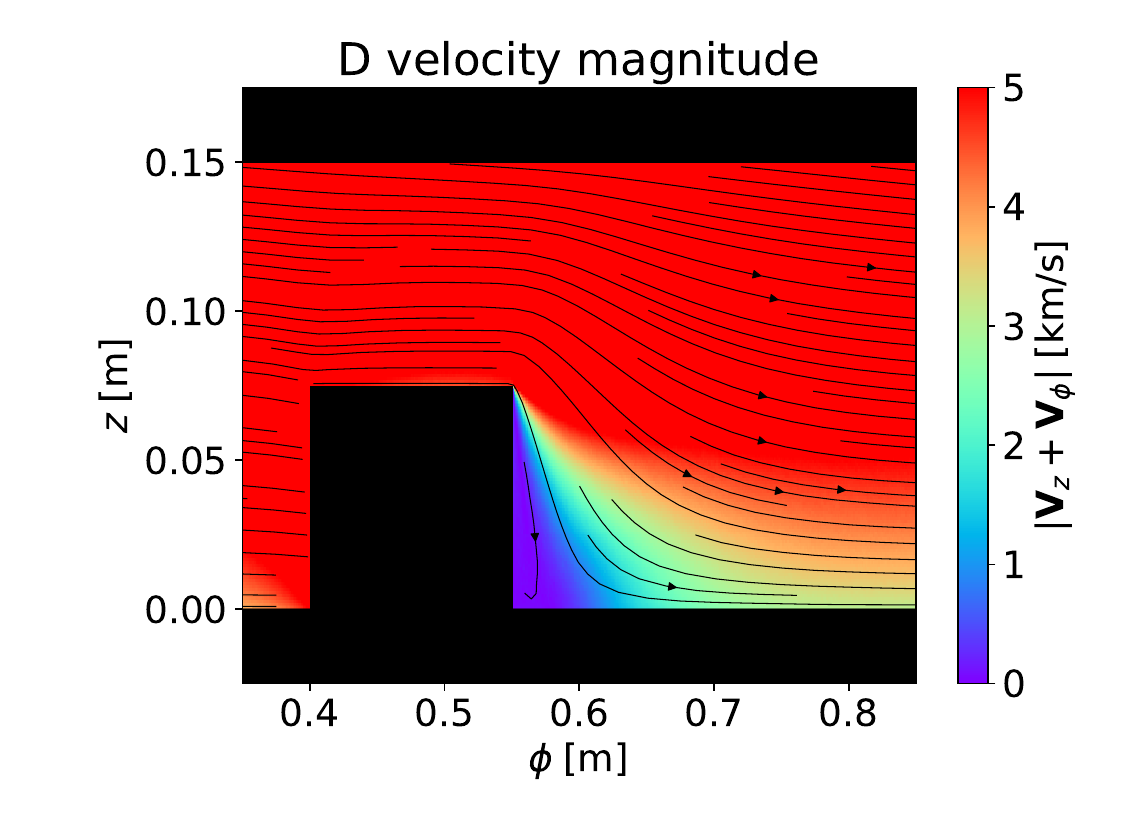}%
        \label{fig:flow_past_obstacle_vmag}}
    \\
    \subfloat[]{%
        \includegraphics[width=0.475\textwidth]{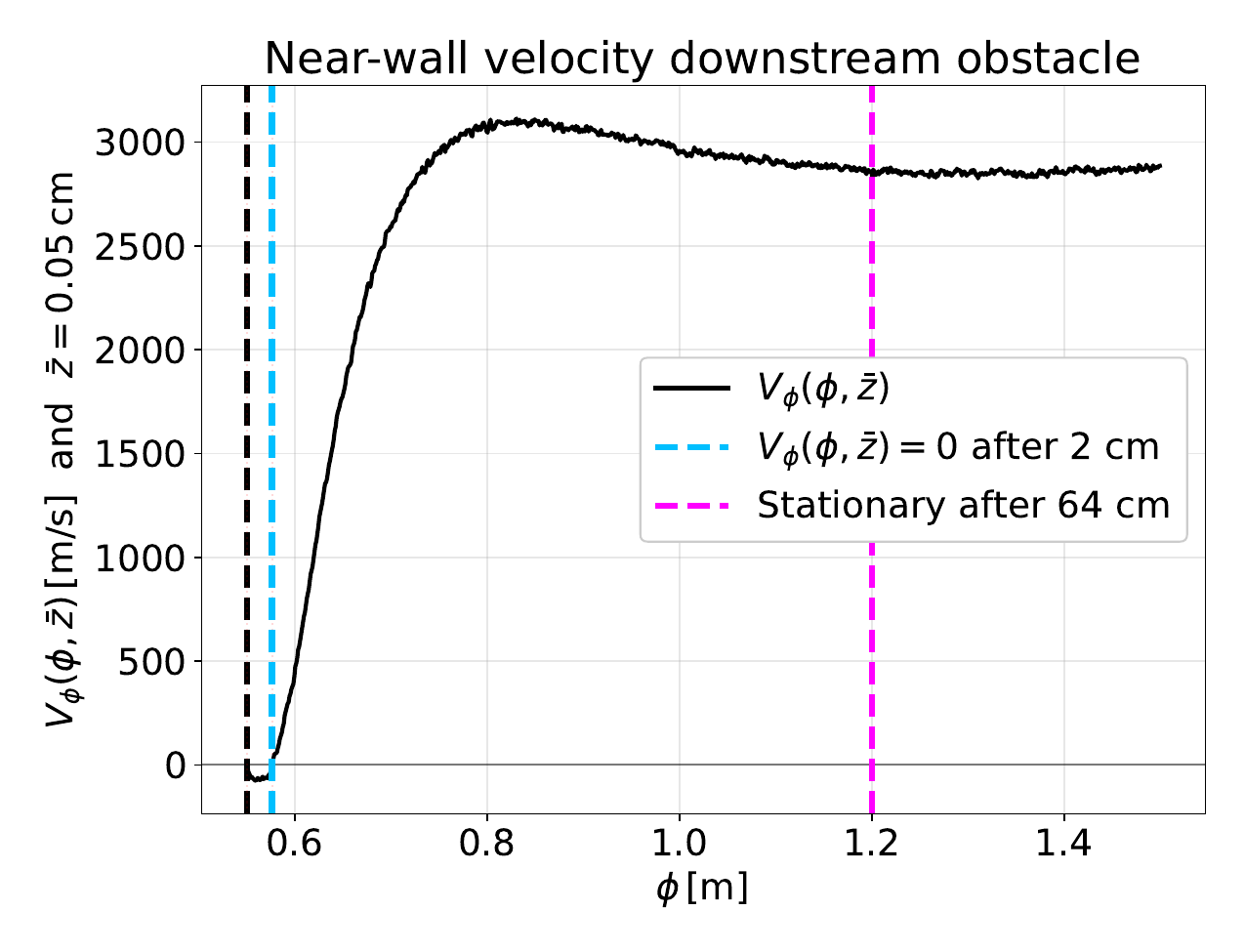}%
        \label{fig:flow_past_obstacle_vphi}}
    \caption{
    Preliminary two-dimensional SPARTA assessment of rarefied neutral flow past a TFP-like obstacle in the $(\phi,z)$ plane. 
    (a) Neutral pressure with iso-contours. 
    (b) Velocity magnitude with streamlines, used to identify the near-obstacle recirculation and reattachment region. 
    (c) Toroidal velocity, showing the downstream recovery of the near-wall flow.}
    \label{fig:flow_past_obstacle}
\end{figure}

The computational domain and resulting flow fields are shown in figure~\ref{fig:flow_past_obstacle}. The displayed region is a zoom of the full domain, which extends to periodic boundaries $0.4\,\mathrm{m}$ upstream and $0.95\,\mathrm{m}$ downstream of the obstacle. The top boundary forces a toroidal velocity $V_\phi=5\,\mathrm{km\,s^{-1}}$ from left to right.

The obstacle is rectangular, with dimensions $15\,\mathrm{cm}\times7.5\,\mathrm{cm}$, and its upstream-facing surface is treated as perfectly absorbing. This idealisation isolates the rarefied flow past the obstacle, while avoiding additional recirculation associated with imperfect absorption. Such effects may become important for realistic partially-transmitting inlets, but are beyond the scope of this preliminary test. The remaining wall and plasma-like boundary conditions are the same as those described in section~\ref{sec:methods__pop_phys}.

For the case shown in figure~\ref{fig:flow_past_obstacle}, the domain-averaged neutral density and pressure are $\langle n \rangle = (1.2\pm0.2)\times10^{20}\,\mathrm{m^{-3}}$ and $\langle p \rangle = 16\pm3\,\mathrm{Pa}$. The pressure distribution is shown in figure~\ref{fig:flow_past_obstacle_p}. The streamline pattern overlaid on the velocity-magnitude map, figure~\ref{fig:flow_past_obstacle_vmag}, indicates reattachment within a few centimetres downstream of the obstacle. Consistently, figure~\ref{fig:flow_past_obstacle_vphi} shows that the near-wall toroidal velocity changes sign and recovers to zero approximately $2\,\mathrm{cm}$ downstream of the obstacle trailing edge. The flow then relaxes toward its unperturbed upstream state over a longer distance, of order $64\,\mathrm{cm}$ downstream of the obstacle.

A similar recovery distance is obtained in a lower-density case with $\langle n\rangle\sim10^{19}\,\mathrm{m^{-3}}$ and $\langle p\rangle\sim2\,\mathrm{Pa}$. At higher density, $\langle n\rangle\sim10^{21}\,\mathrm{m^{-3}}$, the nominal reattachment length increases to approximately $6\,\mathrm{cm}$, while the distance required to recover the unperturbed state decreases to circa 40 cm.

These calculations are consistent with the results in figure 15a-b of Varade \textit{et al.} \cite{Varade_Agrawal_Pradeep_2014} and with \cite{Mahdavi_2022} (commented upon in section 5.12.2 of \cite{Roohi_2025couette})---where the reattachment length diminishes with increasing Kn---but are not intended as an optimised inlet design. In particular, fully three-dimensional simulations, finite transmission probabilities, and realistic surrounding divertor geometry remain to be assessed. Nevertheless, the present results indicate that the wake induced by a TFP-like obstacle can reattach over centimetric distances \cite{Varade_Agrawal_Pradeep_2014}, while the broader perturbation relaxes over sub-metre scales. This supports the plausibility of the toroidal inlet spacing estimated in section~\ref{sec:discussion__integration}. Moreover, staggered inlet layouts, rather than strictly toroidally-aligned apertures, could further reduce direct wake overlap and relax the constraint on the reattachment length. In a realistic divertor, radial curvature of the toroidal streamlines would also help decorrelate neighbouring downstream wakes and upstream plumes, so that coherent perturbations are expected to degrade faster than in the simplified planar calculation considered here.

\bibliographystyle{IEEEtran}
\bibliography{00_references}

\end{document}

%% file: 00_PoP_Phys_domain.tex
\begin{figure*}[t]
\centering
\resizebox{\textwidth}{!}{%
\begin{tikzpicture}[
    font=\large,
    >=Latex,
    line cap=round,
    line join=round,
    geom/.style={draw=black, line width=1.2pt},
    annot/.style={align=left},
    colA/.style={align=left, text width=5.8cm},
    colB/.style={align=left, text width=5.6cm}
]

\coordinate (X)  at (0,3.10);
\coordinate (LT) at (-4.10,1.25);
\coordinate (LB) at (-4.10,-0.95);
\coordinate (D)  at (0,0.90);
\coordinate (RT) at ( 4.10,1.25);
\coordinate (RB) at ( 4.10,-0.95);

\fill[gray!30] (LB) -- (LT) -- (X) -- (RT) -- (RB) -- (D) -- cycle;

\draw[geom] (LB) -- (LT);
\draw[geom, dashed] (LT) -- (X) -- (RT);
\draw[geom] (RT) -- (RB) -- (D) -- (LB);

\fill ($(LT)!0.20!(X)$) circle (2.4pt);
\fill ($(RT)!0.20!(X)$) circle (2.4pt);

\fill (LT) circle (2.4pt);
\fill (X)  circle (2.4pt);
\fill (RT) circle (2.4pt);

\node at (0,3.55) {X-point};

\node[align=center, text=blue] at (-3.00,2.65) {\textbf{Inner SEP leg}\\[0.2em]Plasma};
\node[align=center, text=red]  at ( 3.00,2.65) {\textbf{Outer SEP leg}\\[0.2em]Plasma};

\node[align=center] at (-5.10,0.20) {Inner\\target};
\node[align=center] at ( 5.10,0.20) {Outer\\target};

\node at (0,1.9) {PFR};
\node at (0,0.30) {Dome};


\node[anchor=south east, text=blue] at ($(LT)!0.22!(X)+(-0.45,-0.25)$) {Detached};
\node[anchor=south west, text=red]  at ($(RT)!0.22!(X)+( 0.45,-0.25)$) {Detached};

\coordinate (O) at (-0.95,-1.95);
\draw[->, thick] (O) -- ++(1.55,0) node[below right=-1pt and 2pt] {$r$};
\draw[->, thick] (O) -- ++(0,1.45) node[left] {$z$};
\node at (-1.22,-2.08) {$\phi$};

\node[colA, anchor=north west] at (6.20,3.45) {%
Inner/outer SEP legs:\\[0.5em]
$\bullet\;$ Temperature $\TplasmaBC$\\
$\bullet\;$ Toroidal speed $\vTorBC$\\
$\bullet\;$ Recycling probability $\BEIZ$\\
$\bullet\;$ Net recycling source $\GammaCirc$
};

\node[colB, anchor=north west] at (12.40,3.45) {%
Walls:\\[0.5em]
$\bullet\;$ Temperature $T_{\rm w}$\\
$\bullet\;$ TMAC $\alpha_t$\\
$\bullet\;$ NEAC $\alpha_n$\\
$\bullet\;$ 100\% reflection\\[0.5em]
$\bullet\;$ Width $L_r$\\
$\bullet\;$ Height $L_z$\\
$\bullet\;$ Equiv.\ diameter $d_h$
};

\end{tikzpicture}%
}
\caption{Schematic of the private-flux-region-like domain of the physics proof-of-principle simulations. Associated boundary conditions and main details are annotated. Only dashed boundaries can be crossed by the particles.}
\label{fig:pop_phys_domain_sketch}
\end{figure*}

%% file: 00_scaling_PoP_Phys_summary.tex
\begin{table*}[t]
    \centering
    \caption{Summary of nominal parameters and scanned ranges used in the 124 physics proof-of-principle simulations.}
    \label{tab:pop_phys_database}
    \begin{tabular}{lcccc}
        \toprule
        \textbf{Parameter} & \textbf{Symbol} & \textbf{Units} & \textbf{Baseline} & \textbf{Scan} \\
        \midrule
        \multicolumn{5}{l}{\textit{Deuterium}} \\
        \midrule
        
        Radial extent & $L_r$ & $\mathrm{m}$ & $0.40$ & $0.20$--$0.80$ \\
        
        Vertical extent & $L_z$ & $\mathrm{m}$ & $0.10$ & $0.05$--$0.60$ \\
        
        Temperature BC & $\TplasmaBC$ & $\mathrm{eV}$ & $1.5$ & $1$--$50$ \\
        
        Drift speed BC & $\vTorBC$ & $\mathrm{km\,s^{-1}}$ & $5$ & $0.75$--$20$ \\
        
        Neutral density & $\langle n \rangle$ & $\mathrm{m^{-3}}$ & $10^{20}$ & $10^{19}$--$10^{21}$\\

        Neutral pressure & $\langle p \rangle$ & $\mathrm{Pa}$ & 9 & 0.3--70\\

        Neutral temperature & $\langle T \rangle$ & $\mathrm{eV}$ & 0.6 & 0.3--2.6\\
        
        Knudsen number & $\langle \mathrm{Kn} \rangle$ & $-$ & $0.20$ & $0.02$--$1.95$ \\

        Recycling probability & $\BEIZ$ & $-$ & $0.50$ & $0.50$--$0.95$ \\
        
        Recycling flux & $\GammaCirc$ & $\mathrm{s^{-1}}$ & $3.3\times10^{23}$ & $1.5\times10^{22}$--$2.9\times10^{24}$ \\

        
        \bottomrule
    \end{tabular}
\end{table*}

%% file: 00_PoP_Exh_domain.tex
\begin{figure*}[t]
\centering
\begin{tikzpicture}[
    x=14cm,
    y=14cm,
    font=\small,
    >=Latex,
    line cap=round,
    line join=round,
    geom/.style={
        draw=black,
        line width=0.9pt,
        line cap=butt,
        line join=miter
    },
    lab/.style={
        align=center,
        inner sep=1pt
    },
    flowarrow/.style={
        ->,
        draw=black,
        line width=0.9pt
    },
    axisarrow/.style={
        ->,
        draw=black,
        line width=0.9pt
    },
    scalebar/.style={
        draw=black,
        line width=0.9pt
    },
    dashbox/.style={
        fill=magenta,
        fill opacity=0.25,
        draw=magenta,
        draw opacity=1,
        dotted,
        line width=0.9pt
    }
]


\fill[gray!30]
    (0.000,0.355) --
    (0.170,0.355) --
    (0.170,0.000) --
    (0.200,0.000) --
    (0.200,0.355) --
    (0.350,0.355) --
    (0.350,0.385) --
    (0.705,0.385) --
    (0.705,0.415) --
    (0.350,0.415) --
    (0.350,0.450) --
    (0.000,0.450) --
    cycle;

\draw[geom] (0.000,0.355) -- (0.000,0.450);

\draw[geom] (0.170,0.355) -- (0.170,0.000);
\draw[geom] (0.200,0.000) -- (0.200,0.355);

\draw[geom] (0.350,0.355) -- (0.350,0.385) -- (0.705,0.385);
\draw[geom] (0.705,0.415) -- (0.350,0.415) -- (0.350,0.450);

\draw[geom, dashed] (0.000,0.450) -- (0.350,0.450);
\draw[geom, dashed] (0.000,0.355) -- (0.170,0.355);
\draw[geom, dashed] (0.200,0.355) -- (0.350,0.355);

\def\dlone{0.015}
\def\dltwo{0.060}

\coordinate (Pmid) at ({0.5*(0.350+0.705)}, {0.5*(0.385+0.415)});
\begin{scope}[shift={(Pmid)}, rotate=90]
    \draw[dashbox]
        (-0.5*\dltwo,-0.5*\dlone)
        rectangle
        ( 0.5*\dltwo, 0.5*\dlone);
\end{scope}

\coordinate (Tmid) at ({0.5*(0.170+0.200)}, {0.5*(0.000+0.355)});
\begin{scope}[shift={(Tmid)}, rotate=0]
    \draw[dashbox]
        (-0.5*\dltwo,-0.5*\dlone)
        rectangle
        ( 0.5*\dltwo, 0.5*\dlone);
\end{scope}

\node[lab, anchor=east, align=right]  at (-0.005,0.402) {$\leftarrow$ to X-point\\(symmetry)};
\node[lab, anchor=south] at (0.175,0.455) {Inlet};

\node[lab] at (0.175,0.405) {PFR};

\node[lab, anchor=north] at (0.085,0.349) {Outlet};
\node[lab, anchor=north] at (0.275,0.349) {Outlet};

\node[lab, anchor=west] at (0.360,0.441) {Outer target};

\node[lab, align=left] at (0.535,0.337) {Poloidal\\duct ($\pPol$)};
\node[lab, anchor=west, align=left] at (0.225,0.175) {Toroidal\\duct ($\pTor$)};

\node[lab, anchor=west, align=left] at (0.225,0.020) {Perfect\\sink};
\node[lab, align=right] at (0.665,0.445) {Perfect\\sink};

\draw[flowarrow] (0.090,0.442) -- (0.090,0.372);
\draw[flowarrow] (0.265,0.442) -- (0.265,0.372);


\draw[flowarrow] (0.640,0.400) -- (0.700,0.400);

\draw[flowarrow] (0.185,0.070) -- (0.185,0.005);

\coordinate (O) at (0.470,0.125);

\draw[axisarrow] (O) -- (0.540,0.125);
\draw[axisarrow] (O) -- (0.470,0.195);

\node[lab, anchor=south east] at (0.460,0.121) {$z$};
\node[lab, anchor=west]       at (0.546,0.121) {$r$};
\node[lab, anchor=north]      at (0.470,0.225) {$\phi$};

\draw[scalebar] (0.425,0.055) -- (0.525,0.055);
\draw[scalebar] (0.425,0.050) -- (0.425,0.060);
\draw[scalebar] (0.525,0.050) -- (0.525,0.060);
\node[lab, anchor=north] at (0.475,0.048) {10 cm};

\end{tikzpicture}
\caption{Schematic of the domain of the exhaust proof-of-principle simulations. Associated boundary conditions and main details are annotated. Only dashed/open boundaries can be crossed by the particles. The reference sampling locations are in magenta.}
\label{fig:pop_exh_domain_sketch}
\end{figure*}

%% file: 00_scaling_PoP_Exh_summary.tex
\begin{table*}[t]
    \centering
    \caption{Summary of baseline parameters and scanned ranges used in the 90 exhaust proof-of-principle simulations, with indicative Kn number for the mixture.}
    \label{tab:pop_exh_database}
    \begin{tabular}{llccc}
        \toprule
        \textbf{Parameter} & \textbf{Symbol} & \textbf{Units} & \textbf{Baseline} & \textbf{Scan range} \\
        \midrule
        \multicolumn{5}{l}{\textit{Deuterium}} \\
        \midrule
        Neutral density & $\langle n \rangle$ & $\mathrm{m^{-3}}$ & $1.6\times 10^{20}$ & $3.4\times 10^{18}$--$6.2\times 10^{20}$ \\
        Pressure & $\langle p \rangle$ & $\mathrm{Pa}$ & 8.8 & $0.69$--$200$ \\
        Temperature & $\langle T \rangle$ & $\mathrm{eV}$ & 0.4 & $0.34$--$3.6$ \\
        Toroidal drift speed & $\langle |V_{\phi}| \rangle$ & $\mathrm{km\,s^{-1}}$ & $2.0$ & $1.1$--$4.7$ \\
        Knudsen number & $\langle \mathrm{Kn} \rangle$ & $-$ & $0.11$ & $0.03$--$5.8$ \\
        \midrule
        \multicolumn{5}{l}{\textit{Helium}} \\
        \midrule
        Neutral density & $\langle n \rangle$ & $\mathrm{m^{-3}}$ & $2.3\times 10^{19}$ & $5.7\times 10^{17}$--$9.8\times 10^{19}$ \\
        Pressure & $\langle p \rangle$ & $\mathrm{Pa}$ & $1.3$ & $0.11$--$32$ \\
        Temperature & $\langle T \rangle$ & $\mathrm{eV}$ & 0.4 & $0.35$--$3.7$ \\
        Toroidal drift speed & $\langle |V_{\phi}| \rangle$ & $\mathrm{km\,s^{-1}}$ & $1.8$ & $0.83$--$4.6$ \\
        Knudsen number & $\langle \mathrm{Kn} \rangle$ & $-$ & $0.10$ & $0.03$--$5.5$ \\
        \bottomrule
    \end{tabular}
\end{table*}